\documentclass[journal]{IEEEtran}
\usepackage{graphicx, graphics, times, amsfonts, amsmath, amssymb, cite}
\usepackage{amsthm}
\usepackage{tikz}
\usetikzlibrary{shapes,arrows}
\usepackage{color}
\usepackage{hyperref}
\usepackage{multirow}
\usepackage{rotating}
\usepackage{cleveref, subcaption}
\usepackage{bm}
\input{mysymbol.sty}
\usepackage[utf8]{inputenc}
\usepackage{enumitem}
\usepackage[ruled,linesnumbered]{algorithm2e}
\usepackage{booktabs}

\newtheorem{lemma}{Lemma}

\newtheorem{theorem}{Theorem}

\newtheorem{assumption}{Assumption}

\newcommand\relu{\mathrm{ReLU}}

\def \sqJacob {\boldsymbol{\mathcal{X}}}
\def \barsqJacob {\bar{\boldsymbol{\mathcal{X}}}}
\def \ccalbA {\boldsymbol{\mathcal{A}}}
\def \ccaltbA {\boldsymbol{\mathcal{\tilde{A}}}}
\def \ccalbH {\boldsymbol{\mathcal{H}}}
\def \ccaltbH {\boldsymbol{\mathcal{\tilde{H}}}}
\def \ccalbU {\boldsymbol{\mathcal{U}}}
\def \ccalbC {\boldsymbol{\mathcal{C}}}
\def \ccaltbC {\boldsymbol{\mathcal{\tilde{C}}}}
\renewcommand{\baselinestretch}{1.0}

\begin{document}

\title{Untrained Graph Neural Networks for Denoising}

\author{Samuel Rey,~\IEEEmembership{Student Member,~IEEE},
        Santiago Segarra,~\IEEEmembership{Senior Member,~IEEE}, 
        Reinhard Heckel,~\IEEEmembership{Member,~IEEE}, 
        and~Antonio G. Marques,~\IEEEmembership{Senior Member,~IEEE}
\thanks{S. Rey and A. G. Marques are with the Department
of Signal Theory and Comms., King Juan Carlos University, Madrid, Spain, \{samuel.rey, antonio.garcia.marques\}@urjc.es. S. Segarra is with the ECE Department, Rice University, Houston, USA, segarra@rice.edu. R. Heckel is with the ECE Department, Technical University of Munich, Munich, Germany,  reinhard.heckel@tum.de. Work in this paper was partially supported by the Spanish AEI Grants SPGRAPH (PID2019-105032GB-I00/AEI/10.13039/501100011033), FPU17/04520 and EST21/00420,  F661-MAPPING-UCI CAM-URJC, F663-AAGNCS CAM-URJC, F861 AUTO-BA-GRAPH CAM-URJC, and the USA NSF award CCF-2008555,
and the Institute of Advanced Studies at the Technical University of Munich, and the Deutsche Forschungsgemeinschaft (DFG, German Research Foundation) -
456465471, 464123524.}}

\maketitle

\begin{abstract}
A fundamental problem in signal processing is to denoise a signal.
While there are many well-performing methods for denoising signals defined on regular domains, including images defined on a two-dimensional pixel grid, many important classes of signals are defined over \emph{irregular} domains that can be conveniently represented by a graph.
This paper introduces two untrained graph neural network architectures for graph signal denoising, develops theoretical guarantees for their denoising capabilities in a simple setup, and
provides empirical evidence in more general scenarios.
The two architectures differ on how they incorporate the information encoded in the graph, with one relying on graph convolutions and the other employing graph upsampling operators based on hierarchical clustering. 
Each architecture implements a different prior over the targeted signals. 
Finally, we provide numerical experiments with synthetic and real datasets that i) asses the denoising behavior predicted by our theoretical results and ii) compare the denoising performance of our architectures with that of existing alternatives.
\end{abstract}

\begin{IEEEkeywords}
Geometric Deep Learning, Graph Decoder, Graph Signal Denoising, Graph Signal Processing 
\end{IEEEkeywords}

\section{Introduction}\label{S:intro}
\IEEEPARstart{V}{ast} amounts of data are generated and stored every day, propelling the deployment of data-driven solutions to address a wide variety of real-world problems.
Unfortunately, the input data suffers from imperfections and is corrupted with noise, oftentimes associated with the data-collection process.
Noisy signals appear in a gamut of applications, with examples including the processing of voice and images, the measurements in electric, sensor, social and transportation networks, or the monitoring of biological signals\cite{kolaczyk2014statistical,ortega2018graph,rey2019sampling}.
The presence of noise entails a detrimental influence on the quality of the data, which may become unusable when the noise power is comparable to that of the signal.
As a result, separating the signal from the noise, which is referred to as signal denoising, is a critical and ubiquitous task in contemporary data science applications.
While most existing works focus on the denoising of signals defined over regular domains (time and space), signals with irregular supports are becoming pervasive.
In particular, signals obtained from sensors deployed across different positions, such as voltage in power networks, temperature in weather stations, or neurological activity on the brain, have largely benefited from graph signal denoising since sensor measurements are typically corrupted by noise~{\cite{wu2020probabilistic}}.
Hence, designing (nonlinear) denoising schemes for signals defined over irregular domains constitutes a relevant problem both from a theoretical and practical point of view.

A versatile and tractable approach to overcome the challenges inherent to data supported on irregular domains consists of representing the underlying structure as a graph, with nodes representing variables and edges encoding levels of similarity, influence, or statistical dependence among nodes.
Successful examples of this approach can be found in the subareas of network analytics, machine learning over graphs, and graph signal processing (GSP) \cite{jordan1998learning, kolaczyk2014statistical, djuric2018cooperative}, with graph neural networks (GNNs) and GSP being particularly relevant for the architectures presented in this paper~\cite{shuman2013emerging,bronstein2017geometric,tenorio2021robust}.
Note that traditional data-processing architectures are designed to deal with data defined over regular domains, such as images, and hence, they may incur difficulties when learning and exploiting the more complex structure present in many contemporary applications.
Nonetheless, GSP provides a principled approach to handling this issue~\cite{shuman2013emerging,djuric2018cooperative,ortega2018graph}
Assuming that the structure of the signals can be modeled by a graph, GSP uses the information encoded in the graph topology to analyze, process, and learn from the data.
As a result, it is not surprising that GSP has been successfully applied to design and analyze GNNs \cite{bronstein2017geometric,scarselli2008graph,gama2018convolutional,wu2020comprehensive}, a class of neural network (NN) architectures that incorporate the graph topology information to enhance their performance when the data is composed of signals defined over a graph.

The importance of leveraging the graph influence when using deep nonlinear architectures is reflected in the wide range of GNNs that co-exist in the literature, including graph convolutional NNs (GCNNs)~\cite{sakhavi2018learning,kipf2016semi,li2018deeper}, graph recurrent NNs~\cite{cui2019traffic}, graph autoencoders~\cite{wang2017mgae,rey2019deep,rey2021overparametrized}, graph generative adversarial networks~\cite{wang2018graphgan,liu2019learning}, and simplicial NNs~\cite{schaub2021signal, roddenberry2021principled}, to name a few.
Incorporating the graph structure into deep nonlinear models involves a wide range of options when designing the architecture.
For example, GCNNs can be defined with or without pooling layers and the convolution over a graph can be implemented in several ways (vertex vs frequency), each leading to architectures with different properties and performances.
In fact, one of the key questions when designing a GNN is to decide the particular way in which the graph is incorporated into the architecture.

Considering the preceding paragraphs, the goal of this work is twofold.
First, we explore different ways of incorporating the information encoded in the graph and propose new graph-based NN architectures to denoise graph signals.
Second, we provide theoretical guarantees for the denoising capabilities of this approach and show that it is directly influenced by the properties of the graph.
The mathematical analysis, performed on particular instances of these architectures, provides guarantees on their denoising performance under specific assumptions for the original signal and its underlying graph.
In addition, we provide empirical evidence about the denoising performance of our method for scenarios more general  than those strictly covered by our theory, further illustrating the value of our graph-aware untrained architectures.

The proposed architectures are \textit{untrained} NNs, meaning that the parameters of the network are optimized using only the signal observation that we want to denoise, avoiding the dependency on a training set with multiple observed graph signals.
The underlying assumption behind this \textit{untrained} denoising architecture is that, due to the graph-specific structure incorporated into the different layers, when tuning the network parameters using stochastic gradient steps, the NNs are capable of learning (matching) the structure of the signal faster than that of the noise.
Hence, the denoising process is carried out separately for each individual observation by fitting the weights of the NN and stopping the updates after a few iterations.
This same phenomenon has been observed to hold true in non-graph deep learning architectures\cite{ulyanov2018deep,heckel2018deep} and constitutes a framework that is closely related to that of zero-shot learning~\cite{liu2018generalized,yaman2021zero}.
In the context of signal denoising, the consideration of an overparametrized graph-aware architecture along with early stopping avoids overfitting to the noise.

To incorporate the topology of the graph, the first architecture multiplies the input at each layer by a fixed (non-learnable) graph filter \cite{segarra2017optimal}, which can be seen as a generalization of the convolutional layer
in~\cite{kipf2016semi}.
The second architecture performs graph upsampling operations that, starting from a low-dimensional latent space, progressively increase the size of the input until it matches the size of the signal to denoise.
The sequence of upsampling operators are designed based on hierarchical clustering algorithms~\cite{jain_1988_algorithms, carlsson_2018_hierarchical, carlsson_2013_axiomatic, rey2019deep} so that, in contrast to~\cite{do2020graph}, matrix inversions are not required, avoiding the related numerical issues.
Our work is substantially different from~\cite{rey2019deep,rey2021overparametrized}, which deal with graph encoder-decoder architectures. 
On top of our theoretical analysis and extensive numerical simulations, additional differences to prior work are that: (a)~our graph decoder is an untrained network, and thus, it does not need a training phase; (b)~we only require a decoder-like architecture for denoising graph signals, so it is not necessary to jointly design and train two different architectures as carried out in, \cite{rey2019deep,rey2021overparametrized}. 

\vspace{1mm}
\noindent
{\bf Contributions and outline.}
In summary, the contributions of the paper are the following: (i)~we present two new overparametrized and untrained GNNs for solving graph-signal denoising problems; (ii)~mathematical analysis is conducted for each architecture offering bounds for their performance, improving our understanding of nonlinear architectures and the influence of incorporating graph structure into NNs; and (iii)~the proposed architectures are evaluated and compared to other denoising alternatives through numerical experiments carried out with synthetic and real-world data. These contrast with the contributions of our preliminary work in \cite{rey2019underparametrized}, which only considered a single underparametrized denoising architecture, did not provide mathematical analysis, and focused on synthetic datasets.
Moreover, moving to the overparametrized regime not only endows the proposed architectures with a larger learning capacity, but it also opens the door to a more thorough theoretical analysis.
 
The remainder of the paper is organized as follows. 
Sec.~\ref{S:related_works} reviews related works dealing with graph-signal denoising.
Sec.~\ref{S:fundamentals} explains fundamental concepts leveraged along the paper. Sec.~\ref{S:GNN_for_inverse_problems} formally introduces the problem at hand and presents our general approach.
Secs.~\ref{S:conv_dec} and~\ref{S:ups_dec} detail the proposed architectures and provide the mathematical analysis for each of them.
Numerical experiments are presented in Sec.~\ref{S:experiments} and concluding remarks are provided in Sec.~\ref{S:conclusion}. 

\subsection{Related works}\label{S:related_works}
Untrained NNs enable the recovery of signals without the need of training over large (or any) datasets by carefully incorporating prior information of the signals~\cite{ulyanov2018deep,mataev2019deepred,heckel2019denoising,heckel2018deep}.
In \cite{ulyanov2018deep}, it is shown that fitting a standard convolutional autoencoder to only one noisy signal using early stopping enables the effective denoising of an image.
For this approach to work, it is key that the signal class (images) matches the NN architecture (2D convolutional NN with particular filters).

Previous approaches to the graph-signal denoising task included a graph-regularization term that promoted desired properties on the estimated signals~\cite{chen2014signal}.
Some existing works minimize the graph total variation pushing the signal value at neighboring nodes to be close~\cite{chen2014signal,wang2015trend}.
Later on, total generalized variation extended this idea to promote similar values of higher-order terms~\cite{ono2015total}.
A related approach assumes that the signals are smooth on the graph and add a regularization parameter based on the quadratic form of the graph Laplacian~\cite{pang2017graph}.
Also, in \cite{onuki2016graph}, the authors propose a spectral graph trilateral filter as a regularizer, based on the prior assumption that the gradient is smooth over the graph.
It is worth noting that these alternatives rely on imposing some notion of smoothness on the original graph signal.
Furthermore, classical denoising methods typically assume that the signal and the graph are related by a linear or a quadratic mapping.
Nonetheless, the actual relation between the signal and the graph may be of a different nature and, in fact, in many relevant applications the actual prior is more complex than that represented by linear and quadratic terms, motivating the development of nonlinear models.

More recently, nonlinear solutions for denoising graph signals have been proposed to tackle the aforementioned issues.
In \cite{tay2020time}, a median graph filter~\cite{segarra2017designmedian} is used to denoise time-varying graph signals defined over dynamic graphs.
A different nonlinear approach is followed in \cite{do2020graph}, where a graph autoencoder is trained to recover the denoised signals.
To change the size of the graph, the autoencoder relies on Kron reduction operations \cite{dorfler2012kron}.
However, since the Kron reduction is based on the inverse of a submatrix of the graph Laplacian, it could fall into numerical issues if the submatrix is  singular.
Moreover, both architectures need several observations to recover the noiseless signals. 
Later on, \cite{chen2021graph} proposes a graph \textit{unrolling} architecture based on GCNNs to approach the denoising task.
The architecture is trained in an unsupervised fashion and relies on regularizing the objective function to avoid learning the noise.
Differently, our proposed solution implicitly encodes the regularization in the architectures enabling them to learn the signal faster than the noise.

\section{Processing architectures for graph signals}\label{S:fundamentals}
This section introduces mathematical notation and the fundamentals of GSP and GNNs.
In addition, the main notation is summarized in Table~\ref{T:symbols}.
Readers familiar with these concepts may give a quick pass and move on to Sec. III.

\begin{table}
\begin{center}
    \caption{{Summary of main notation.}}
    \label{T:symbols}
    \begin{tabular}{ l l }
        \textbf{Symbol} & \textbf{Explanation}  \\ \hline
        $\ccalG,N$ & Graph, number of nodes.  \\
        $\ccalV$, $\ccalE$ & {Set of nodes and edges}. \\
        $\bbA, \ccalbA \in \reals^{N \times N}$ & Adjacency matrix, expectation of $\bbA$. \\ 
        $\bbV\in \reals^{N \times N}$ & Matrix containing the eigenvectors of $\bbA$. \\
        $\bbLambda \in \reals^{N \times N}$ & Diagonal matrix collecting the eigenvalues of $\bbA$. \\
        \hline
        $F$, $\bbTheta$ & Number of features of the GNN, learnable \\
            & parameters. \\
        $\bbZ \in \reals^{N^{(0)}\times F^{(0)}}$ & Random input of the GNN. \\
        $\ccalT_{\bbTheta^{(\ell)}}^{(\ell)}\{\cdot|\ccalG\}$ & Linear graph-dependent transformation of the \\
            & GNN at layer $\ell$. \\
        $f_{\bbTheta}(\bbZ|\ccalG)$ & GNN architecture. \\
        \hline
        $\bbx, \bbn \in \reals^N$ & Noisy signal observation, noise vector. \\
        $\bbx_0, \hbx_0\in \reals^N$ & Original signal, denoised estimate.  \\\hline
        $\sqJacob\in\reals^{N \times N}$ & Expected squared Jacobian of $f_{\bbTheta}(\bbZ|\ccalG)$ with \\ &respect to $\bbTheta$. \\
        $\bbW\in\reals^{N \times N}$ & Matrix containing the eigenvectors of $\sqJacob$. \\
        $\bbSigma\in\reals^{N \times N}$ & Diagonal matrix containing the eigenvalues of $\sqJacob$. \\
        $\ccalM(\ccalbA)$ & Set of SBMs with expected adjacency matrix $\ccalbA$. \\
        $\ccalM_N(\beta_{min},\rho)$ & Set of SBMs with minimum expected degree \\ 
        & increasing with  $N$. \\ \hline
        $f_{\bbTheta}(\bbH)$, $f_{\bbTheta}(\bbU)$ & Two-layer GCG,  two-layer GDec. \\
        $\bbH\in\reals^{N\times N}$ & Graph filter. \\
        $\bbP^{(\ell)} \! \in \! \reals^{N^{(\ell)}\!\times\! N^{(\ell\!-\!1)}} $ & Membership matrix at layer $\ell$. \\
         $\bbU^{(\ell)} \! \in \! \reals^{N^{(\ell)}\!\times\! N^{(\ell\!-\!1)}} $ & Upsampling matrix at layer $\ell$. \\
        \hline
    \end{tabular}%
\end{center}
\vspace{-5mm}
\end{table}

\subsection{Fundamentals of GSP}
Let $\ccalG=(\ccalV, \ccalE)$ denote an undirected\footnote{Although our theoretical results assume that the graph is undirected, the proposed architectures can tackle signals defined on directed graphs~\cite{marques2020signal}.} graph, where $\ccalV$ is the set of $N$ nodes, and $\ccalE$ is the set of links such that $(i,j)$ belong to $\ccalE$ if nodes $i$ and $j$ are connected.
For a given graph $\ccalG$, the symmetric adjacency matrix $\bbA\in\reals^{N\times N}$ has non-zero entries $A_{ij}$ only if $(i,j)\in\ccalE$.
The value of $A_{ij}$ captures the strength of the link between nodes $i$ and $j$.
Define the degree matrix as $\bbD=\diag(\bbA\textbf{1})$, where $\textbf{1}$ is the vector of all ones and $\diag(\cdot)$ is the diagonal operator that turns a vector into a diagonal matrix.
A popular alternative to $\bbA$ is the degree normalized adjacency matrix $\tbA:=\bbD^{-\frac{1}{2}}\bbA \bbD^{-\frac{1}{2}}$.
Indeed, in the subsequent discussions, we assume that the rows and columns of $\bbA$ are normalized by its degree, so that $\bbA=\tbA$.
Finally, when the adjacency matrix is symmetric, we can write $\bbA=\bbV\bbLambda\bbV^\top$, where $\bbV$ is an orthonormal $N\times N$ matrix collecting the eigenvectors of $\bbA$ and $\bbLambda$ is diagonal matrix collecting its eigenvalues.

\vspace{1mm}
\noindent\textbf{Graph signals.} In this paper, we focus on the processing of graph signals which are defined on $\ccalV$. Graph signals can be represented as a vector $\bbx=[x_1,\ldots,x_N]^\top\in\reals^N$, where the $i$-th entry represents the value of the signal at node $i$.
Since the signal $\bbx$ is defined on $\ccalG$, the core assumption of GSP is that either the values or the properties of $\bbx$ depend on the topology of $\ccalG$~\cite{sandryhaila2013discrete}. For instance, consider a graph that encodes similarity. If the value of $A_{ij}$ is high, then one expects the signal values $x_i$ and $x_j$ to be akin to each other.
This rationale helps to explain the success of GNNs since the incorporation of $\ccalG$ into the architectures amounts to including prior information about the signals to process.

\vspace{1mm}
\noindent\textbf{Graph filtering.}
{Graph filters, an important tool of GSP, play a fundamental role in the definition of our GNN architectures.}
Graph filters are linear operators $\reals^N \rightarrow \reals^N$ that can be expressed as a polynomial of the adjacency matrix of the form
\begin{equation}\label{eq:graph_filter}
    \bbH:=\sum_{m=0}^{M-1}h_m\bbA^m,
\end{equation}
where $\bbH$ is the graph filter, $h_m$ are the filter coefficients, and $M\leq N$~\cite{segarra2017optimal}.
Since $\bbA^m$ encodes the $m$-hop neighborhoods of the graph, graph filters can be used to diffuse input graph signals $\bbx$ across the graph as $\bby=\sum_{m=0}^{M-1}h_m\bbA^m\bbx=\bbH\bbx$.
Because graph filters diffuse signals across $(M\!-\!1)$-hop neighborhoods, they are widely used to generalize the convolution operation to signals defined over graphs.

\vspace{1mm}
\noindent\textbf{Frequency representation.}
{The theoretical analysis developed in this paper leverages the notion of bandlimited graph signals, a widely-used definition that links the properties of a signal to those of the (spectrum  of) the supporting graph \cite{shuman2013emerging}. 
To be specific, the frequency representation of the signal $\bbx$ is given by the $N$-dimensional vector $\tbx=\bbV^\top\bbx$, with $\bbV^\top$ acting as the graph Fourier transform (GFT)\cite{sandryhaila2014discrete}. Then, a graph signal is said to be bandlimited if $\tbx$ satisfies that $\tilde{x}_k=0$ for $k>K$, where $K\leq N$ is referred to as the bandwidth of the signal $\bbx$.}
If $\bbx$ is bandlimited with bandwidth $K$ it holds that
\begin{equation}\label{E:bl_signals}
    \bbx=\bbV_K\tbx_K,
\end{equation}
with $\tbx_K=[\tilde{x}_1,\cdots,\tilde{x}_K]$ collecting the active frequency components and $\bbV_K$ collecting the corresponding $K$ eigenvectors.
This reduced-dimensionality representation, which can be generalized to graph filters as well, has been shown to bear practical relevance in real-world datasets and it is exploited in denoising and other inverse problems~\cite{chen2015discrete}.

\subsection{Fundamentals of GNNs}
Generically, we represent a GNN using a parametric nonlinear function $f_{\bbTheta}(\bbZ|\ccalG):\reals^{N^{(0)} \times F^{(0)}} \rightarrow \reals^N$ that depends on the graph $\ccalG$. 
The parameters of the architecture are collected in $\bbTheta$, and the matrix $\bbZ \in \reals^{N^{(0)} \times F^{(0)}}$ represents the input of the network.
{Despite the many possibilities for defining a GNN, a broad range of such architectures recursively apply a graph-aware linear transformation followed by an entry-wise nonlinearity.}
Then, a generic deep architecture $f_{\bbTheta}(\bbZ|\ccalG)$ with $L$ layers can be described as
\begin{align}
    \hbY^{(\ell)}&=\ccalT_{\bbTheta^{(\ell)}}^{(\ell)}\left\{ \bbY^{(\ell-1)}|\ccalG\right\}, \;\; 1 \leq\ell\leq L,  \label{eq:graph_aware_lin_trans_generic_GNN} \\
    Y^{(\ell)}_{ij}&=g^{(\ell)}\left( \hat{Y}^{(\ell)}_{ij} \right),  \;\; 1 \leq\ell\leq L,  \label{eq:nonlinear_trans} 
\end{align}
where $\bbY^{(0)}=\bbZ$ and $\bby=\bbY^{(L)}$ denote the input and output of the architecture, $\ccalT^{(\ell)}_{\bbTheta^{(\ell)}}\{\cdot|\ccalG\}\colon \reals^{N^{(\ell-1)} \times F^{(\ell-1)}} \rightarrow \reals^{N^{(\ell)}\times F^{(\ell)}}$ is a \textit{graph-aware} linear transformation, $\bbTheta^{(\ell)} \in \reals^{F^{(\ell-1)} \times F^{(\ell)}}$ are the parameters that define such a transformation, and $g^{(\ell)}\colon \reals\rightarrow\reals$ is a \textit{scalar} nonlinear transformation (e.g., the ReLU function), which is oftentimes omitted in the last layer.
Moreover, $N^{(\ell)}$ and $F^{(\ell)}$ represent the number of nodes and features at layer $\ell$, $\bbTheta=\{\bbTheta^{(\ell)}\}_{\ell=1}^L$ collects all the parameters of the architecture, and $\bby$ is the output of the GNN.
Note that although $f_{\bbTheta}(\bbZ|\ccalG)$ generates output signals defined in $\reals^N$, which is the case of interest for this paper, it can be easily adapted to output graph signals with more than one feature.

\section{GNNs for graph-signal denoising}\label{S:GNN_for_inverse_problems}

We now formally introduce the problem of graph-signal denoising within the GSP framework, and present our approach to tackle it using untrained GNN architectures.
Given the graph $\ccalG$, let us consider the observed graph signal $\bbx \in \reals^N$, which is a noisy version of the original graph signal $\bbx_0$. With $\bbn\in \reals^N$ being a noise vector, the relation between $\bbx$ and $\bbx_0$ is
\begin{equation}\label{eq:noise_model}
    \bbx = \bbx_0 + \bbn.
\end{equation}
Then, the goal of graph-signal denoising is to remove as much noise as possible from the observed signal $\bbx$ to estimate the original signal $\bbx_0$, which is performed by exploiting the information encoded in $\ccalG$.

A traditional approach for the graph-signal denoising task is to solve an optimization problem of the form
\begin{alignat}{2}\label{eq:denoising_reg}
    \!\!&\! \hbx_0 = \mathrm{arg min}_{\check\bbx_0} \
    && \|\bbx-\check\bbx_0\|_2^2 + \alpha R(\check\bbx_0|\ccalG).
\end{alignat}
The first term promotes fidelity to the signal observations, the regularizer $R(\cdot|\ccalG)$ promotes denoised signals with desirable properties over the given graph $\ccalG$, and $\alpha>0$ controls the influence of the regularization.
Common choices for the regularizer include the quadratic Laplacian $R(\bbx|\ccalG)=\bbx^\top\bbL\bbx$~\cite{pang2017graph},
or regularizers involving high-pass graph filters $R(\bbx|\ccalG)=\|\bbH\bbx\|_2^2$ that foster smoothness on the estimated signal\cite{sandryhaila2014discrete,chen2014signal}.

\begin{algorithm}[tb]
\SetKwInOut{Input}{Inputs}
\SetKwInOut{Output}{Outputs}
\Input{$\bbx$ and $\ccalG$}
\Output{$\hbx_0$ and $\hbTheta(\bbx)$}
\SetAlgoLined
Set $f_{\bbTheta}(\bbZ|\ccalG)$ as explained in Sec. \ref{S:conv_dec} or \ref{S:ups_dec} \\
Generate $\bbZ$ from iid zero-mean Gaussian distribution \\
Initialize $\bbTheta_{(0)}$ from iid zero-mean Gaussian \\
\For{$t=1$ \KwTo $T$}{
 Update $\bbTheta_{(t)}$ minimizing \eqref{E:nonlinear_denoising} with SGD
}
$\hbTheta(\bbx)=\bbTheta_{(T)}$ \\
$\hbx_0=f_{\hbTheta(\bbx)}(\bbZ|\ccalG)$ \\
\caption{Proposed graph-signal denoising method}
\label{A:denoising}
\end{algorithm}

While those traditional approaches exhibit a number of advantages (including interpretability, mathematical tractability, and convexity),
they may fail to capture more complex relations between $\ccalG$ and $\bbx_0$, motivating the development of nonlinear graph-denoising approaches. 

As summarized in Algorithm~\ref{A:denoising}, in this paper we advocate handling the graph-signal denoising task by employing an overparametrized GNN (denoted by $f_{\bbTheta}(\bbZ|\ccalG)$) as described in \eqref{eq:graph_aware_lin_trans_generic_GNN}-\eqref{eq:nonlinear_trans}.
The weights of the architecture, collected in $\bbTheta$, are learned by minimizing the loss function
\begin{align}
\label{E:nonlinear_denoising}
\ccalL(\bbx,\bbTheta) = \frac{1}{2}\|\bbx-f_{\bbTheta}(\bbZ|\ccalG)\|_2^2,
\end{align}
applying stochastic gradient descent (SGD) in combination with early stopping to avoid overfitting the noise.
The entries of the parameters $\bbTheta$ and the input matrix $\bbZ$ are initialized at random using an iid zero-mean Gaussian distributions, and the weights learned after a few iterations of denoising the observation $\bbx$ are denoted as $\hbTheta(\bbx)$.
Note that $\bbZ$ is fixed to its random initialization.
Finally, the denoised graph signal estimate is computed as 
\begin{equation}
    \hbx_0=f_{\hbTheta(\bbx)}(\bbZ|\ccalG).    
\end{equation}

The intuition behind this approach is as follows: since the architecture is overparametrized it can in principle fit any signal, including noise.
However, as shown formally later, both empirically and theoretically, the proposed architectures fit graph signals faster than the noise, and therefore with early stopping they fit most of the signal and little of the noise, enabling signal denoising.

{
\vspace{1mm}\noindent
\textbf{Remark 1.} 
    The proposed architectures are described as \textit{untrained} NNs because, when minimizing \eqref{E:nonlinear_denoising}, the weights in $\bbTheta$ are learned to fit \emph{each observation} $\bbx$, with the denoised signal $\hbx_0$ being the output for those particular weights.
    This implies that each noisy-denoised signal pair $(\bbx,\hbx_0)$ is associated with a particular value of the weights $\bbTheta$, in contrasts with trainable NNs, where the weights $\bbTheta$ are first learned by fitting the signals in a \emph{training set} and later used (unchanged) to denoise signals that were not in the training set. }

Regarding the specific implementation of the untrained network $f_{\bbTheta}(\bbZ|\ccalG)$, there are multiple possibilities for selecting the linear and nonlinear transformations $\ccalT^{(\ell)}_{\bbTheta^{(\ell)}}$ and $g^{(\ell)}$ defined in equations~\eqref{eq:graph_aware_lin_trans_generic_GNN} and~\eqref{eq:nonlinear_trans}, respectively.
{As is customary} in NNs dealing with signals defined in $\reals^N$, we select the $\relu$ operator, defined as $\relu(x)=\max(0,x)$, to be the entrywise nonlinearity $g^{(\ell)}$.
Then, we focus on the design of the linear transformation, which is responsible for incorporating the structure of the graph.
The two following sections postulate the implementation of two particular linear transformations $\ccalT^{(\ell)}_{\bbTheta^{(\ell)}}$ (each giving rise to a different GNN) and analyze the resulting architectures.

\section{Graph convolutional generator}\label{S:conv_dec}
Our first architecture to address the graph-signal denoising task is a graph-convolutional generator (GCG) network that incorporates the topology of the graph into the NN pipeline via vertex-based graph convolutions.
Then, leveraging the fact that convolutions of a graph signal on the vertex domain can be represented by a graph filter $\bbH \in \reals^{N \times N}$ \cite{segarra2017optimal}, we define the linear transformation for the convolutional generator as 
\begin{equation}\label{E:linear_trans_gcg}
      \ccalT^{(\ell)}_{\bbTheta^{(\ell)}}\{\bbY^{(\ell-1)}|\ccalG\} = \bbH\bbY^{(\ell-1)}\bbTheta^{(\ell)}.
\end{equation}
Remember that the $F^{(\ell-1)} \times F^{(\ell)}$ matrix $\bbTheta^{(\ell)}$ collects the learnable weights of the $\ell$-th layer, and the graph filter $\bbH$ is given by \eqref{eq:graph_filter}. {The coefficients $\{h_m\}_{m=0}^{M-1}$ are fixed a priori so that $\bbH$ promotes desired properties on the estimated signal.}
Using the linear transformation defined in \eqref{E:linear_trans_gcg}, the output of the GCG with $L$ layers is given by the recursion
\begin{align}
      \bbY^{(\ell)} &=\relu(\bbH\bbY^{(\ell-1)}\bbTheta^{(\ell)}),\;\; \mathrm{for}\;\; \ell=1,...,L-1, \label{E:gcg1}\\ 
      \bby^{(L)} &= \bbH\bbY^{(L-1)}\bbTheta^{(L)}, \label{E:gcg2}
\end{align}
where $\bbY^{(0)}=\bbZ$ {denotes the random input} and the $\relu$ is not applied in the last layer of the architecture.
With the proposed linear transformation, the GCG learns to combine the features within each node by fitting the weights of the matrices $\bbTheta^{(\ell)}$ while the graph filter $\bbH$ interpolates the signal by mixing features from $M-1$ neighborhoods.

Even though the proposed GCG exploits graph convolutions to incorporate the graph topology into the architecture, it is intrinsically different from other GCNNs.
The linear transformation proposed in \cite{kipf2016semi}, arguably one of the most popular implementations of GCNNs, is given by
\begin{equation}\label{E:gcnn_kipf}
    \ccalT^{(\ell)}_{\bbTheta^{(\ell)}}\{\bbY^{(\ell-1)}|\ccalG\} = (\bbA+\bbI)\bbY^{(\ell-1)}\bbTheta^{(\ell)}.
\end{equation}
Recalling the definition of graph filters in \eqref{eq:graph_filter}, it is evident that \eqref{E:gcnn_kipf} is a particular case of our proposed linear transformation, obtained by setting the generative graph filter to $\bbH=\bbA+\bbI$, a low-pass graph filter of degree one.
In addition to representing a more general scenario, \eqref{E:gcg1} endows the GCG with two main advantages.
First, the graph filter $\bbH$ allows us to incorporate prior information on the signals to denoise, making our GCG architecture more suitable to denoise a (high-) low-frequency signal by employing a (high-) low-pass filter.
Second, in \eqref{E:gcnn_kipf} there is an equivalence between the depth of the network and the radius of the considered neighborhood, so that gathering information from nodes that are $M$ hops apart requires a GNN with $M$ layers. In contrast, with the architecture considered in \eqref{E:gcg1}, the same can be achieved by considering a GCG with $L$ layers and a graph filter $\bbH$
of degree $M/L$~\cite{segarra2017optimal}, reducing the number of learnable parameters and bypassing some of the well-known over-smoothing problems associated with \eqref{E:gcnn_kipf}~\cite{chen2020measuring}.

Next, we adopt some simplifying assumptions to provide theoretical guarantees on the denoising capability of the GCG (Sec.~\ref{S:analysis_GCG}). Then, we rely on numerical tests to demonstrate that the results also hold in more general settings (Sec.~\ref{S:analyze_deep_gcg}).

\subsection{Guaranteed denoising with the GCG}\label{S:analysis_GCG}
To formally prove that the proposed architecture can successfully denoise the observed graph signal $\bbx$, we consider a two-layer GCG given by
\begin{equation}\label{eq:2layer_gen}
    f_{\bbTheta}(\bbZ|\ccalG) = \relu(\bbH\bbZ\bbTheta^{(1)})\bbtheta^{(2)},
\end{equation}
where $\bbTheta^{(1)}\in\reals^{F\times F}$ and $\bbtheta^{(2)}\in\reals^{F}$ are the learnable coefficients. With $F$ denoting the number of features, we consider the overparametrized regime where $F \geq 2N$, and analyze the behavior and performance of denoising with the untrained network defined in~\eqref{eq:2layer_gen}. 

We start by noting that scaling the $i$-th entry of $\bbtheta^{(2)}$ is equivalent to scaling the $i$-th column of $\bbTheta^{(1)}$, so that, without loss of generality, we can set the weights to $\bbtheta^{(2)}=\bbb$, where $\bbb$ is a vector of size $F$ with half of its entries set to $1/\sqrt{F}$ and the other half to $-1/\sqrt{F}$.
Furthermore, since $\bbZ$ is a random matrix of dimension $N \times F$, the column space of $\bbZ$ spans $ \reals^N$, and hence, minimizing over $\bbZ\bbTheta^{(1)}$ is equivalent to minimizing over $\bbTheta\in\reals^{N\times F}$.
With these considerations in place, the optimization over \eqref{E:nonlinear_denoising} can be performed replacing the two-layer GCG described in \eqref{eq:2layer_gen} by its simplified form
\begin{equation}\label{E:2layer_gcg_simp}
    f_{\bbTheta}(\bbH) = f_{\bbTheta}(\bbZ|\ccalG) = \relu(\bbH\bbTheta)\bbb.
\end{equation}
Note that we replaced $f_{\bbTheta}(\bbZ|\ccalG)$ with  $f_{\bbTheta}(\bbH)$ since the graph influence is modeled by the graph filter $\bbH$, and the influence of the matrix $\bbZ$ is absorbed by the learnable weights $\bbTheta$.
{Also note that the behavior of the optimization algorithm of \eqref{eq:2layer_gen} and \eqref{E:2layer_gcg_simp} may differ and the upcoming theoretical analysis is focused on the latter case.}

The denoising capability of the two-layer architecture is related to the eigendecomposition of its expected squared Jacobian\cite{heckel2019denoising}.
However, to understand which signals can be effectively denoised with the proposed architecture, we need to connect the spectral domain of the expected squared Jacobian with the spectrum of the graph, given by the eigenvectors of the adjacency matrix.

To that end, we next compute the expected squared Jacobian of the two-layer architecture in \eqref{E:2layer_gcg_simp}.
Denote as $\ccalJ_{\bbTheta}(\bbH)\in~\reals^{N \times NF}$ the Jacobian matrix of $f_{\bbTheta}(\bbH)$  with respect to $\bbTheta$, which is given by 
\begin{align}\label{E:jacobian_gcg}
    \ccalJ^\top_{\bbTheta}(\bbH)
    =
    \begin{bmatrix}
    \mathrm{b}_1 \bbH^\top \diag( \relu'(\bbH \bbtheta_1) ) \\
    \vdots \\
    \mathrm{b}_F \bbH^\top \diag( \relu'(\bbH \bbtheta_F) ) \\
    \end{bmatrix}
    \in \reals^{NF \times N},
\end{align}
where $\bbtheta_i$ represents the $i$-th column of $\bbTheta$, and $\relu'$ is the derivative of the $\relu$, which is the Heaviside step function.
Then, define the $N\times N$ expected squared Jacobian matrix as
\begin{align}\label{eq:sqJac_step1}
    \sqJacob&:=\mathbb{E}_{\bbTheta}[\ccalJ_{\bbTheta}(\bbH)\ccalJ^\top_{\bbTheta}(\bbH)] \nonumber \\ 
    &= {\sum_{i=1}^F b_i^2 \mathbb{E}\left[\relu'(\bbH \bbtheta_i)\relu'(\bbH\bbtheta_i)^\top\right] \odot \bbH\bbH^\top. }
\end{align}
Moreover, from the work in \cite[Sec. 3.2]{daniely2016toward}, we note that $\mathbb{E}\left[\relu'(\bbH \bbtheta_i)\relu'(\bbH\bbtheta_i)^\top\right]$ is in fact the so-called dual activation of the step function.
Therefore, combining the expression for the dual activation of the step function from \cite[Table~1]{daniely2016toward} with \eqref{eq:sqJac_step1}, we obtain that
%
\begin{equation}\label{eq:expected_jac}
    \sqJacob = 0.5 \left( \mathbf{1} \mathbf{1}^\top - {\pi}^{-1} \arccos(\bbC^{-1} \bbH^2 \bbC^{-1})\right) \odot \bbH \bbH^\top,
\end{equation}
where $\odot$ represents the Hadamard (entry-wise) product, $\arccos(\cdot)$ is computed entry-wise, $\bbh_i$ represents the $i$-th column (row) of $\bbH$, $\bbC=\diag([\|\bbh_1\|_2,...,\|\bbh_N\|_2])$ is a normalization term so that $\bbC^{-1} \bbH^2 \bbC^{-1}$ is the autocorrelation of the graph filter $\bbH$.

Since $\sqJacob$ is symmetric and positive (semi) definite, it has an eigendecomposition $\sqJacob=\bbW\bbSigma\bbW^\top$.
Here, the columns of the orthonormal matrix $\bbW=[\bbw_1,\ldots,\bbw_N]$ are the $N$ eigenvectors, and the nonnegative eigenvalues in the diagonal matrix $\bbSigma$ are assumed to be ordered as $\sigma_1 \geq \sigma_2 \geq ... \geq \sigma_N$.

After defining the two-layer GCG  $f_{\bbTheta}(\bbH)$ and its expected square Jacobian $\sqJacob$, we formally analyze its performance when denoising bandlimited graph signals.
This is particularly relevant given the importance of (approximate) bandlimited graph signals both from analytical and practical points of view~\cite{djuric2018cooperative}.
For the sake of clarity, we first introduce the main result (Th.~\ref{theorem_denoising_gcg}) and then we detail a key intermediate result (Lemma~\ref{lemma_eigs_gcg}) that provides additional insight.

Formally, consider the $K$-bandlimited graph signal $\bbx_0$ as described in \eqref{E:bl_signals}, and let the architecture $f_{\bbTheta}(\bbH)$ have a sufficiently large number of features $F$:
\begin{equation}\label{bound_on_F}
    F \geq \left( \frac{\sigma_1^2}{\sigma_N^2}\right)^{26} \xi^{-8}N,\;\;\text{with}\;\xi \in (0,(2\log (2N/\phi))^{-\frac{1}{2}})
\end{equation}
being an error tolerance parameter for some prespecified $\phi$.
Then, for a specific set of graphs {with minimum number of nodes ${N_{K,\epsilon,\delta}}$} that is introduced later in the section (cf. Ass.~\ref{A:sbm}), if we solve \eqref{E:nonlinear_denoising} running gradient descent with a step size $\eta\!\leq\frac{1}{\sigma_1^2}$, the following result holds (see App.~\ref{proof_theorem_denoising_gcg}).

\begin{theorem}\label{theorem_denoising_gcg}
    Let $f_{\bbTheta}(\bbH)$ be the network defined in equation~\eqref{E:2layer_gcg_simp}, and assume it is sufficiently wide, i.e., it satisfies condition \eqref{bound_on_F} for some error tolerance parameter $\xi$. 
    Let $\bbx_0$ be a $K$-bandlimited graph signal spanned by the eigenvectors $\bbV_K$, and let $\bbw_i$ and $\sigma_i$ be the $i$-th eigenvector and eigenvalue of $\sqJacob$. 
    Let $\bbn$ be the noise present in $\bbx$, set $\phi$ and $\epsilon$ to small positive numbers, and let the conditions from Ass.~\ref{A:sbm} hold.
    Then, for any $\epsilon$, $\delta$, there exists some ${N_{K,\epsilon,\delta}}$ such that if $N>{N_{K,\epsilon,\delta}}$,
    the error for each iteration $t$ of gradient descent with stepsize $\eta$ used to fit the architecture is bounded as
    \begin{align}\label{eq_bound_theorem_fitting_eigs_Jacobian_for_t}
        & \|\bbx_0-f_{\bbTheta_{(t)}}(\bbH)\|_2 \leq  \left((1-\eta\sigma_K^2)^t+\delta(1-\eta\sigma_N^2)^t\right)\|\bbx_0\|_2 \nonumber \\
        & +\xi\|\bbx\|_2 +\sqrt{\textstyle\sum_{i=1}^N((1-\eta\sigma_i^2)^t-1)^2(\bbw_i^\top\bbn)^2},
    \end{align}
    with probability at least $1-e^{-F^2}-\phi-\epsilon$.
\end{theorem}

As explained next, the fitting (denoising) bound provided by the theorem first decreases and then increases with the number of iterations $t$. To be more precise, let us analyze separately each of the three terms in the right hand side of \eqref{eq_bound_theorem_fitting_eigs_Jacobian_for_t}. 
The first term captures the part of the signal $\bbx_0$ that is fitted after $t$ iterations while accounting for the misalignment of the eigenvectors $\bbV_K$ and $\bbW_K$.
This term decreases with $t$ and, since $\delta$ can be made arbitrary small (cf. Lemma~\ref{lemma_eigs_gcg}), vanishes for moderately low values of $t$.
The second term is an error term that is negligible if the network is sufficiently wide.
{Therefore,  $\xi$ can be chosen to be sufficiently small by designing the architecture according to the condition in~\eqref{bound_on_F}.}
Finally, the third term, which depends on the noise present in each of the spectral components of the squared Jacobian $(\bbw_i^\top\bbn)^2$, grows with $t$. More specifically, if the $\sigma_i$ associated with a spectral component is very small, the term $(1-\eta\sigma_i^2)$ is close to $1$ and, hence, the noise power in the $i$-th frequency will be small. Only when $t$ grows very large the coefficient $(1-\eta\sigma_i^2)^t$ vanishes and the $i$-th frequency component of the noise is fitted. As a result, if the filter $\bbH$ is designed such that eigenvalues of the squared Jacobian satisfy that $\sigma_K \gg \sigma_{K+1}$, then there will be a range of moderate-to-high values of $t$ for which: i) the first term is zero and ii) only the $K$ strongest components of the noise have been fitted, so that the third term can be approximated as $\sqrt{\sum_{i=1}^K (\bbw_i^\top\bbn)^2 }$. Clearly, as $t$ grows larger, the coefficient $((1-\eta\sigma_i^2)^t-1)$ will also be close to one for $i>K$, meaning that additional components of the noise will be fitted as well, deteriorating the performance of the denoising architecture. This implies that if the optimization algorithm is stopped before $t$ grows too large, the original signal is fitted along with the noise that aligns with the signal, but not the noise present in other components.

In other words, Th. \ref{theorem_denoising_gcg} not only characterizes the performance of the two-layer GNN, but also illustrates that, if early stopping is adopted, our overparametrized architecture is able to effectively denoise the bandlimited graph signal.
This result is related to the error bound for denoising images presented in~\cite{heckel2019denoising}, where $\bbx_0$ is assumed to lie in the span of $\bbW_K$.
However, when dealing with graphs, it is unclear which signals would satisfy this requirement.
Motivated by this, we assume that $\bbx_0$ is a bandlimited signal (i.e., lies in the span of $\bbV_K$), which is a natural condition employed in many applications.

As a consequence, a critical step to attain Th.~\ref{theorem_denoising_gcg} is to relate the eigenvectors of $\sqJacob$ with those of the adjacency matrix $\bbA$, denoted as $\bbV$.
To achieve this, we assume that $\bbA$ is random and provide high-probability bounds between the leading eigenvectors of $\bbA$ and $\sqJacob$.
More specifically, consider a graph $\ccalG$ drawn from a stochastic block model (SBM)~\cite{newman2018networks} with $K$ communities.
Also, denote by $\ccalM(\ccalbA)$ the SBM with expected adjacency matrix ${\ccalbA}=\mathbb{E}[\bbA]$, and by $\beta_{min}$ the minimum expected degree $\beta_{min}:=\mathrm{min}_i[\ccalbA\textbf{1}]_i$.
Given some $\rho > 0$, {we define as $\ccalM_N(\beta_{min},\rho)$ the class of SBMs $\ccalM(\ccalbA)$ with $N$ nodes for which the minimum expected degree is $\beta_{min}$ or higher.}
Then, the condition of $\ccalG$ being drawn from this SBM whose expected minimum degree increases with $N$ is formally expressed in the following assumption.
\begin{assumption}\label{A:sbm}
    The model $\ccalM(\ccalbA)$ from which $\bbA$ is drawn satisfies $\ccalM(\ccalbA)\in\ccalM_N(\beta_{min},\rho)$, with $\beta_{min} = \omega (\mathrm{ln}(N/\rho))$.
\end{assumption}
{Here, $\omega(\cdot)$ denotes the (conventional) asymptotic dominance.}
We note that, as discussed in~\cite{schaub2020blind}, the minimal degree condition considered in Ass.~1 ensures that nodes belonging to the same community also belong to the same connected component with high probability, which is {helpful to relate} $\bbA$ and $\ccalbA$.
Under these conditions, the following result holds.
\begin{lemma}\label{lemma_eigs_gcg}
    Let the matrix $\sqJacob$ be defined as in \eqref{eq:expected_jac}, set $\epsilon$ and $\delta$ to small positive numbers, and denote by $\bbV_K$ and $\bbW_K$ the $K$ leading eigenvectors in the respective eigendecompositions of $\bbA$ and $\sqJacob$. Under Ass.~\ref{A:sbm}, there exists an orthonormal matrix $\bbQ$ and an integer ${N_{K,\epsilon,\delta}}$ such that, for $N > {N_{K,\epsilon,\delta}}$, the bound
$$\| \bbV_K - \bbW_K \bbQ \|_{\text{F}} \leq \delta,$$
holds with probability at least $1-\epsilon$.
\end{lemma}
%
The proof is provided in App.~\ref{proof_lemma_eigs_gcg}, and it leverages Ass.~1 to relate the eigenvectors $\bbV_K$ and $\bbW_K$ based on the eigenvectors of the expected values of $\bbA$ and $\sqJacob$.

For a given $K$, Lemma~\ref{lemma_eigs_gcg} bounds the difference between the subspaces spanned by the $K$ leading eigenvectors of $\bbA$ and $\sqJacob$ when graphs are big enough, a result that is key in obtaining Th.~\ref{theorem_denoising_gcg}. 
Moreover, the lemma shows that if the lower bound ${N_{K,\epsilon,\delta}}$ increases, then the error encoded $\delta$ becomes arbitrary small.
Also note that, if a larger value of $K$ is considered, then the minimum required  graph size ${N_{K,\epsilon,\delta}}$ will also be larger.
An inspection of~\eqref{eq:expected_jac} reveals that the result in Lemma~\ref{lemma_eigs_gcg} is not entirely unexpected.
Indeed, since $\bbH$ is a polynomial in $\bbA$, so is $\bbH^2$.
This implies that $\bbV$ are also the eigenvectors of $\bbH^2$, and because $\bbH^2$ appears twice on the right hand side of~\eqref{eq:expected_jac}, a relationship between the eigenvectors of $\sqJacob$ and $\bbV$ can be anticipated.
However, the presence of the Hadamard product and the (non Lipschitz continuous) nonlinearity $\arccos$ renders the exact analysis of the eigenvectors a challenging task.
Consequently, we resorted to a stochastic framework in deriving Lemma~\ref{lemma_eigs_gcg}.

\begin{figure*}[!t]
	\centering
	\begin{subfigure}{0.24\textwidth}
		\centering
		    \includegraphics[width=1\textwidth]{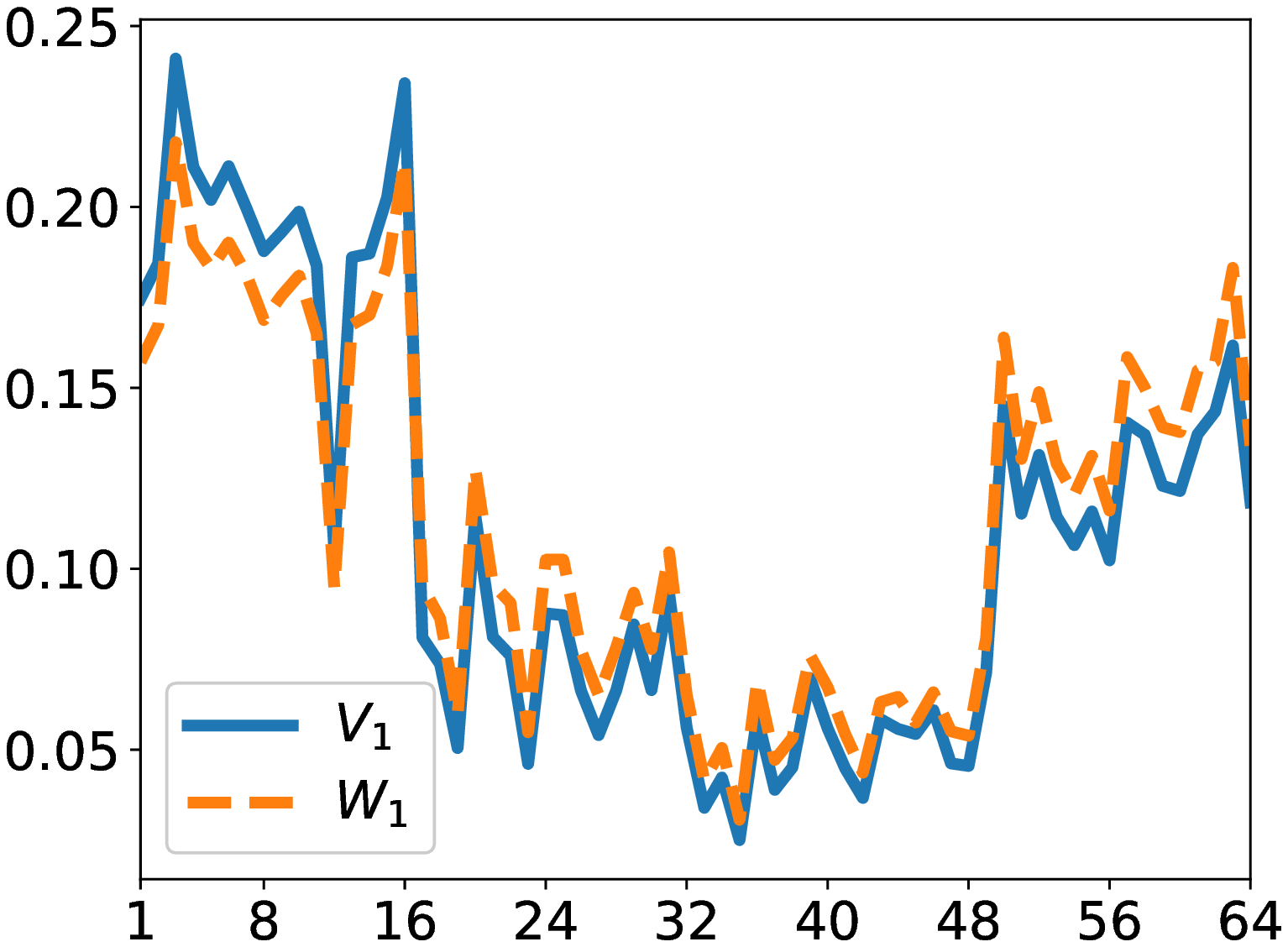}
	\end{subfigure}
	\begin{subfigure}{0.24\textwidth}
		\centering
		    \includegraphics[width=1\textwidth]{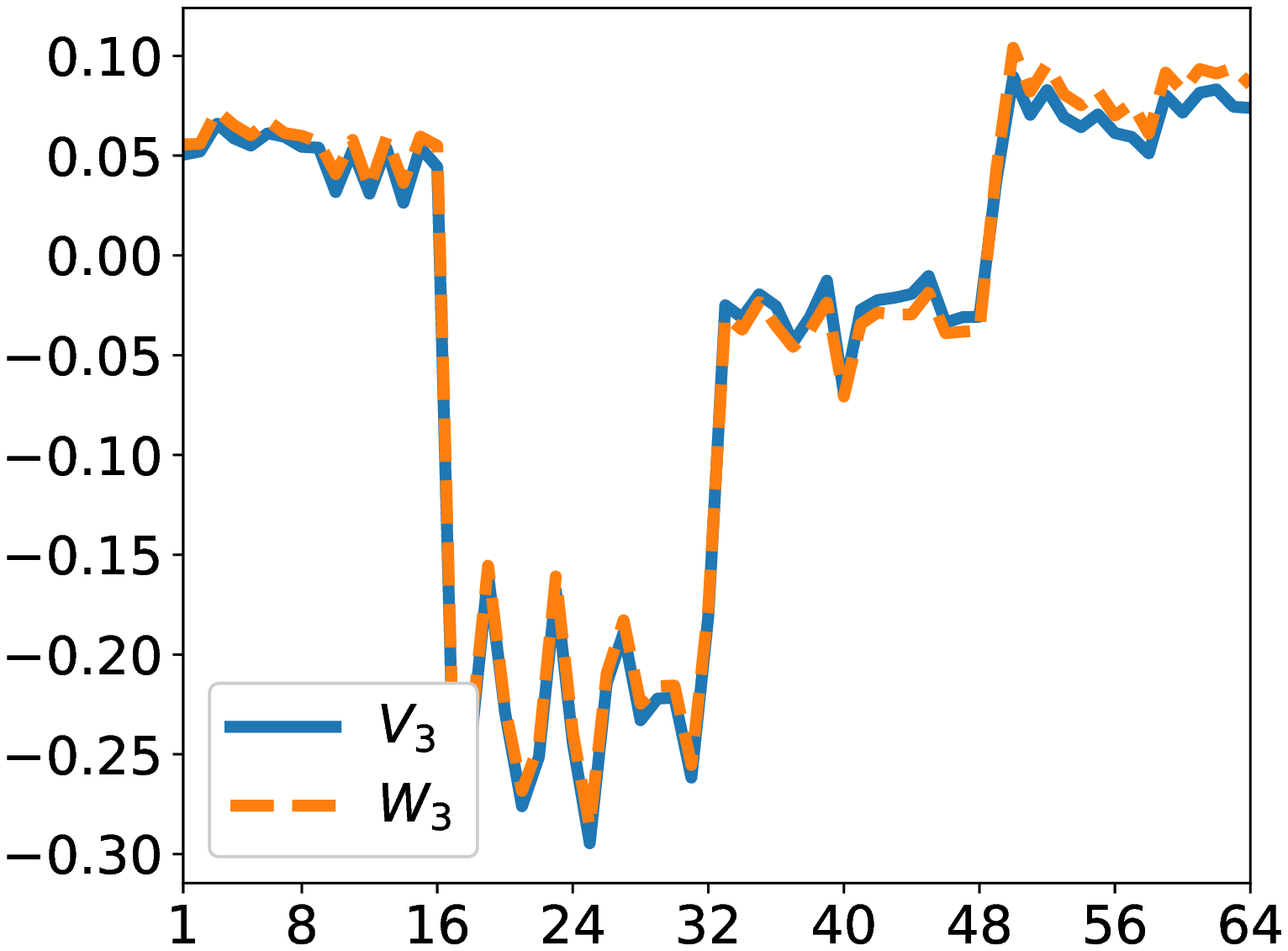}
	\end{subfigure}
	\begin{subfigure}{0.24\textwidth}
		\centering
		    \includegraphics[width=1\textwidth]{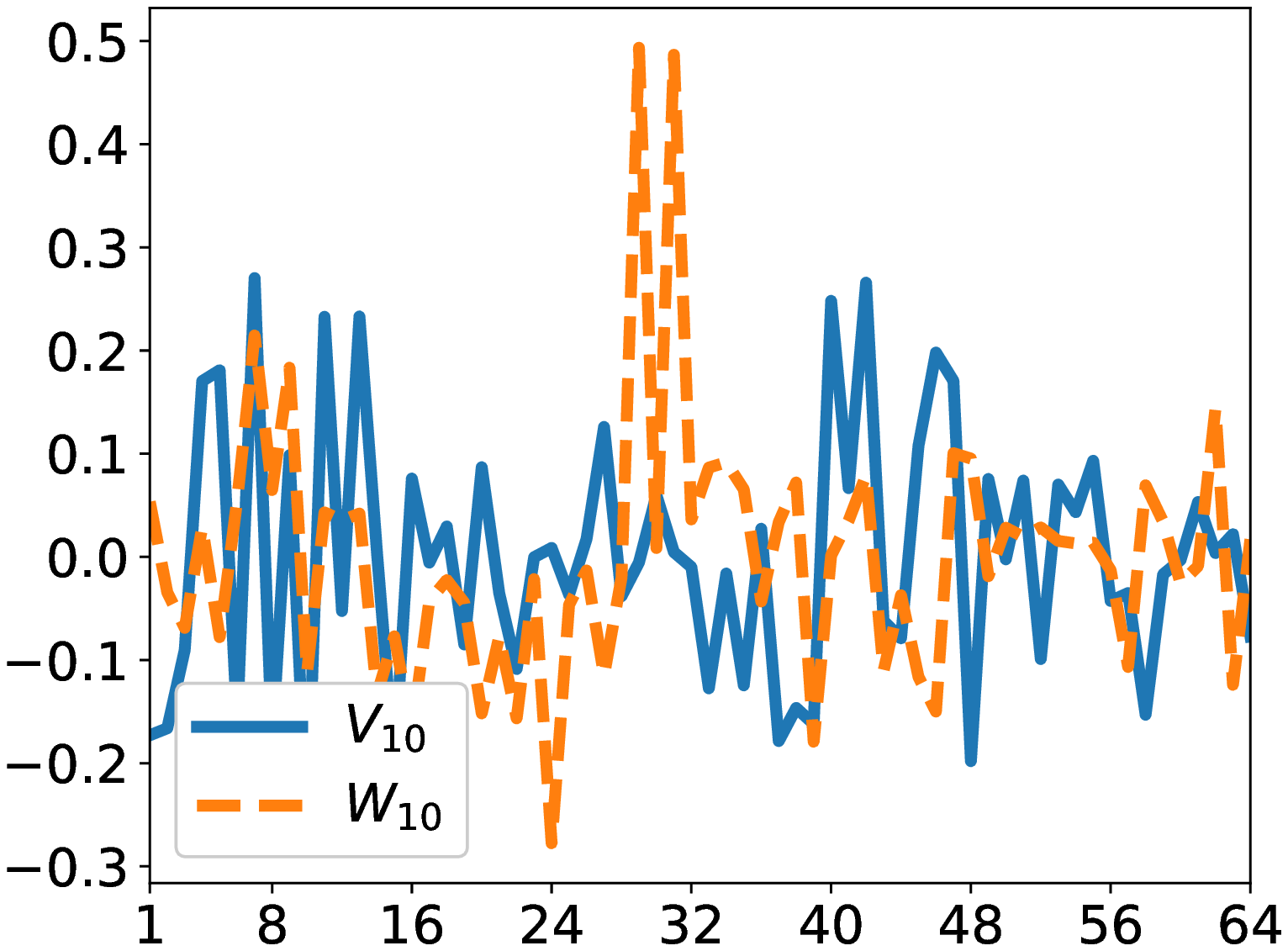}
	\end{subfigure}
	\begin{subfigure}{0.24\textwidth}
		\centering
		    \includegraphics[width=1\textwidth]{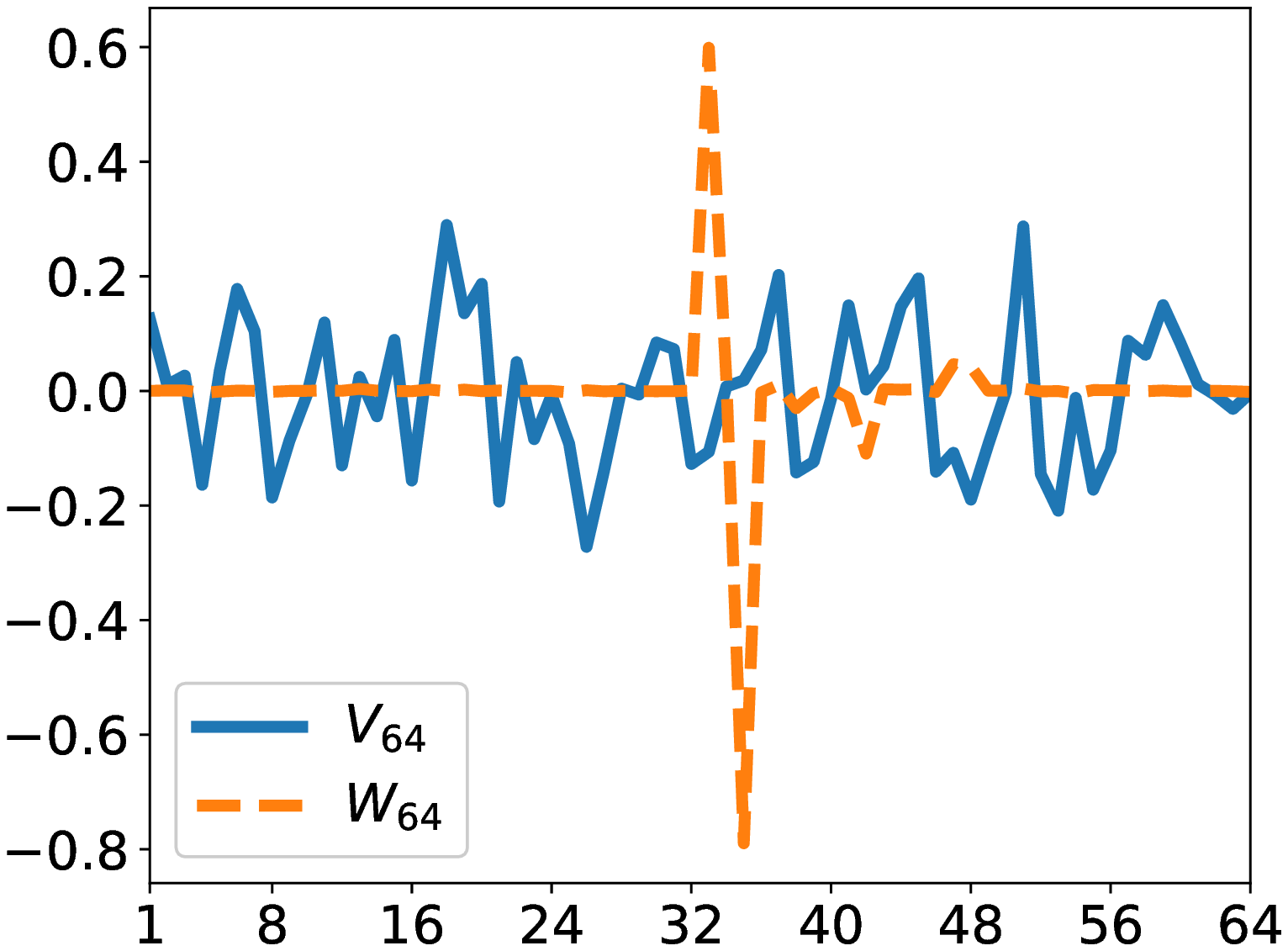}
	\end{subfigure}
	\vspace{-.15cm}
	\caption{Comparison between the eigenvectors of the matrices $\bbA$ and $\sqJacob$ for an SBM graph with $N=64$ nodes and $K=4$ communities, and for a GCG of $L=5$ layers. From left to right, the figures represent the first, third, tenth, and last eigenvectors.} \label{fig:generalizing_deep}
	\vspace{-4mm}
\end{figure*}

\begin{figure}
    \centering
    \includegraphics[width=0.3\textwidth]{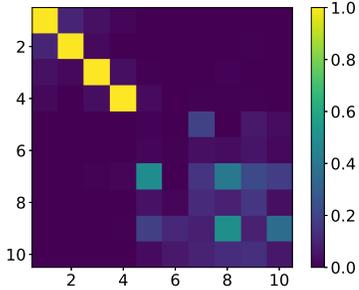}
    \caption{Heatmap representation of the matrix product $\bbV_K^\top\bbW_K$. The low values of the off-diagonal entries illustrate the orthogonality between both sets of eigenvectors. These eigenvectors are the same as those depicted in Fig.~\ref{fig:generalizing_deep}.}
    \label{fig:orthogonality_WV}
    \vspace{-4mm}
\end{figure}

\begin{figure*}[!t]
	\centering
	\begin{subfigure}{0.2\textwidth}
		\centering
		    \includegraphics[width=1\textwidth]{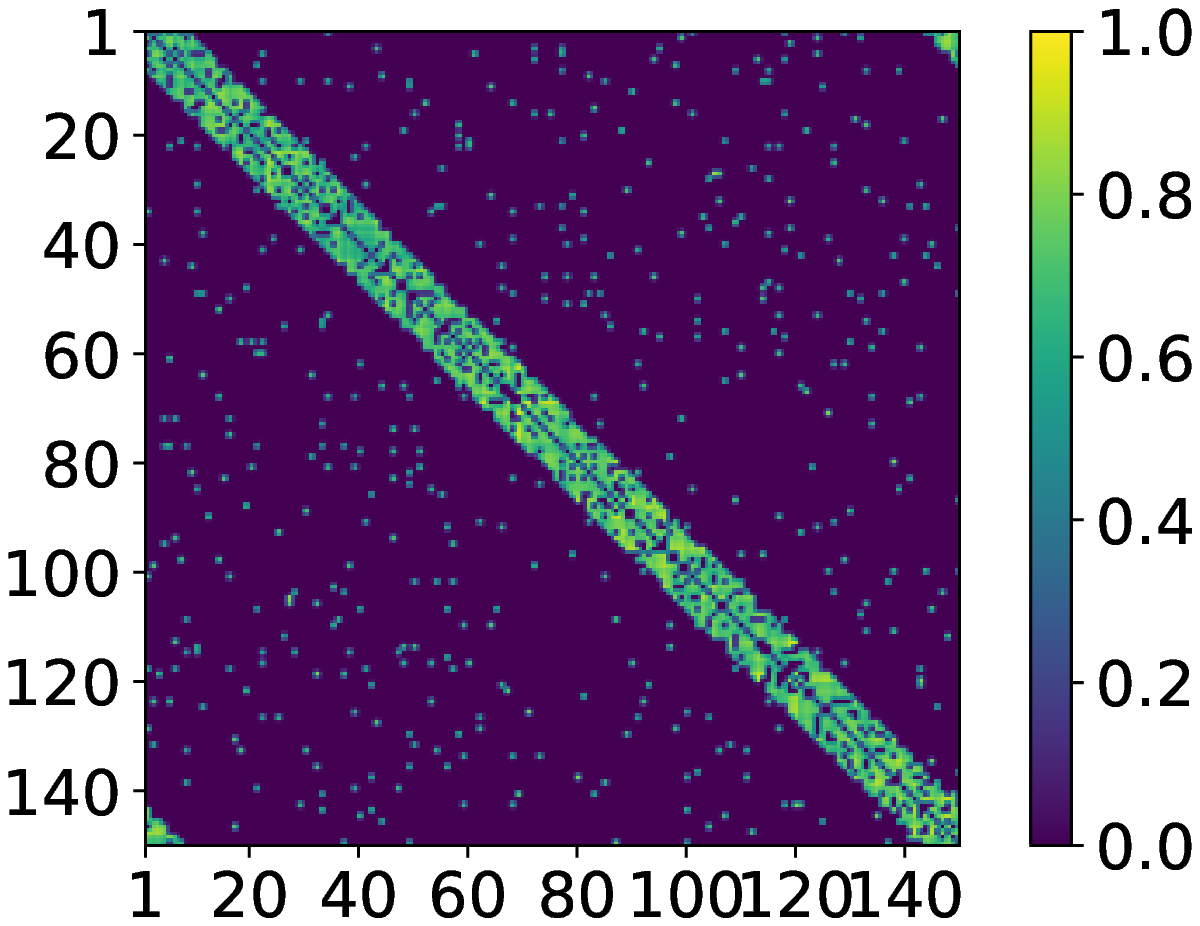}
	\end{subfigure}
	\begin{subfigure}{0.2\textwidth}
		\centering
		    \includegraphics[width=1\textwidth]{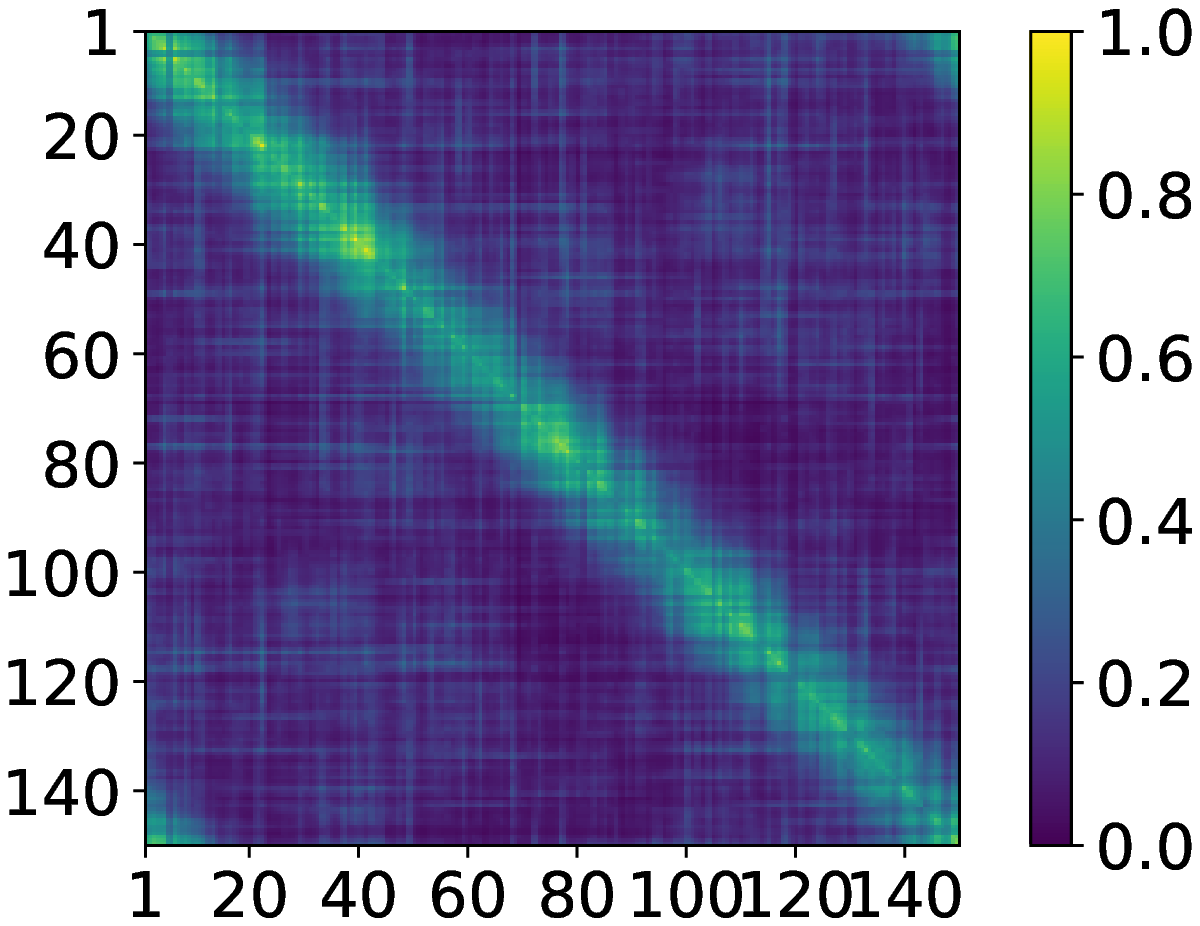}
	\end{subfigure}
	\begin{subfigure}{0.2\textwidth}
		\centering
		    \includegraphics[width=1\textwidth]{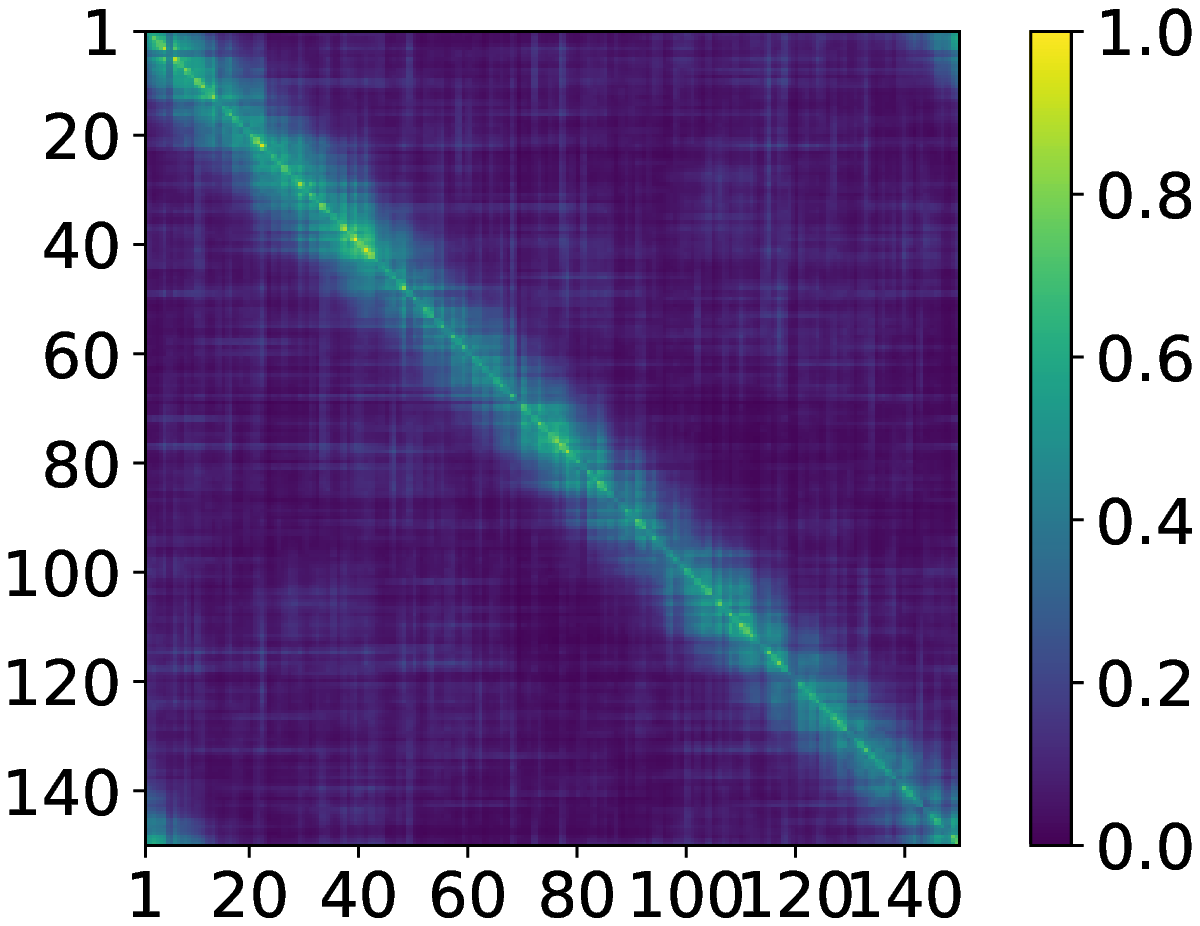}
	\end{subfigure}
	\begin{subfigure}{0.2\textwidth}
		\centering
		    \includegraphics[width=1\textwidth]{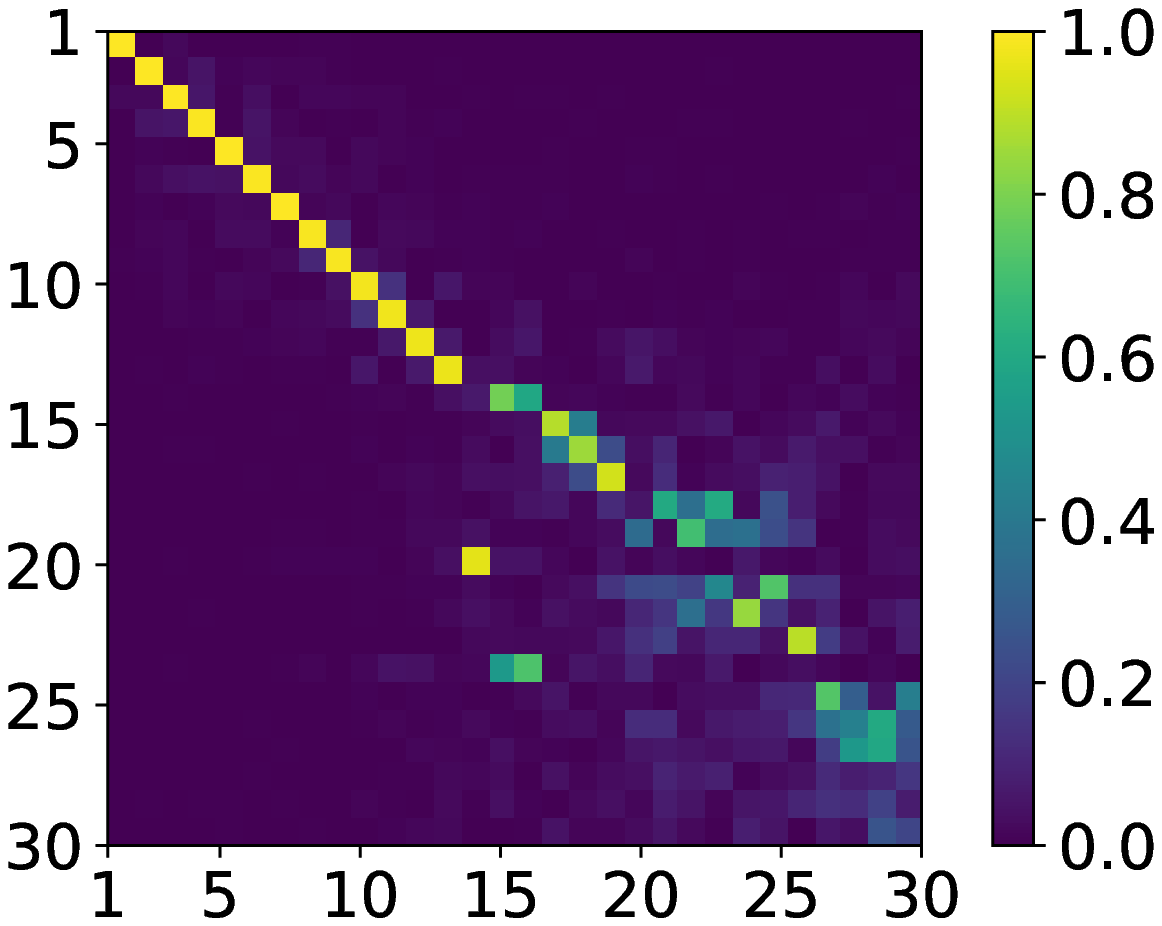}
	\end{subfigure}
	
	\begin{subfigure}{0.2\textwidth}
		\centering
		    \includegraphics[width=1\textwidth]{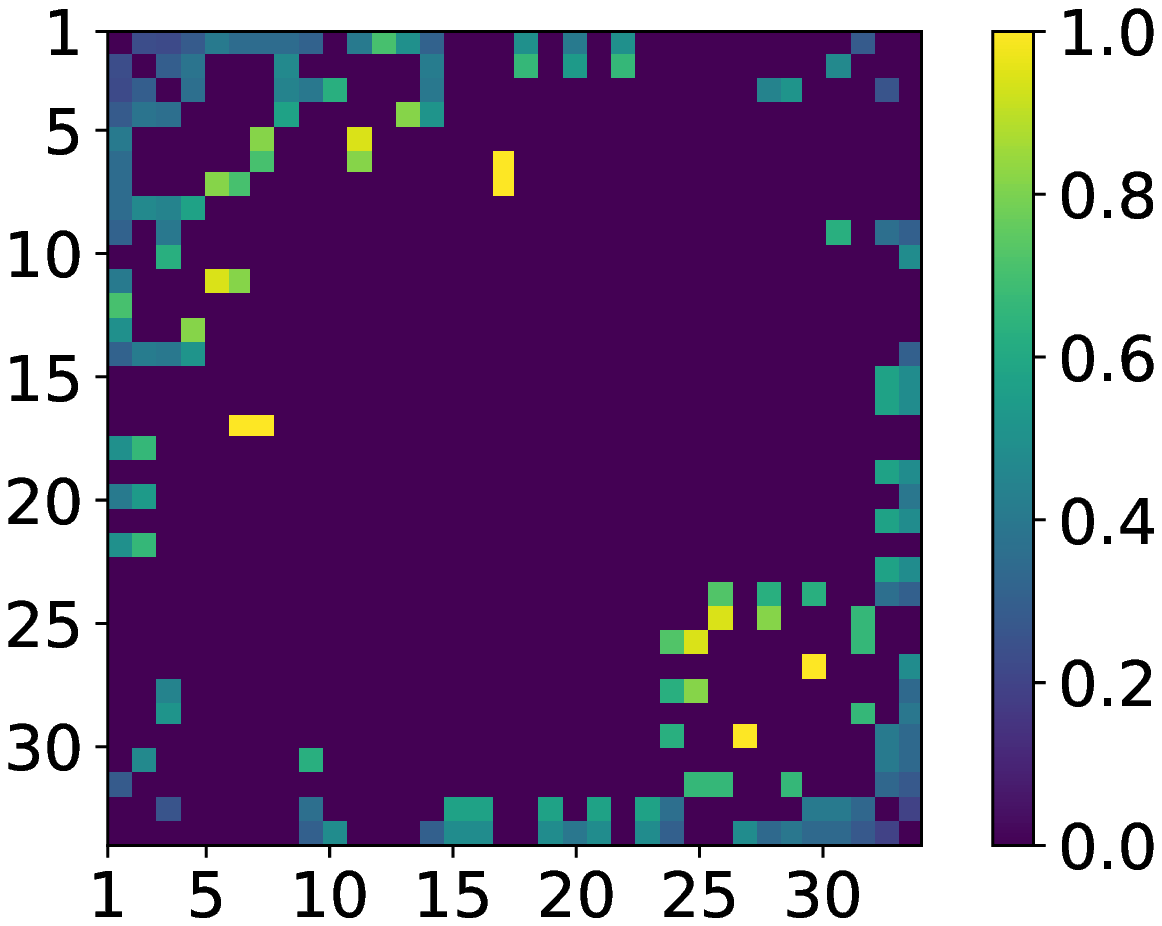}
	\end{subfigure}
	\begin{subfigure}{0.2\textwidth}
		\centering
		    \includegraphics[width=1\textwidth]{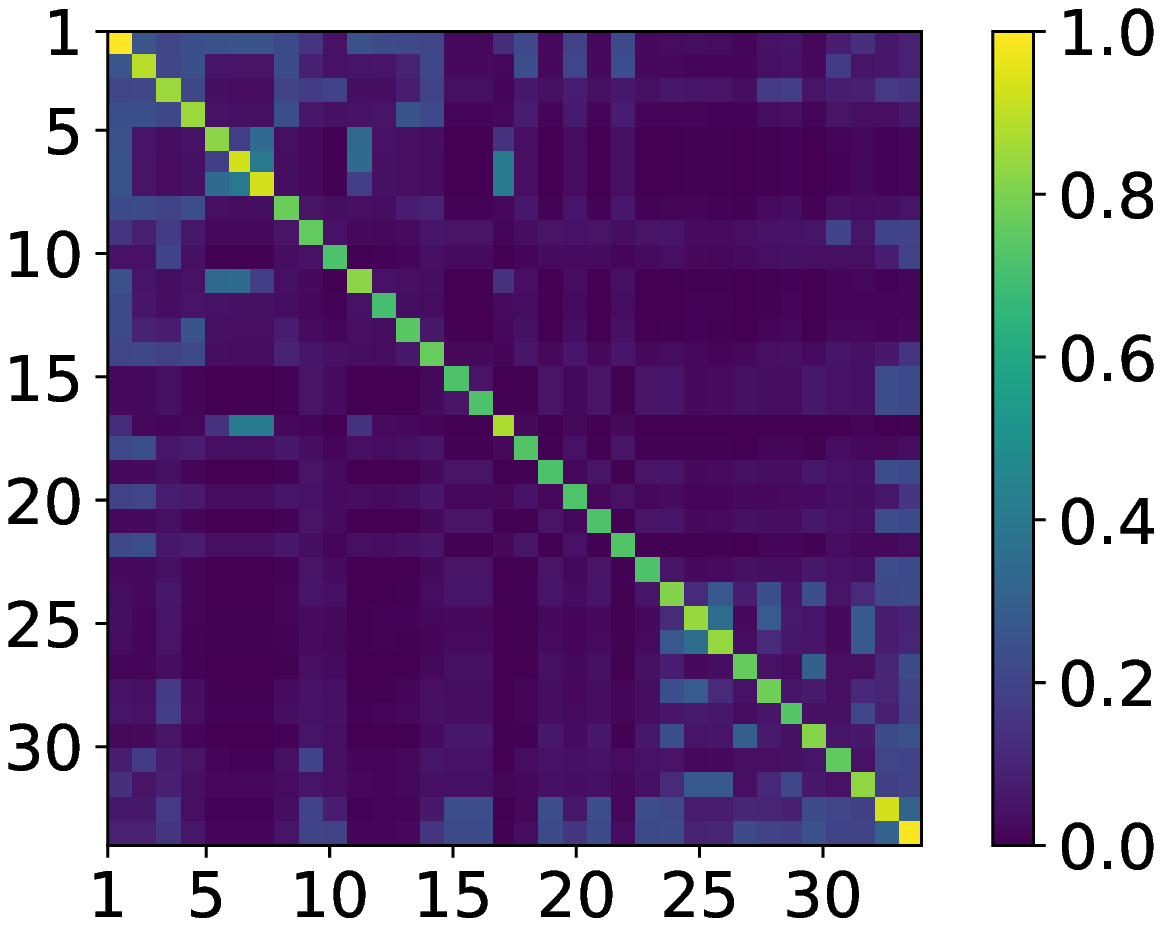}
	\end{subfigure}
	\begin{subfigure}{0.2\textwidth}
		\centering
		    \includegraphics[width=1\textwidth]{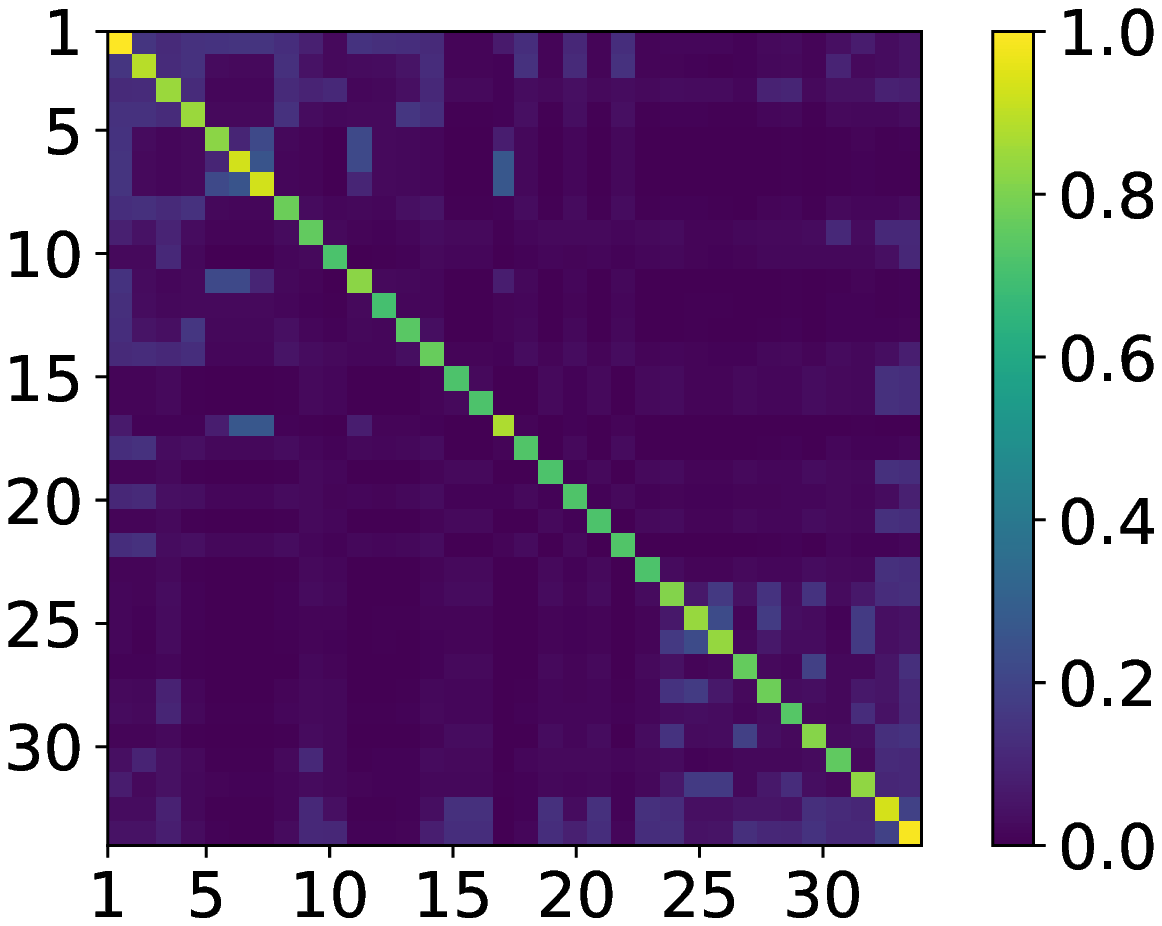}
	\end{subfigure}
	\begin{subfigure}{0.2\textwidth}
		\centering
		    \includegraphics[width=1\textwidth]{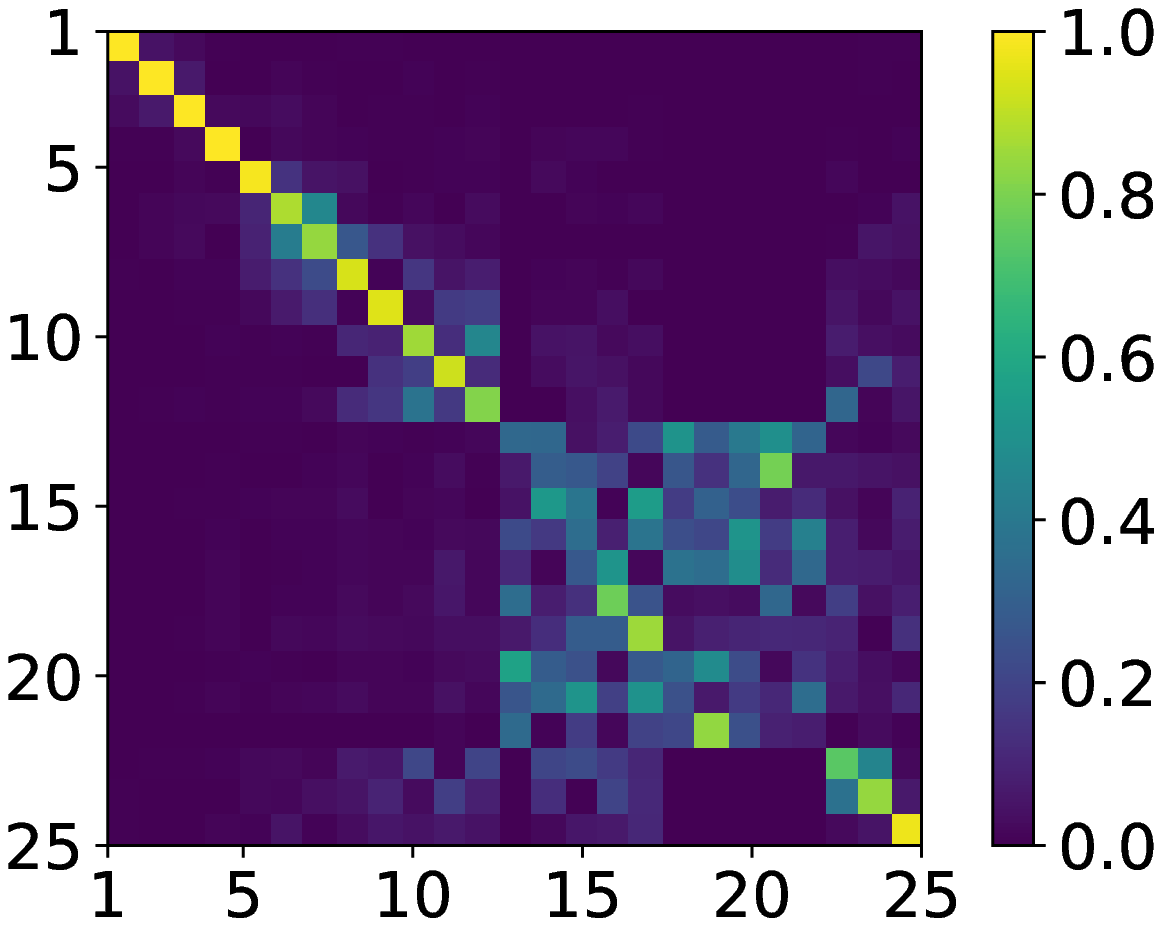}
	\end{subfigure}
	
	\begin{subfigure}{0.2\textwidth}
		\centering
		    \includegraphics[width=1\textwidth]{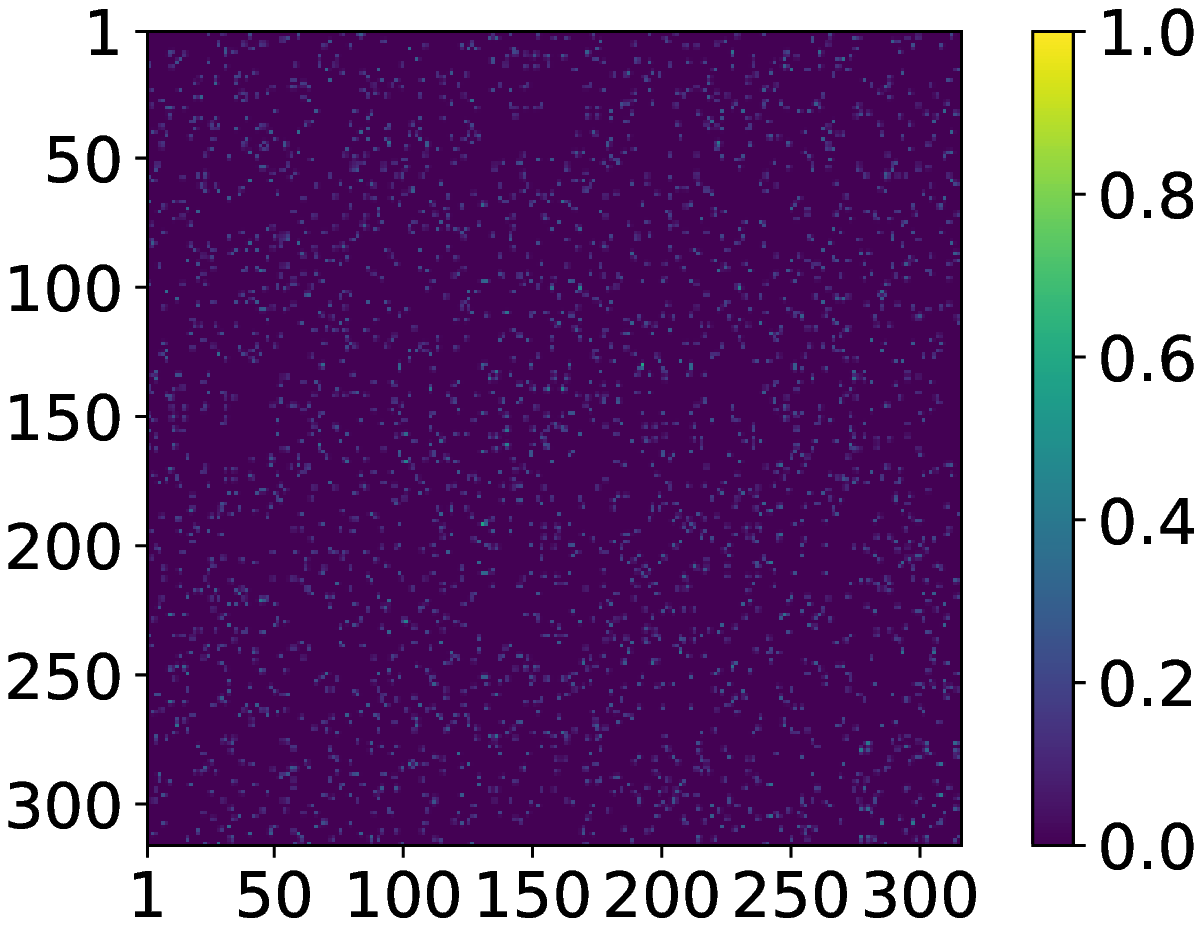}
	\end{subfigure}
	\begin{subfigure}{0.2\textwidth}
		\centering
		    \includegraphics[width=1\textwidth]{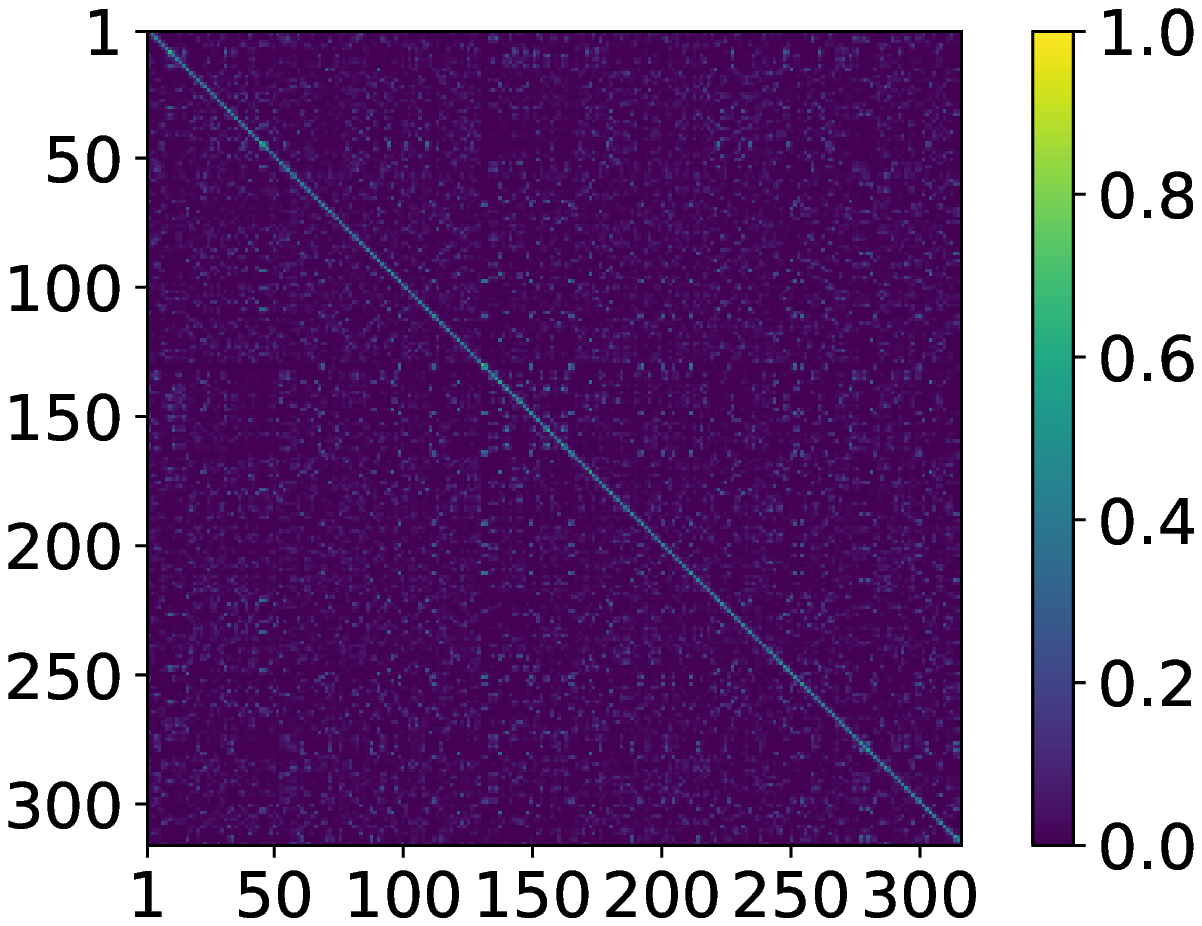}
	\end{subfigure}
	\begin{subfigure}{0.2\textwidth}
		\centering
		    \includegraphics[width=1\textwidth]{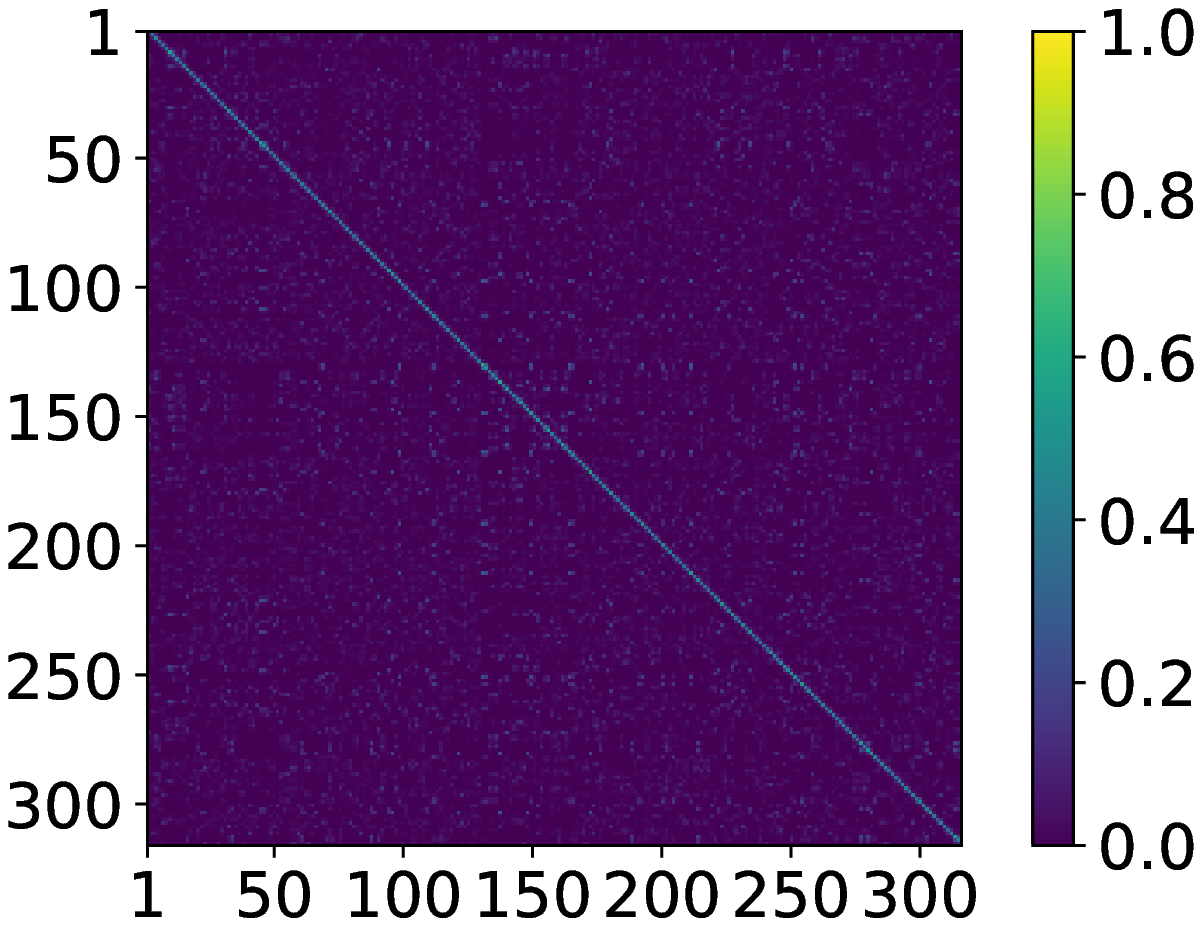}
	\end{subfigure}
	\begin{subfigure}{0.2\textwidth}
		\centering
		    \includegraphics[width=1\textwidth]{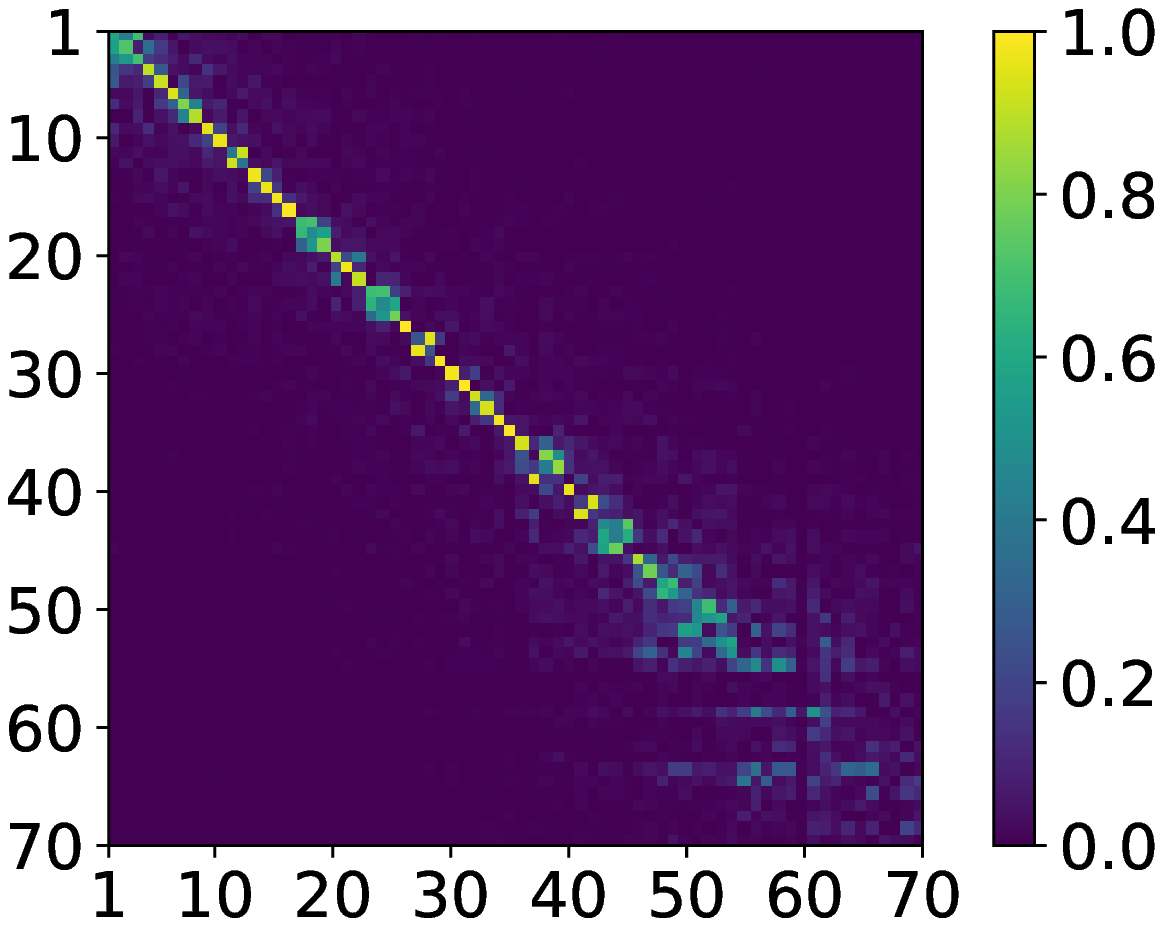}
	\end{subfigure}
	\vspace{-1.7mm}
	\caption{Illustration of matrices $\bbA$, $\bbH^2$, $\sqJacob$, and $\bbV_K^\top\bbW_K$, shown in columns 1, 2, 3, and 4, respectively, for different types of graphs. The rows 1, 2, and 3 correspond to a small world, the Zachary's Karate, and the weather stations graph. The graph filter $\bbH^2$ is created as a square graph filter with coefficients drawn from a uniform distribution and with unitary $\ell_1$ norm. For each graph (rows), it can be seen that the matrices $\bbA$, $\bbH^2$, and $\sqJacob$ are related, and that 
	$\bbV_K$ and $\bbW_K$ are close to orthogonal.} \label{fig:geralizing_graphs}
	\vspace{-5mm}
\end{figure*}

\subsection{Numerical inspection of the deep GCG spectrum}\label{S:analyze_deep_gcg}
While for convenience, the previous section focused on analyzing the GCG architecture with $L = 2$ layers, in practice we often work with a larger number of layers.
In this section, we provide numerical evidence showing that the relation between matrices $\bbA$ and $\sqJacob$ described in Lemma~\ref{lemma_eigs_gcg} also holds when $L > 2$.

To that end, Fig. \ref{fig:generalizing_deep} shows the pairs of eigenvectors $\bbv_i$ and $\bbw_i$ for the indexes $i=\{1,3,10,64\}$, for a given graph $\ccalG$ drawn from an SBM with $N=64$ nodes and 4 communities.
The GCG is composed of $L=5$ layers and, to obtain the eigenvectors of the squared Jacobian matrix, the Jacobian is computed using the \textit{autograd} functionality of PyTorch.  
The nodes of the graph are sorted by communities, i.e., the first $N_1$ nodes belong to the first community and so on.
It can be clearly seen that, even for moderately small graphs, the leading eigenvectors of $\bbA$ and $\sqJacob$ are almost identical, becoming more dissimilar as the eigenvectors are associated with smaller eigenvalues.
It can also be observed how leading eigenvectors have similar values for entries associated with nodes within the same community.
Moreover, Fig. \ref{fig:orthogonality_WV} depicts the matrix product $\bbV^\top\bbW$, where it is observed that the $K=4$ leading eigenvectors of both matrices are orthonormal.
The presented numerical results strengthen the argument that the analytical results obtained for the two-layer case can be extrapolated to deeper architectures. 

Another key assumption of Lemma~\ref{lemma_eigs_gcg} is that $\ccalG$ is drawn from the SBM described in $\ccalM_N(\beta_{min},\rho)$.
This assumption facilitates the derivation of a bound relating the spectra of $\bbA$ and $\sqJacob$ (i.e., the subspaces spanned by the eigenvectors $\bbV_K$ and $\bbW_K$). 
However, the results reported in Fig.~\ref{fig:geralizing_graphs} suggest that such a relation exists for other type of graphs, even though its analytical characterization is more challenging.
The figure has 12 panels (3 rows and 4 columns).
Each of the rows corresponds to a different graph, namely: 1) a realization of a small-world (SW) graph~\cite{watts1998collective} with $N=150$ nodes, 2) the Zachary's Karate graph~\cite{zachary1977information} with $N=34$ nodes, and 3) a graph of $N=316$ weather stations across the United States~\cite{temperatures2020}.
Each of the three first columns correspond to an $N\times N$ matrix, namely: 1) the normalized adjacency matrix $\bbA$, 2) $\bbH^2$, the squared version of a low pass graph filter and whose coefficients are drawn from a uniform distribution and set to unit $\ell_1$ norm, and 3) the squared Jacobian matrix $\sqJacob$. 
Although we may observe some similarity between $\bbA$ and $\sqJacob$, the relation between $\sqJacob$ and the graph $\ccalG$ becomes apparent when comparing the matrices $\bbH^2$ and $\sqJacob$.
The matrix $\bbH$ is a random graph filter used in the linear transformation of the convolutional generator $f_{\bbTheta}(\bbH)$, and it is clear that the vertex connectivity pattern of $\sqJacob$ is related to that of $\bbH^2$.
Since $\sqJacob$ and $\bbH^2$ are closely related and we know that the eigenvectors of $\bbH^2$ and those of $\bbA$ are the same, we expect $\bbW$ (the eigenvectors of $\sqJacob$) and $\bbV$ (the eigenvectors of $\bbA$) to be related as well. To verify this, the fourth column of Fig. \ref{fig:geralizing_graphs} represents $\bbV_K^\top\bbW_K$, i.e., the pairwise inner products of the $K$ leading eigenvectors of $\bbA$ and those of $\sqJacob$. It can be observed that the $K$ leading eigenvectors are close to orthogonal, which means that the relation observed in the vertex domain carries over to the spectral domain and $\bbV_K$ and $\bbW_K$ expand the same subspace.
These results suggest that a deep GCG could be able to {denoise} signals living in the subspace spanned by $\bbV_K$.
However, because the bound in Th.~\ref{theorem_denoising_gcg} assumed a 2-layer GCG, we address this hypothesis numerically in Sec.~\ref{S:experiments}.

To summarize, the presented results illustrate that the analytical characterization provided in Sec. \ref{S:analysis_GCG}, which considered a 2-layer GCG operating over SBM graphs, carries over to more general setups. 

\section{Graph upsampling decoder}\label{S:ups_dec}
The GCG architecture presented in Sec. \ref{S:conv_dec} incorporated the topology of $\ccalG$ via the vertex-based convolutions implemented by the graph filter $\bbH$.
In this section, we introduce the graph decoder (GDec) architecture.
{In contrast to the GCG and other GCNNs, this novel graph-aware denoising NN incorporates the topology of $\ccalG$ via a (nested) collection of graph upsampling operators~\cite{rey2019underparametrized}.} 
Specifically, we propose the linear transformation for the GDec denoiser to be given by
\begin{equation}\label{eq:graph_decoder}
      \ccalT_{\bbTheta^{(\ell)}}^{(\ell)}\{\bbY^{(\ell-1)}|\ccalG\} = \bbU^{(\ell)}\bbY^{(\ell-1)}\bbTheta^{(\ell)},
\end{equation}
where $\bbU^{(\ell)} \in \reals^{N^{(\ell)} \times N^{(\ell-1)}}$, with $N^{(\ell)}\geq N^{(\ell-1)}$, are graph upsampling matrices to be defined soon.
Note that, compared to \eqref{E:linear_trans_gcg}, the graph filter $\bbH$ is replaced with the upsampling operator $\bbU^{(\ell)}$ that \emph{depends} on $\ell$.
Adopting the proposed linear transformation, the output of the GDec with $L$ layers is given by the recursion
\begin{align}
      \bbY^{(\ell)}\! &=\relu(\bbU^{(\ell)}\bbY^{(\ell-1)}\bbTheta^{(\ell)}),\;\; \mathrm{for}\; \ell=1,...,L\!-\!1, \label{E:gd1}\\ 
      \bby^{(L)} \!&= \bbU^{(L)}\bbY^{(L-1)}\bbTheta^{(L)}, \label{E:gd2}
\end{align}
where the $\relu$ is also removed from the last layer.

Similar to the GCG, the proposed GDec learns to combine the features within each node.
{However, the interpolation of the signals in this case is determined by the graph upsampling operators $\{\bbU^{(\ell)}\}_{\ell=1}^L$, rather than by employing convolutions.}
The size of the input $N^{(0)}$ is now a design parameter that will determine the implicit degrees of freedom of the architecture.
Note that, from the GSP perspective, the input feature matrix $\bbY^{(\ell-1)} \in \reals^{N^{(\ell-1)} \times F^{(\ell-1)}}$ represents $F^{(\ell-1)}$ graph signals, each of them defined over a graph $\ccalG^{(\ell-1)}$ with $N^{(\ell-1)}$ nodes.
Therefore, even though the input $\bbY^{(0)}=\bbZ$ is still a random white matrix across rows and columns, since $N^{(\ell)} \geq N^{(\ell-1)}$, the dimensionality of the input is progressively increasing. 

{A closer comparison with the GCG reveals that the smaller dimensionality of the input $\bbZ$ endows the GDec architecture with fewer degrees of freedom, rendering the architecture more robust to noise.
Not only that, but the graph information is now included via the graph upsampling operators $\bbU^{(\ell)}$ instead of relying on graph filters.}
Clearly, the method used to design the graph upsampling matrices, which is the subject of the next section, will have an impact on the type of graph signals that can be efficiently denoised using the GDec architecture.

\subsection{Graph upsampling operator from hierarchical clustering}\label{S:upsampling_operator}

\begin{figure}
    \centering
    \includegraphics[width=0.38\textwidth]{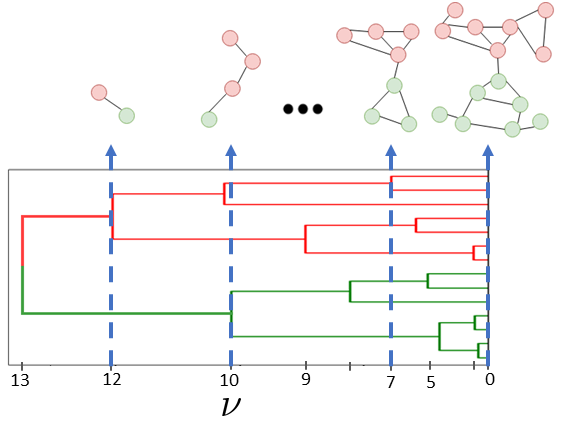}
    \vspace{-2mm}
    \caption{Dendrogram of an agglomerative hierarchical clustering algorithm and the resulting graphs with 2, 4, 7 and 14 nodes.}
    \label{fig:dendrogram}
    \vspace{-5.5mm}
\end{figure}

Regular upsampling operators have been successfully used in NN architectures to denoise signals defined on regular domains \cite{heckel2019denoising}. While the design of upsampling operators in regular grids is straightforward, when the signals are defined on irregular domains the problem becomes substantially more challenging. The approach that we put forth in this paper  is to use agglomerative hierarchical clustering methods~\cite{jain_1988_algorithms, carlsson_2018_hierarchical, carlsson_2013_axiomatic} to design a graph upsampling operator that leverages the graph topology.
These methods take a graph as an input and return a dendrogram; see Fig.~\ref{fig:dendrogram}.
A dendrogram can be interpreted as a rooted-tree structure that shows different clusters at the different levels of resolution $\nu$. 
At the finest resolution ($\nu=0$) each node forms a cluster of its own.
Then, as $\nu$ increases, nodes start to group together (agglomerate) in bigger clusters and, when the resolution becomes large (coarse) enough, all nodes end up being grouped in the same cluster.

By cutting the dendrogram at $L+1$ resolutions, including $\nu=0$, we obtain a collection of node sets with parent-child relationships inherited by the refinement of clusters.
Since we are interested in performing graph upsampling, note that the dendrogram is interpreted from left to right.
This can be observed in the example shown in Fig.~\ref{fig:dendrogram}, where the three red nodes in the second graph ($\nu=10$, layer $\ell=1$) are children of the red parent in the coarsest graph ($\nu=12$, layer $\ell=0$).
{In this sense, the graph upsampling operator is given by the inverse operation of the clustering algorithm.}
We leverage these parent-children relations to define the membership matrices $\bbP^{(\ell)} \in \{0,1\}^{N^{(\ell)} \times N^{(\ell-1)}}$, where the entry $P_{ij}^{(\ell)}=1$ only if the $i$-th node in layer $\ell$ is the child of the $j$-th node in layer $\ell-1$.
{Moreover, we can further exploit the dendrogram to obtain  coarser-resolution versions of the original graph $\ccalG$.
To that end, note that the clusters at layer $\ell$ can be interpreted as nodes of a graph $\ccalG^{(\ell)}$ with $N^{(\ell)}$ nodes and adjacency matrix $\bbA^{(\ell)}$.}
There are several ways of defining $\bbA^{(\ell)}$ based on the original adjacency matrix $\bbA$. While our architecture does not focus on a particular form, in the simulations we set $A^{(\ell)}_{ij} \neq 0$ only if, in the original graph $\ccalG$, there is at least one edge between nodes belonging to the cluster $i$ and nodes from cluster $j$.
In addition, the weight of the edge depends on the number of existing edges between the two clusters.

With the definition of the membership matrix $\bbP^{(\ell)}$ and the adjacency matrix $\bbA^{(\ell)}$, the upsampling operator of the $\ell$-th layer is given by
\begin{equation}\label{eq:upsampling_operator}
    \bbU^{(\ell)} = \left(\gamma\bbI+\left(1-\gamma\right)\bbA^{(\ell)}\right)\bbP^{(\ell)},
\end{equation}
where $\gamma\in[0, 1]$ is a pre-specified constant. 
Notice that $\bbU^{(\ell)}$ first copies the signal value from the parents to the children by applying the matrix $\bbP^{(\ell)}$, and then every child performs a convex combination between this value and the average signal value of its neighbors.
This design promotes that nodes descending from the same parent have similar (related) values, which conveys a notion (prior) of smoothness on the targeted graph signals.
As we show in Sec.~\ref{S:experiments}, the implicit smoothness prior results in a better performance when denoising smooth signals but, on the other hand, makes the architecture more sensitive to model mismatch.
Therefore, when dealing with high-frequency signals, a worth-looking approach left as a future research direction is to rely on algorithms that cluster the nodes considering not only the topology of $\ccalG$ but also the properties of the graph signals.

Because the membership matrices $\bbP^{(\ell)}$ are designed using a clustering algorithm over $\ccalG$, and the matrices $\bbA^{(\ell)}$ capture how strongly connected the clusters of layer $\ell$ are in the original graph, these two matrices are responsible for incorporating the information of $\ccalG$ into the upsampling operators $\bbU^{(\ell)}$.
Furthermore, we remark that the upsampling operator $\bbU^{(\ell)}$ can be reinterpreted as the application of $\bbP^{(\ell)}$ followed by the application of a graph filter
\begin{equation} \label{eq:upsampling_filter_second_step}
\tbH^{(\ell)}=\gamma\bbI+(1-\gamma) \bbA^{(\ell)},    
\end{equation} 
which sets the filter coefficients as $h_0=\gamma$ and $h_1=1-\gamma$.

\subsection{Guaranteed denoising with the GDec}
As we did for the GCG, our goal is to theoretically characterize the denoising performance of the GNN architecture defined by \eqref{E:gd1}-\eqref{eq:upsampling_operator}.
To achieve that goal, we replicate the approach implemented in Sec.~\ref{S:analysis_GCG}.
We first derive the matrix $\sqJacob$ and provide theoretical guarantees when denoising a $K$-bandlimited graph signal with the GDec.
Then, to gain additional insight, we detail the relation between the subspace spanned by the eigenvectors $\bbW$ and the spectral domain of $\bbA$.
This relation is key in deriving the theoretical analysis.

We start by introducing the 2-layer GDec
\begin{equation}\label{eq:gdec_2lay_orig}
    f_{\bbTheta}(\bbZ|\ccalG) = \relu(\bbU\bbZ\bbTheta^{(1)})\bbtheta^{(2)}.
\end{equation}
%
{Then, following a similar reasoning to that provided after \eqref{E:2layer_gcg_simp}, instead of employing the architecture in \eqref{eq:gdec_2lay_orig} we can optimize \eqref{E:nonlinear_denoising} over its simplifying version}
\begin{equation}\label{E:2layer_gd_simp}
   f_{\bbTheta}(\bbU) = f_{\bbTheta}(\bbZ|\ccalG) = \relu(\bbU\bbTheta) \bbb.
\end{equation}
%
An important difference with respect to the GCG presented in \eqref{E:2layer_gcg_simp} is that the matrix $\bbTheta$ has a dimension of $N^{(0)} \times F$, so it spans $\reals^{N^{(0)}}$ instead of $\reals^{N}$.
Since $N^{(0)} < N$, the smaller subspace spanned by the weights of the GDec renders the architecture more robust to fitting noise, but, on the other hand, the number of degrees of freedom to learn the graph signal of interest are reduced. As a result, the alignment between the targeted graph signals and the low-pass vertex-clustering architecture becomes more important.

The expected squared Jacobian $\sqJacob=\mathbb{E}_{\bbTheta}[\ccalJ_{\bbTheta}(\bbU) \ccalJ^\top_{\bbTheta}(\bbU)]$ is obtained following the procedure used to derive~\eqref{eq:expected_jac}, arriving at the expression
\begin{equation}\label{eq:expected_jac_dec}
    \sqJacob = 0.5 \left( \mathbf{1} \mathbf{1}^\top - \frac{1}{\pi} \arccos(\tbC^{-1} \bbU\bbU^\top \tbC^{-1})\right) \odot \bbU\bbU^\top,
\end{equation}
where $\bbu_i$ represents the $i$-th row of $\bbU$, and $\tbC=\diag([\|\bbu_1\|_2,...,\|\bbu_N\|_2])$ is a normalization matrix.

Then, let $\bbx_0$ be a $K$-bandlimited graph signal and let $f_{\bbTheta}(\bbU)$ have a number of features $F$ satisfying \eqref{bound_on_F}.  
If we solve \eqref{E:nonlinear_denoising} running gradient descent with a step size $\eta\leq\frac{1}{\sigma_1^2}$, the following result holds.

\begin{theorem}\label{theorem_denoising_gd}
    Let $f_{\bbTheta}(\bbU)$ be the network defined in equation~\eqref{E:2layer_gd_simp}.
    Consider the conditions described in Th.~\ref{theorem_denoising_gcg} and let $N^{(0)}$ match the number of communities $K$ (see Ass.~\ref{A:sbm}).
    Then, for any $\epsilon$, $\delta$, there exists some ${N_{K,\epsilon,\delta}}$ such that if $N>{N_{K,\epsilon,\delta}}$, then the error for each iteration $t$ of gradient descent with stepsize $\eta$ used to fit the architecture is bounded as \eqref{eq_bound_theorem_fitting_eigs_Jacobian_for_t}, with  probability at least $1-e^{-F^2}-\phi-\epsilon$.
\end{theorem}

The proof is analogous to the one provided in App.~\ref{proof_theorem_denoising_gcg} but exploiting Lemma~\ref{lemma_eigs_gd} instead of Lemma~\ref{lemma_eigs_gcg}.
Lemma~\ref{lemma_eigs_gd} is fundamental in attaining Th.~\ref{theorem_denoising_gd} and is presented later in the section.

Th.~\ref{theorem_denoising_gd} formally establishes the denoising capability of the GDec when $\bbx_0$ is a $K$-bandlimited graph signal and  $K=N^{(0)}$ matches the number of communities in the SBM graph.
When compared with the GCG, the smaller dimensionality of the input $\bbZ$, and thus the smaller rank of the matrix $\bbTheta$, constrains the learning capacity of the architecture, making it more robust to the presence of noise.
However, this additional robustness also implies that the architecture is more sensitive to model mismatch, since its capacity to learn arbitrary signals is smaller.
Intuitively, the GDec represents an architecture tailored for a more specific family of graph signals than the GCG.
Moreover, employing the GDec instead of the GCG has a significant impact on the relation between the subspaces spanned by $\bbV_K$ and $\bbW_K$.

To establish the new relation between $\bbV_K$ and $\bbW_K$, assume that the adjacency matrix is drawn from an SBM $\ccalM(\ccalbA)$ with $K$ communities such that $\ccalM(\ccalbA)\in\ccalM_N(\beta_{min},\rho)$, so that the SBM follows Ass. \ref{A:sbm}.
In addition, set the size of the latent space to the number of communities so $N^{(0)} = K$.
Under this setting, the counterpart to Lemma~\ref{lemma_eigs_gcg} for the case where $f_{\bbTheta}(\bbU)$ is a GDec architecture follows.
\begin{lemma}\label{lemma_eigs_gd}
    Let the matrix $\sqJacob$ be defined as in \eqref{eq:expected_jac_dec}, set $\epsilon$ and $\delta$ to small positive numbers, and denote by $\bbV_K$ and $\bbW_K$ the $K$ leading eigenvectors in the respective eigendecompositions of $\bbA$ and $\sqJacob$. Under Ass.~\ref{A:sbm}, there exist an orthonormal matrix $\bbQ$ and an integer ${N_{K,\epsilon,\delta}}$ such that for $N > {N_{K,\epsilon,\delta}}$ the bound
$$\| \bbV_K - \bbW_K \bbQ \|_{\text{F}} \leq \delta,$$
holds with probability at least $1-\epsilon$.
\end{lemma}
%

Lemma~\ref{lemma_eigs_gd} asserts that the difference between the subspaces spanned by $\bbV_K$ and $\bbW_K$ becomes arbitrarily small as the size of the graph increases. 
The proof is provided in App.~\ref{proof_lemma_eigs_gd} and the intuition behind it arises from the fact that the upsampling operator can be understood as $\bbU=\tbH\bbP$, where $\tbH$ is a graph filter of the specific form described in \eqref{eq:upsampling_filter_second_step}.
Remember that $\bbP$ is a binary matrix encoding the cluster in the layer $\ell-1$ to which the nodes in the layer $\ell$ belong.
Since we are only considering two layers, and we have that $N^{(0)}=K$, the matrix $\bbP$ is encoding the node-community membership of the SBM graph and, hence, the product $\bbP\bbP^\top$ is a block matrix with constant entries matching the block pattern of $\ccalbA$.
As shown in the proof, this property can be leveraged to bound the eigendecomposition of $\bbA$ and $\sqJacob$.

\subsection{Analyzing the deep GDec}
The deep GDec composed of $L>2$ layers can be constructed following the recursion presented in \eqref{E:gd1} and \eqref{E:gd2}.
In this case, by stacking more layers we perform the upsampling of the input signal in a progressive manner and, at the same time, we add more nonlinearities, which helps alleviating the rank constraint related to the input size $N^{(0)}$.
In the absence of nonlinear functions, the maximum rank of the weights would be $N^{(0)}$, and thus, only signals in a subspace of size $N^{(0)}$ could be learned.
By properly selecting the number of layers and the input size when constructing the network, we can obtain a trade-off between the robustness of the architecture and its learning capability.

\begin{figure*}[!t]
	\centering
	\begin{subfigure}{0.27\textwidth}
		\centering
        \includegraphics[width=1\textwidth]{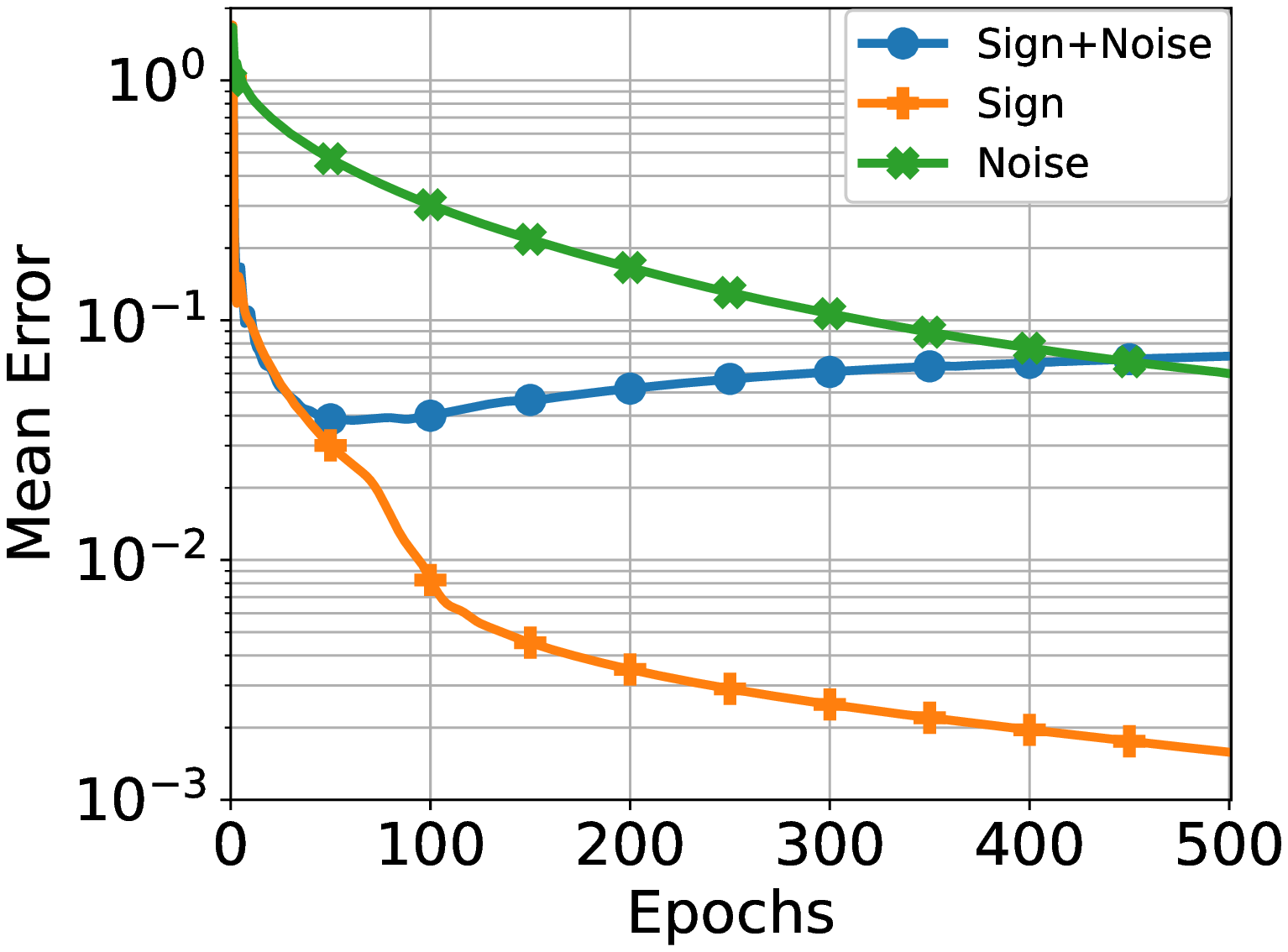}
	\end{subfigure}
	\begin{subfigure}{0.27\textwidth}
		\centering
		\includegraphics[width=1\textwidth]{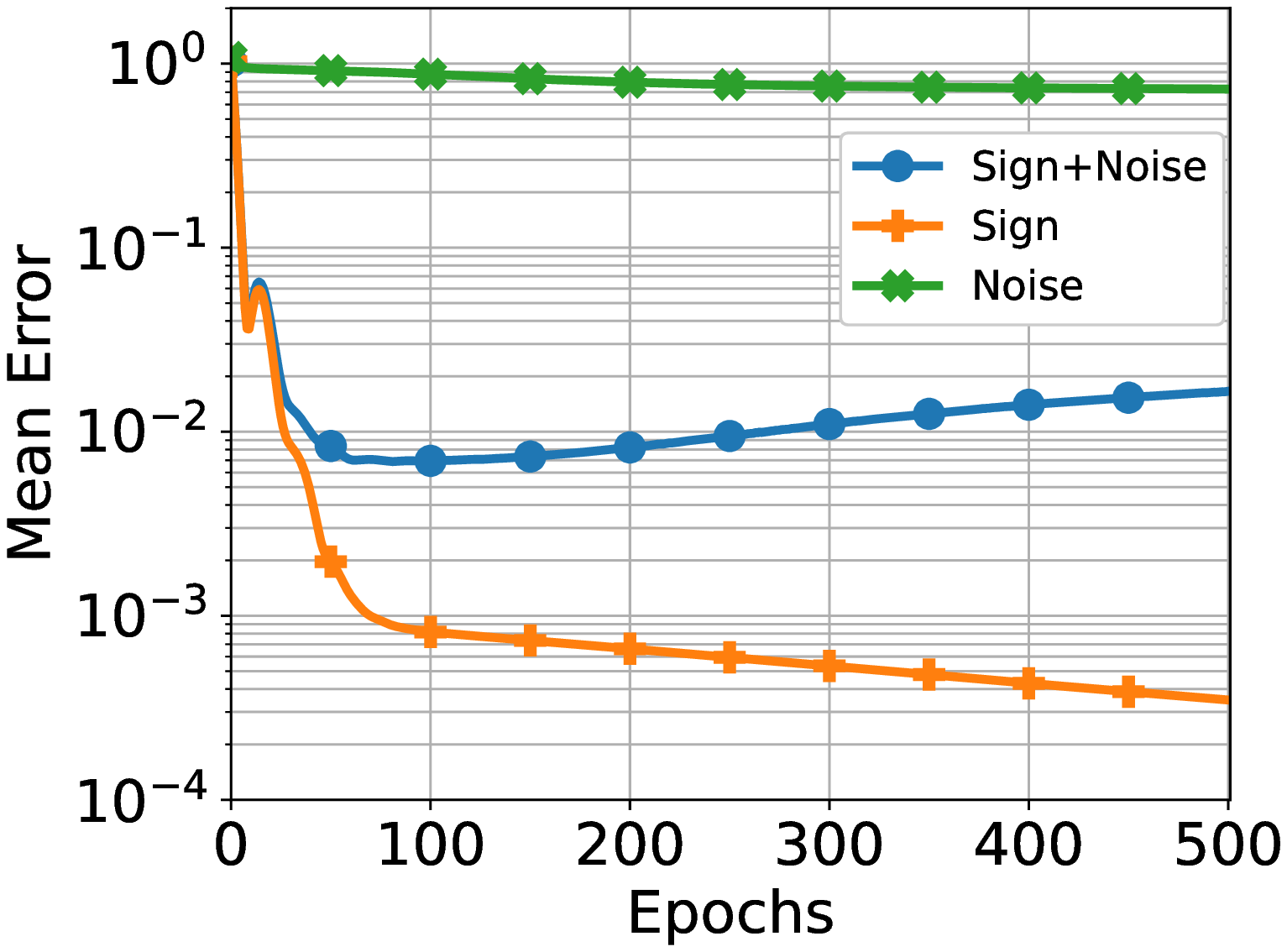}
	\end{subfigure}
	\begin{subfigure}{0.27\textwidth}
		\centering
		    \includegraphics[width=1\textwidth]{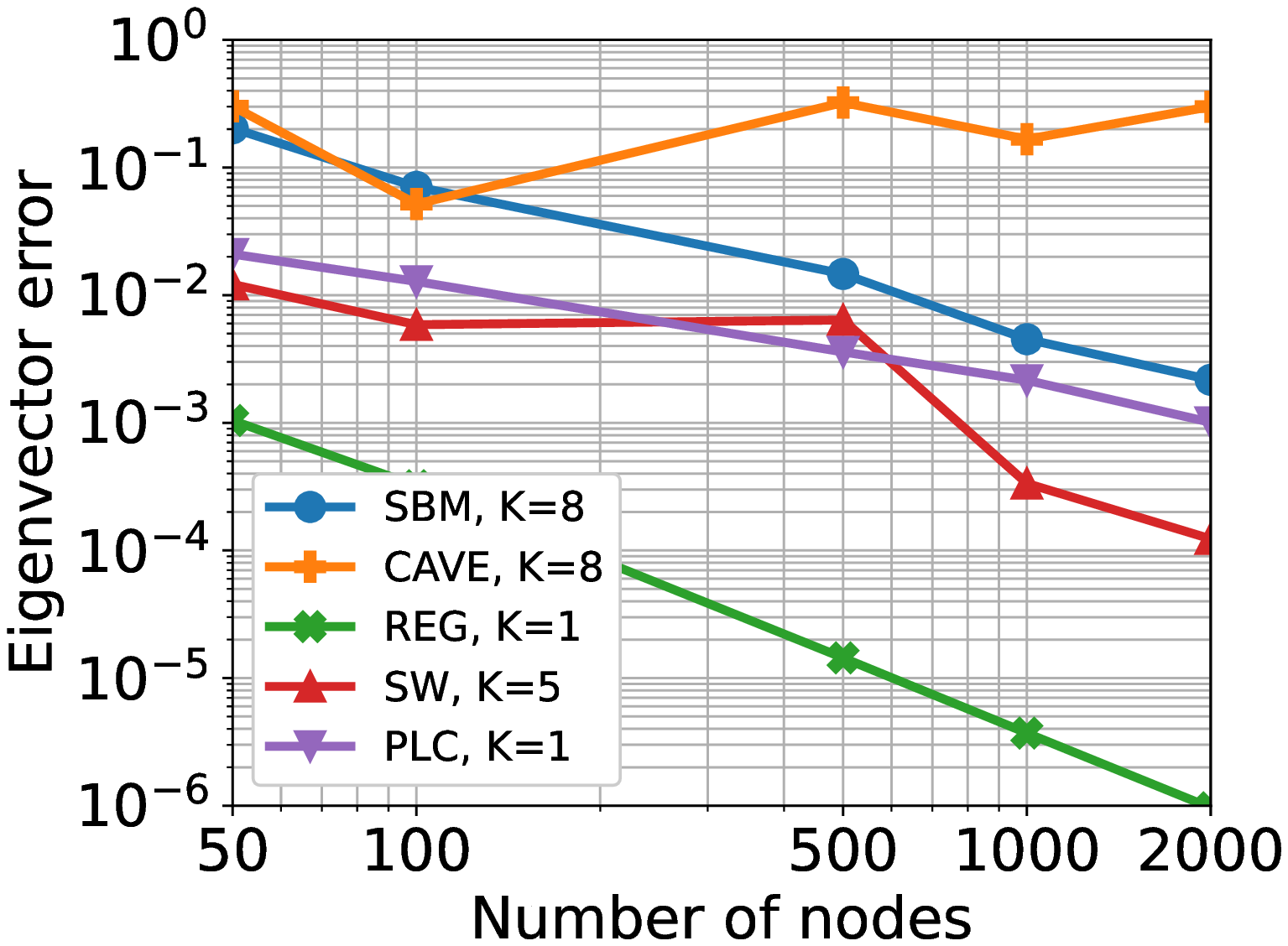}
	\end{subfigure}
	\vspace{-.2cm}
	\caption{a)~Error of the 2-layer GCG when fitting a piece-wise constant signal, noise, and a noisy signal, as a function of the number of epochs. The graph is drawn from an SBM with 64 nodes and 4 communities, and the normalized noise power is $P_n=0.1$.
	b)~Counterpart of a) but for the 2-layer GDec architecture.
	c)~Mean distance between the $K$ leading eigenvectors of the adjacency matrix and $\sqJacob$ as a function of the graph size for several graph models.}
	\label{fig:experiments1}
	\vspace{-.2cm}
\end{figure*}

In addition, the effect of adding more layers is also reflected on the smoothness assumption inherited from the construction of the upsampling operator.
Adding more layers is related to less smooth signals, since the number of nodes in $\ccalG$ with a common parent, and thus, with similar values, is smaller. 

We note that numerically illustrating that the bound between $\bbV_K$ and $\bbW_K$ holds true for the deep GDec, and that its denoising capability is not limited to signals defined over SBM graphs provide results similar to those in Sec.~\ref{S:analyze_deep_gcg}.
Therefore, instead of replicating the previous section, we directly illustrate the performance of the deep GDec under more general settings in the following section, where we present the numerical evaluation of the proposed architectures.

\begin{figure*}[!t]
	\centering
	\begin{subfigure}{0.27\textwidth}
		\centering
 \includegraphics[width=1\textwidth]{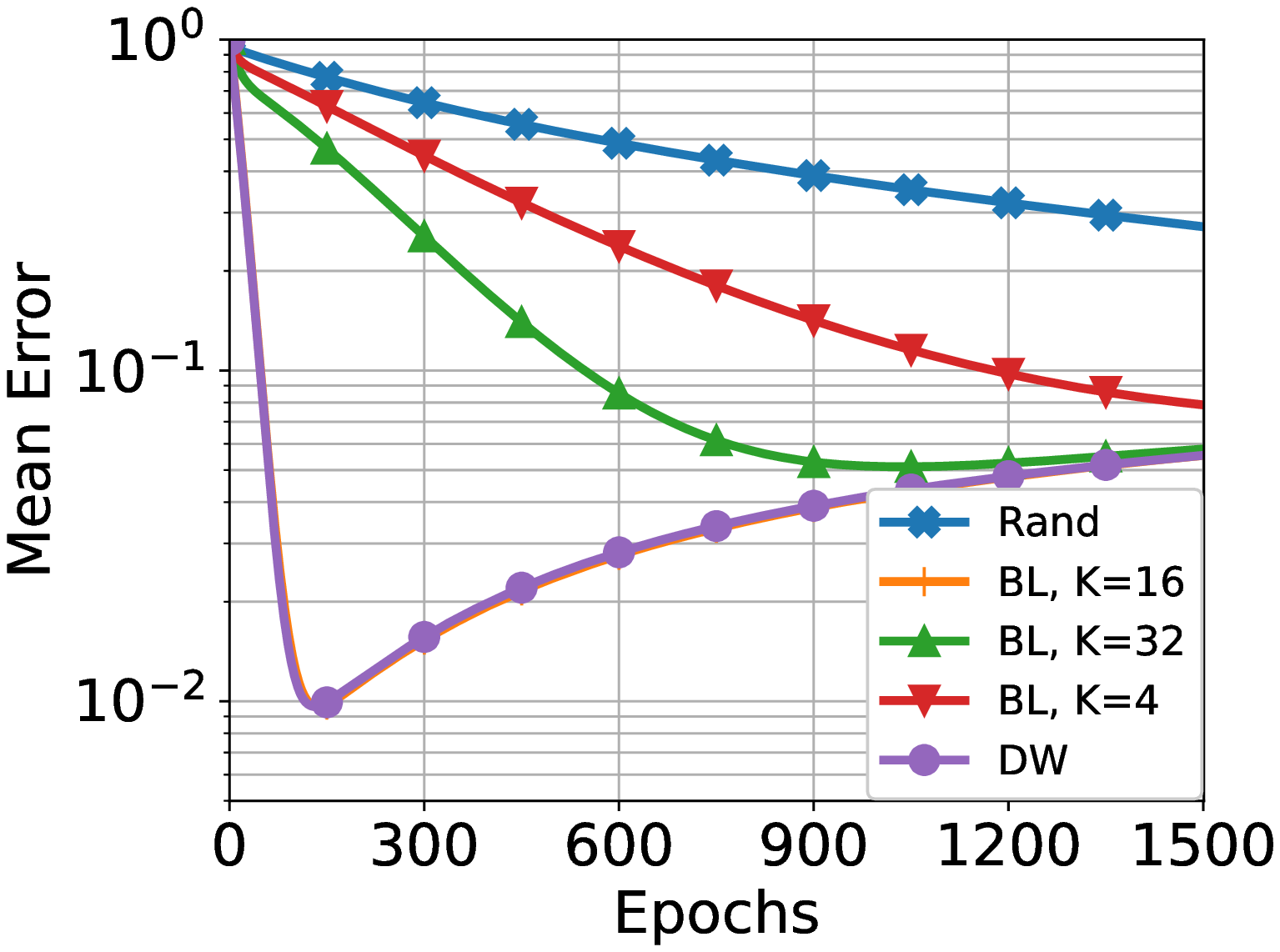}
	\end{subfigure}
	\begin{subfigure}{0.27\textwidth}
		\centering
		    \includegraphics[width=1\textwidth]{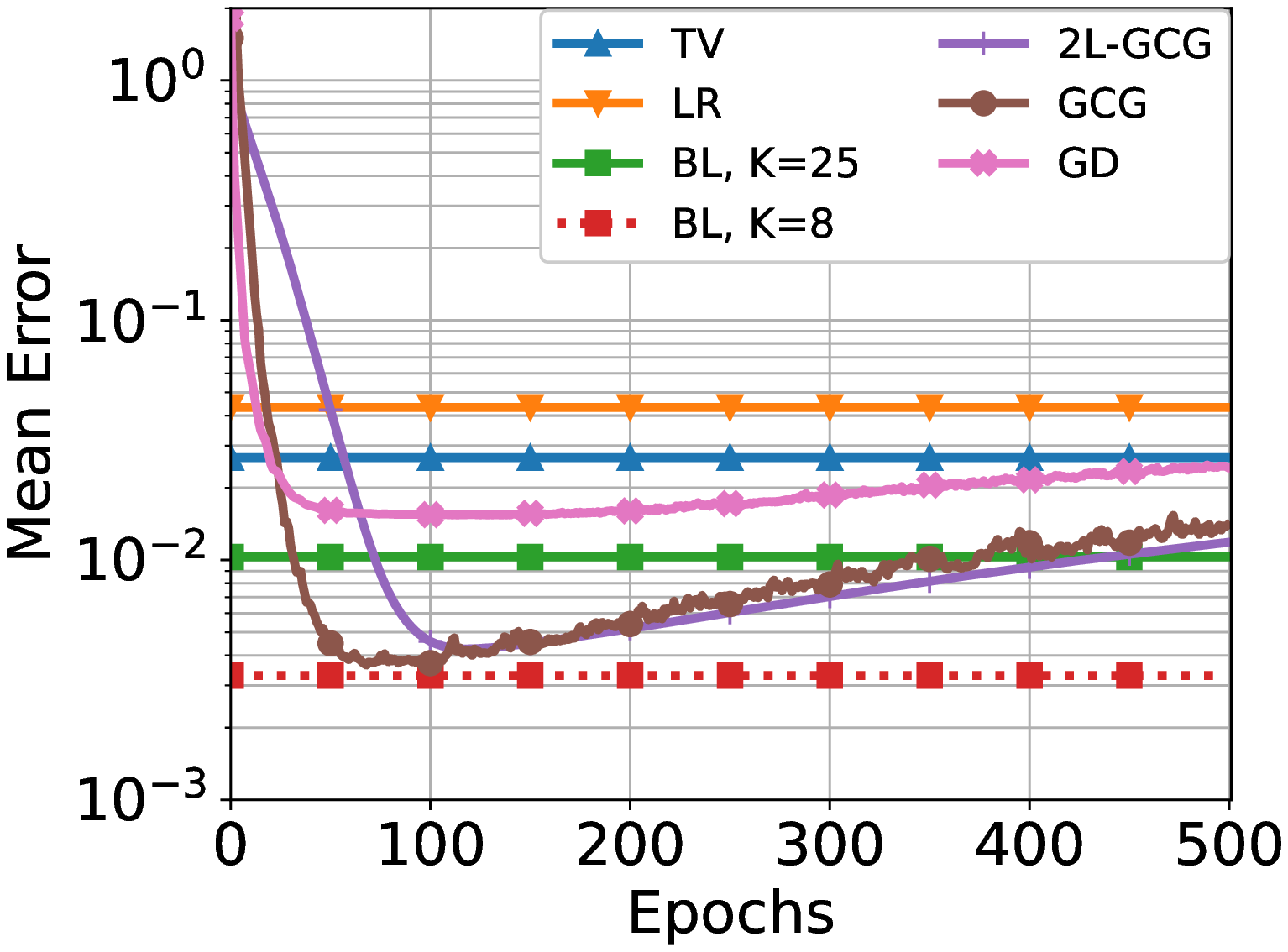}
	\end{subfigure}
	\begin{subfigure}{0.27\textwidth}
		\centering
		    \includegraphics[width=1\textwidth]{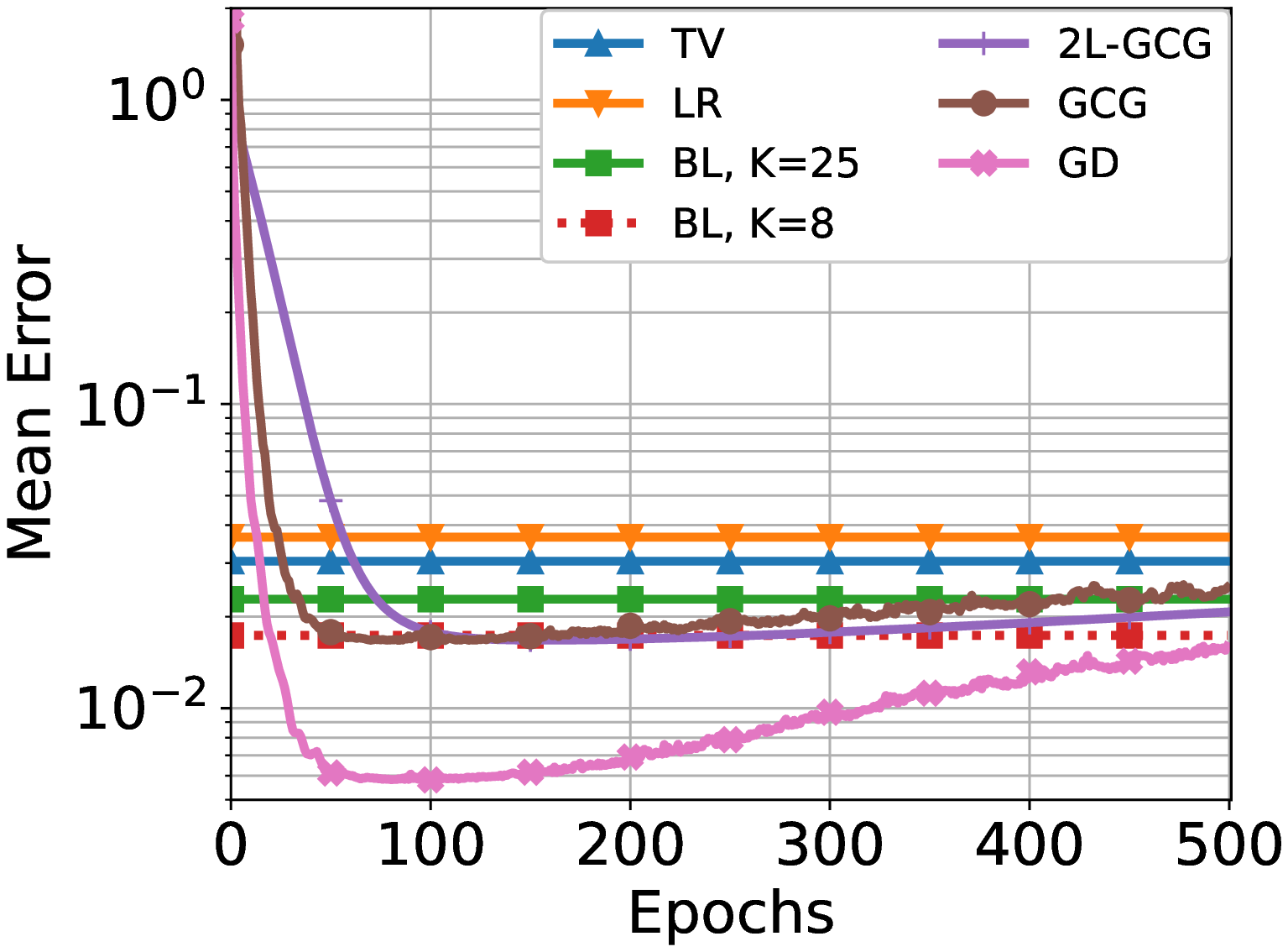}
	\end{subfigure}
	\vspace{-.25cm}
	\caption{Median MSE when denoising a graph signal as a function of the number of epochs. a)~The 2-layer GCG is used to denoise different families of signals. b)~Performance comparison between total variation, Laplacian regularization, bandlimited models, the 2-layer GCG, the deep GCG, and the deep GDec, when the signals are bandlimited.
	c)~Counterpart of b) for the case where signals are diffused white.}
	\label{fig:experiments2}
	\vspace{-.4cm}
\end{figure*}

\section{Numerical results}\label{S:experiments}
This section presents different experiments to numerically validate the theoretical claims introduced in the paper, and to illustrate the denoising performance of the GCG and the GDec.
The experiments are carried out using synthetic and real-world data, and the proposed architectures are compared to other graph-signal denoising alternatives.
The code for the experiments and the architectures is available on GitHub\footnote{\url{https://github.com/reysam93/Graph_Deep_Decoder}}.
For hyper-parameter settings and implementation details the interested reader is referred to the online available code. 

\subsection{Denoising capability of graph untrained architectures}
The goal of the experiment shown in Figs.~\ref{fig:experiments1}a and~\ref{fig:experiments1}b is to illustrate that the proposed graph untrained architectures are capable of learning the structured original signal $\bbx_0$ faster than the noise, which is one of the core claims of the paper.
To that end, we generate an SBM graph with $N=64$ nodes and $K=4$ communities, and define 3 different signals: (i)~``Signal'': a piece-wise constant signal $\bbx_0$ with the value of each node being the label of its community; (ii)~``Noise'': zero-mean white Gaussian noise $\bbn$ with unit variance; and (iii)~``Signal + Noise'': a noisy observation $\bbx=\bbx_0+\bbn$ where the noise has a normalized power of $0.1$.
Figs.~\ref{fig:experiments1}a and~\ref{fig:experiments1}b show the normalized mean squared error (NMSE), with the error for each realization being $\|\bbx_0-\nolinebreak \hbx_0\|_2^2/\|\bbx_0\|_2^2$.
The mean is computed for 100 realizations of the noise as the number of epochs increases when the different signals are fitted by the 2-layer GCG and the 2-layer GDec, respectively.
It can be seen how, in both cases, the error when fitting the noisy signal $\bbx$ decreases for a few epochs until it reaches a minimum, and then starts to increase.
This is because the proposed untrained architectures learn the signal $\bbx_0$ faster than the noise, but if they fit the observation for too many epochs, they start learning the noise as well and, hence, the MSE increases.
As stated by Th.~\ref{theorem_denoising_gcg} and Th.~\ref{theorem_denoising_gd}, this result illustrates that, if early stopping is applied, both architectures are capable of denoising the observed graph signals without a training step. 
It can also be noted that, under this setting, the GDec learns the signal $\bbx_0$ faster than the GCG and, at the same time, is more robust to the presence of noise.
This can be seen as a consequence of GDec implicitly making stronger assumptions about the smoothness of the targeted signal. 

The goal of the second test case is two-fold.
First, it illustrates that the result presented in Lemma~\ref{lemma_eigs_gcg} is not constrained to the family of SBM (as specified by Ass.~1), but can be generalized to other families of random graphs as well.
In addition, it measures the influence of the number of nodes in the discrepancies between $\bbV_K$ and $\bbW_K$.
To that end, Fig. \ref{fig:experiments1}c  contains the mean eigenvector similarity measured as $\frac{1}{K}\|\bbV_K-\bbW_K\bbQ\|_F$ as a function of the number of nodes in the graph.
The eigenvector similarity is computed for 50 realizations of random graphs and the presented error is the median of all the realizations. 
The random graph models considered are: the SBM (``SBM''), the connected caveman graph (``CAVE'')\cite{watts1999networks}, the regular graph whose fixed degree increases with its size (``REG''), the small world graph (``SW'')\cite{watts1998collective}, and the power law cluster graph model (``PLC'')\cite{holme2002growing}.
The second term in the legend denotes the number of leading eigenvectors taken into account in each case, which depends on the number of active frequency components of the specific random graph model.   
We can clearly observe that for most of the random graph models, the eigenvector error goes to 0 as $N$ increases and, furthermore, the error is below $10^{-1}$ even for moderately small graphs.
This illustrates that, although the conditions assumed for Lemma~\ref{lemma_eigs_gcg} and Lemma~\ref{lemma_eigs_gd} focus on the specific setting of the SBM, the results can be applied to a wider class of graphs.
Here, the regular graphs are particularly interesting since most classical signals may be interpreted as signals defined over regular graphs.
As a result, this empirical evidence motivates the extension of the proposed theorems to more general settings as a future line of work.

\subsection{Denoising synthetic data}
We now proceed to comment on the denoising performance of the proposed architectures with synthetic data.
The usage of synthetic signals allows us to study how the properties of the noiseless signal influence the quality of the denoised estimate.

The first experiment,  shown in Fig.~\ref{fig:experiments2}a, studies the error of the denoised estimate obtained with the 2-layer GCG as the number of epochs increases.
The reported error is the NMSE of the estimated signal $\hbx_0$, and the figure shows the mean values of 100 realizations of graphs and graph signals.
The normalized power of the noise present in the data is $0.1$.
Graphs are drawn from an SBM with $N=64$ nodes and 4 communities, and the graph signals are generated as: (i)~a  zero-mean  white Gaussian noise with unit variance (``Rand'');
(ii)~a bandlimited graph signal (cf. \ref{E:bl_signals}) using the $K$ leading eigenvectors of $\bbA$ as base (``BL''); and (iii)~a diffused white (``DW'') signal created as $\bby = \mathrm{med}(\bbH\bbw|{\ccalG})$, where $\bbw$ is a white vector whose entries are sampled from $\ccalN(0,1)$, $\bbH$ is a low-pass graph filter, and $\mathrm{med}(\cdot|{\ccalG})$ represents the graph-aware median operator such that the value of the node $i$ is the median of its neighborhood~\cite{tay2020time, segarra2017designmedian}.
The results in Fig.~\ref{fig:experiments2}a show that the best denoising error is obtained when the signal is composed of just a small number of eigenvectors, and the performance deteriorates as the bandwidth (i.e., the number of eigenvectors that span the signal subspace) increases, obtaining the worst result when the signal is generated at random.
This result is aligned with the theoretical claims since it is assumed that the signal $\bbx_0$ is bandlimited.
It is also worth noting that the architecture also achieves a good denoising error with the ``DW'' model, showcasing that the GCG is also capable of denoising other types of smooth graph signals.

Next, Fig.~\ref{fig:experiments2}b compares the performance of the 2-layer GCG (``2L-GCG''), the deep GCG (``GCG'') and the deep GDec (``GDec'') with the baseline models introduced in Sec.~\ref{S:GNN_for_inverse_problems}, which are the total variation (``TV'')\cite{chen2014signal}, Laplacian regularization (``LR'')\cite{pang2017graph}, and bandlimited model (``BL'')\cite{chen2015discrete}.
In this setting, the graphs are SBM with 256 nodes and 8 communities, and the signals are bandlimited with a bandwidth of 8.
Since the ``BL'' model with $K=8$ captures the actual generative model of the signal $\bbx_0$, it achieves the best denoising performance.
However, it is worth noting that the GCG obtains a similar result, outperforming the other alternatives.
On the other hand, the ``LR'' obtains an error noticeably larger than that of ``BL'' and ``GCG'', highlighting that, even though ``BL'' and ``LR'' are related models their different assumptions lead to different performances.
Moreover, the benefits of using the deep GCG instead of the 2-layer architecture are apparent, since it achieves a better performance in fewer epochs.

On the other hand, Fig.~\ref{fig:experiments2}c illustrates a similar experiment but with the graph signals generated as ``DW''.
Under this setting, it is clear that the GDec outperforms the other alternatives.
These results showcase the benefits of employing a nonlinear architecture relative to classical denoising approaches.
Furthermore, this experiment corroborates that the GDec is more robust to the presence of noise when the signals are aligned with the prior implicitly captured by the architecture.

\begin{table*}[t]
\begin{center}
\vspace{-4mm}
\caption{{Denoising error of several datasets with different types of random noise}}
\label{T:real_data}
\begin{tabular}{ c c || c c c c || c c c c c || c c}
\toprule
\begin{tabular}{c}
    DATASET  \\
    (METRIC)
\end{tabular} & METHOD & BL & TV & LR & GTF & MED & GCNN & GAT & K-GAE & GUSC & GCG & GDec \\ \midrule 
TEMPERATURE & Gaussian & 0.062  & 0.117  & 0.095  & 0.066 & 0.053 & 0.123 & 0.045 & 0.134 & 0.044 & 0.056 & \textbf{0.035} \\
 (NMSE)   & Uniform  & 0.063  & 0.117  & 0.094  & 0.064 & 0.053 & 0.118 & 0.047 & 0.136 & 0.049 & 0.057 & \textbf{0.036} \\ \midrule
S\&P 500 & Gaussian & 0.350 & 0.238 & 0.231 & 0.239 & 0.319 & 0.252 & 0.199 & 0.354 & 0.203 & \textbf{0.188} & \textbf{0.188} \\
 (NMSE) & Uniform & 0.216 & 0.246 & 0.161 & 0.298 & 0.340 & \textbf{0.091} & 0.222 & 0.273 & 0.127 & \textbf{0.094} & 0.121 \\ \midrule
CORA & Whole $\ccalG$ & 0.154 & 0.142 & 0.115 & 0.126 & 0.167 & 0.099 & 0.141 & 0.135 & 0.099 & \textbf{0.093} & 0.121 \\
 (ERROR RATE) & Conn. comp. & 0.151 & 0.141 & 0.105 & 0.116 & 0.165 & 0.093 & 0.139 & 0.135 & 0.094 & \textbf{0.088} & 0.125 \\ \bottomrule
\end{tabular}
\end{center}
\vspace{-5mm}
\end{table*}

\subsection{Denoising real-world signals}
{
Finally, we assess the performance of the proposed architectures in several real-world datasets.
To the baselines considered in the previous experiments, we add the following competitive denoising algorithms: graph trend filtering (``GTF'')~\cite{wang2015trend}, a graph-aware median operator (``MED'')~\cite{tay2020time}, a GCNN (``GCNN'') implemented as in~\cite{kipf2016semi}, a graph attention network (``GAT'')~\cite{velivckovic2017graph}, a Kron reduction-based autoencoder (``K-GAE'')~\cite{dorfler2012kron}, and the graph unrolling sparse coding architecture (``GUSC'') in~\cite{chen2021graph}.
Moreover, we consider the following noise distributions: (i) zero-mean Gaussian distribution, which is the noise model typically assumed for sensor measurements in signal processing; (ii) uniform distribution on some interval $[0, a]$, where $a\in\reals_+$ is chosen accordingly to the desired noise power; and (iii) Bernoulli distribution to model errors in binary signals.
Next, we describe the selected datasets and analyze the achieved results, which are summarized in Table~\ref{T:real_data}.

\vspace{1mm}
\noindent\textbf{Temperature.}
We consider a network of 316 weather stations distributed across the United States~\cite{sandryhaila2014discrete}.
Graph signals represent daily temperature measurements in the first three months of the year 2003.
The graph $\ccalG$ represents the geographical distance between weather stations and is given by the 8-nearest neighbors graph.
The first and second rows of Table~\ref{T:real_data} list the NMSE when the noise is drawn from a Gaussian and a uniform distribution, respectively.
In both cases, the noise has a normalized power of 0.3.
It is clear that the GDec architecture outperforms the alternatives in both scenarios.
Furthermore, we can observe that the GCG achieves a better performance than GCNN, showcasing the benefits of being able to use a more general graph filter.

\vspace{1mm}
\noindent\textbf{S\&P 500.}
In this experiment, we have 189 nodes representing stocks belonging to 6 different sectors of the S\&P 500 with the graph signals representing the prices of those stocks at particular time instants.
We follow \cite{cardoso2020algorithms} to estimate the graph $\ccalG$ assuming that the signals are drawn from a multivariate Gaussian distribution and are smooth on $\ccalG$.
We consider the noise specifications described in the previous dataset and provide the NMSE in the third and fourth rows of Table~\ref{T:real_data}.
It is worth noting that considering Gaussian noise in this dataset constitutes a more challenging denoising problem than using uniform noise.
A plausible explanation is that the graph is estimated assuming that the data follows a Gaussian distribution, and hence, it is harder to separate the Gaussian noise from the true signals. 
In the presence of Gaussian noise, the GCG and the GDec outperform the other 8 alternatives.
However, when the noise follows a uniform distribution, the best performance is obtained by the GCG and the GCNN, with GDec being the third best.
In addition, we observe that traditional methods yield an error that is considerably larger than that incurred by the proposed architectures.
This is aligned with our initial intuition about linear and quadratic methods being more limited when the actual relation between $\bbx_0$ and $\ccalG$ is more intricate, as is the case for financial data.

\vspace{1mm}
\noindent\textbf{Cora.}
Lastly, we consider the Cora citation network dataset~\cite{kipf2016semi}.
Nodes represent different scientific documents and edges capture citations among them.
Like in \cite{chen2021graph}, we consider the 7 class labels as binary graph signals encoding if the particular node belongs to that class.
For each signal, we consider 25 realizations of Bernoulli noise that randomly flips 30\% of the binary values of the signals, resulting in a total of 175 noisy graph signals.
With the error rate denoting the proportion of labels correctly recovered after the denoising process, Table~\ref{T:real_data} shows the error metric averaged over all the signals.
Moreover, since the graph is formed by several connected components, we report two results: the error rate when the whole graph is considered (fifth row) and the error rate when only the largest connected component is considered (sixth row).
It can be seen that the GCG yields the best performance in both cases.
}

\section{Conclusion}\label{S:conclusion}
In this paper, we faced the relevant task of graph-signal denoising. 
To approach this problem, we presented two overparametrized and untrained GNNs and provided theoretical guarantees on the denoising performance of both architectures when denoising $K$-bandlimited graph signals under some simplifying assumptions.
Moreover, we numerically illustrated that the proposed architectures are also capable of denoising graph signals in more general settings.  
The key difference between the two architectures resided in the linear transformation that incorporates the information encoded in the graph.
The GCG employs fixed (non-learnable) low-pass graph filters to model convolutions in the vertex domain, promoting smooth estimates.
On the other hand, the GDec relies on a nested collection of graph upsampling operators to progressively increase the input size, limiting the degrees of freedom of the architecture, and providing more robustness to noise.
In addition to the aforementioned analysis, we tested the validity of the proposed theorems and evaluated the performance of both architectures with real and synthetic datasets, showcasing a better performance than other classical and nonlinear methods for graph-signal denoising.
Finally, we consider extending the results from Th.~\ref{theorem_denoising_gcg} and Th.~\ref{theorem_denoising_gd} to more general scenarios as an interesting future line of work.

\appendices
\section{Proof of Th.~\ref{theorem_denoising_gcg}}\label{proof_theorem_denoising_gcg}
Let $\bbx_0$ be a $K$ bandlimited graph signal as described in \eqref{E:bl_signals}, which is spanned by the $K$ leading eigenvectors of the graph $\bbV_K$, with $\tbx_0$ denoting its frequency representation.
Let $\bbQ$ be an orthonormal matrix that aligns the subspaces spanned by $\bbV_K$ and $\bbW_K$, and denote as $\barbx_0=\bbW_K\bbQ\tbx_0$ the bandlimited signal using $\bbW_K$ as basis and whose frequency response is also $\tbx_0$.
Note that $\barbx_0$ can be interpreted as recovering $\bbx_0$ from its frequency response using $\bbW_K$ in lieu of $\bbV_K$.
Also, note that $\bbx_0-\barbx_0=(\bbV_K-\bbW_K\bbQ)\tbx_0$ represents the error between the signal $\bbx_0$ and its approximation inside the subspace spanned by $\bbW_K$.
With these definitions in place, in~\cite[Th. 3]{heckel2019denoising} the authors showed that error when denoising a signal $\bbx = \bbx_0 + \bbn$ is bounded with probability at least $1-e^{-F^2}-\phi$ by
\begin{align}\label{E:bound_theorem_eigs_J}
    & \|\bbx_0-f_{\bbTheta_{(t)}}(\bbZ|\ccalG)\|_2 \leq  \|\bbPsi\bbx_0\|_2+\xi\|\bbx\|_2\\ \nonumber
    & +\sqrt{\textstyle \sum_{i=1}^N((1-\eta\sigma_i^2)^t-1)^2(\bbw_i^\top\bbn)^2},
\end{align}
with $\bbPsi:=\bbW(\bbI_N-\eta\bbSigma^2)^t\bbW^\top$, and $\bbI_N$ the $N\times N$ identity matrix.
However, note that the bound provided for $\|\bbPsi\bbx_0\|_2$ in \cite{heckel2019denoising} requires $\bbx_0$ lying in the subspace spanned by $\bbW_K$, which is not the case.
As a result, we further bound this term as
\begin{align}\label{E:bound_basis_mismatch}
    \|\bbPsi  \bbx_0 \|_2 &= \|\bbPsi(\bbx_0+\barbx_0-\barbx_0)\|_2 \nonumber\\ 
    & \stackrel{(i)}{=} \|\bbPsi_K\barbx_0+\bbPsi(\bbV_K-\bbW_K\bbQ)\tbx_0\|_2 \nonumber \\
    &\stackrel{(ii)}{\leq}\|\bbPsi_K\barbx_0\|_2+\|\bbPsi(\bbV_K-\bbW_K\bbQ)\tbx_0\|_2 \nonumber \\
    &\stackrel{(iii)}{\leq}\|\bbPsi_K\|_2\|\barbx_0\|_2+ \|\bbPsi\|_2\|\bbV_K-\bbW_K\bbQ\|_F\|\tbx_0\|_2 \nonumber \\
    &\stackrel{(iv)}{\leq}\left(\|\bbPsi_K\|_2 + \delta\|\bbPsi\|_2\right)\|\bbx_0\|_2 \nonumber \\
    &\stackrel{(v)}{=} \left((1-\eta\sigma_K^2)^t+\delta(1-\eta\sigma_N^2)^t\right)\|\bbx_0\|_2.
\end{align}
Here, $\bbPsi_K:=\bbW_K(\bbI_K-\eta\bbSigma_K^2)^t\bbW_K^\top$, and $\bbSigma_K$ represents a diagonal matrix containing the first $K$ leading eigenvalues $\sigma_k$.
We have that $(i)$ follows from $\barbx_0$ being bandlimited in $\bbW_K$, so $\bbPsi\barbx_0=\bbPsi_K\barbx_0$.
Then, $(ii)$ follows from the triangle inequality, and $(iii)$ from the $\ell_2$ norm being submultiplicative and using the Frobenius norm as an upper bound for the $\ell_2$ norm.
In $(iv)$ we apply the result of Lemma~\ref{lemma_eigs_gcg}, which holds with probability at least $1-\epsilon$ because $N>{N_{K,\epsilon,\delta}}$, and the fact that, since both $\bbW_K$ and $\bbV_K$ are orthonormal matrices, we have that $\|\bbx_0\|_2=\|\barbx_0\|_2=\|\tbx_0\|_2$.
We obtain $(v)$ from the largest eigenvalues present in $\bbPsi_K$ and $\bbPsi$.

Finally, the proof concludes by combining \eqref{E:bound_basis_mismatch} and \eqref{E:bound_theorem_eigs_J}.

\section{Proof of Lemma \ref{lemma_eigs_gcg}}\label{proof_lemma_eigs_gcg}
Define $\ccaltbA$ as $\ccaltbA:=\mathbb{E}[\tbA]=\mathbb{E}[\bbD]^{-\frac{1}{2}}\ccalbA\mathbb{E}[\bbD]^{-\frac{1}{2}}$ and let $\sqJacob$ be given by \eqref{eq:expected_jac}. 
Denote by $\ccalbH$ a graph filter defined as a polynomial of the expected adjacency matrix $\ccaltbA$, and let $\barsqJacob$ be the expected squared Jacobian using the graph filter $\ccalbH$, i.e.,
\begin{equation}\label{eq:expected_E}
    \barsqJacob = 0.5 \left( \mathbf{1} \mathbf{1}^\top - \frac{1}{\pi} \arccos(\ccalbC^{-1} \ccalbH^2 \ccalbC^{-1})\right) \odot \ccalbH^2,
\end{equation}
where $\ccalbC$ is the counterpart of $\bbC$ in \eqref{eq:expected_jac}, but using $\ccalbH$ instead of $\bbH$.
Given the following eigendecompositions $\tbA=\bbV\bbLambda\bbV^\top$, $\sqJacob=\bbW\bbSigma\bbW^\top$, $\ccaltbA=\barbV\barbLambda\barbV^\top$, and $\barsqJacob=\barbW\barbSigma\barbW^\top$, for arbitrary orthonormal matrices $\bbT$ and $\bbR$, we have that
\begin{align}\label{E:main_triangle_ineq}
& \| \bbV_K - \bbW_K \bbQ \|_{\text{F}} \leq  \\ \nonumber
& \| \bbV_K - \bar{\bbV}_K \bbT \|_{\text{F}} + \| \bar{\bbV}_K \bbT - \bar{\bbW}_K \bbR \|_{\text{F}} + \| \bar{\bbW}_K \bbR - \bbW_K \bbQ \|_{\text{F}}.
\end{align}
To prove the theorem, we bound the three terms on the right hand side of \eqref{E:main_triangle_ineq}.

\vspace{1mm}
\noindent \emph{Bounding $\| \barbV_K \bbT - \barbW_K \bbR \|_{\text{F}}$.}
From the definition of an SBM, it follows that $\ccalbA=\mathbb{E}[\bbA]=\bbB\bbOmega\bbB^\top$, where $\bbB \in \{0,1\}^{N\times K}$ is an indicator matrix encoding the community to which each node belongs, and $\bbOmega$ is a $K \times K$ matrix encoding the link probability between the communities of the graph.
Therefore, $\ccaltbA$ and $\barsqJacob$ are both block matrices whose blocks coincide with the communities in the SBM.
This implies that the eigenvectors associated with non-zero eigenvalues must span the columns of $\bbB$. 
Hence, the leading eigenvectors must be related by an orthonormal transformation, from where it follows that, given $\bbT$, we can always find $\bbR$ such that
\begin{equation}\label{E:first_bound}
\| \bar{\bbV}_K \bbT - \bar{\bbW}_K \bbR \|_{\text{F}} = 0.
\end{equation}

\vspace{1mm}
\noindent \emph{Bounding $\| \bbV_K - \bar{\bbV}_K \bbT \|_{\text{F}}$.} 
Under Ass.~\ref{A:sbm}, as it is shown in~\cite{schaub2020blind}, with probability at least $1-\rho$ we have that
\begin{equation}\label{E:second_bound_aux}
    \| \tilde{\bbA} - \ccaltbA\| \leq 3 \sqrt{\frac{3 \ln(4N/\rho)}{\beta_{\min}}}.
\end{equation}
Then, we combine the concentration \eqref{E:second_bound_aux} with the Davis-Kahan results~\cite[Th. 2]{yu2015useful}, which bound the distance between the subspaces spanned by the population eigenvectors ($\barbV_K$) and their sample version ($\bbV_K$).
Denoting as $\bar{\lambda}_i$ the $i$-th eigenvalue collected in $\barbLambda$, i.e. $\bar{\lambda}_i=\bar{\Lambda}_{ii}$, we obtain that there exists an orthonormal matrix $\bbT$ such that
\begin{align}\label{E:second_bound}
    \|\bbV_K-\barbV_K\bbT\|_F &\leq \frac{\sqrt{8K}}{\bar{\lambda}_{K}-\bar{\lambda}_{K+1}}\|\tbA-\ccaltbA\|_F \nonumber\\
    & \leq \frac{3\sqrt{8K}}{\bar{\lambda}_{K}}\sqrt{\frac{3 \ln(4N/\rho)}{\beta_{\min}}},
\end{align}
where we note that, since $\ccaltbA$ follows an SBM, then $\bar{\lambda}_i=0$ for all $i>K$.

Since $\beta_{\min} = \omega(\ln (N/\rho))$, we obtain that 
\begin{equation}\label{E:second_bound_lim}
\| \bbV_K - \bar{\bbV}_K \bbT \|_{\text{F}} \to 0, \qquad \text{as } N \to \infty.
\end{equation}

\vspace{1mm}
\noindent \emph{Bounding $\| \bar{\bbW}_K \bbR - \bbW_K \bbQ \|_{\text{F}}$.}
If we show that $\|\sqJacob - \barsqJacob\| \to 0$ as $N \to \infty$, we can then mimic the procedure in~\eqref{E:second_bound_aux} and~\eqref{E:second_bound} to show that the difference between the leading $K$ eigenvectors of $\sqJacob$ and $\barsqJacob$ also vanishes. 
Hence, we are left to show that $\|\sqJacob - \barsqJacob\| \to 0$ as $N \to \infty$. From the definitions of $\sqJacob$ and $\barsqJacob$, it follows that
\begin{align}\label{E:third_bound_aux}
& \|\sqJacob - \barsqJacob\| \leq 0.5 \| \bbH^2 - \ccalbH^2 \| \\ \nonumber
& + \frac{1}{2\pi} \| \arccos(\ccalbC^{-1} \ccalbH^2 \ccalbC^{-1}) \odot \ccalbH^2 - \\ \nonumber
&\arccos(\bbC^{-1} \bbH^2 \bbC^{-1}) \odot \bbH^2 \|.
\end{align}
To bound the difference between the sampled and expected filters, we have that
\begin{align}\label{E:third_bound_aux_2}
&\| \bbH^2 - \ccalbH^2 \| = \left\| \left(\sum_{\ell=0}^L h_\ell \tilde{\bbA}^\ell \right)^2 - \left(\sum_{\ell=0}^L h_\ell \ccaltbA^\ell\right)^2 \right\| \\ \nonumber
&= \left\| \sum_{\ell=0}^{2L} \alpha_\ell (\tilde{\bbA}^\ell - \ccaltbA^\ell) \right\| \leq \sum_{\ell=0}^{2L} \alpha_\ell \left\| \tilde{\bbA}^\ell - \ccaltbA^\ell\right\|,
\end{align}
for suitable coefficients $\alpha_\ell$ and recalling that $L=2$. Then,
we can then leverage the fact that $\|\tilde{\bbA}\| = \|\ccaltbA\| = 1$ to see that $\left\| \tilde{\bbA}^\ell - \ccaltbA^\ell\right\| \leq \ell \left\| \tilde{\bbA} - \ccaltbA\right\|$. We thus get that
\begin{equation}\label{E:third_bound_aux_3}
\| \bbH^2 - \ccalbH^2 \| \leq \sum_{\ell=0}^{2L} \ell \alpha_\ell  \left\| \tilde{\bbA} - \ccaltbA\right\| \to 0, \quad \text{as } N \to \infty,
\end{equation}
where the limiting behavior follows from~\eqref{E:second_bound_aux}. Finally, to bound the second term in~\eqref{E:third_bound_aux}, we first note that the argument of the norm can be re-written as $\arccos(\bbC^{-1} \bbH^2 \bbC^{-1}) \odot (\ccalbH^2 - \bbH^2) +  (\arccos(\ccalbC^{-1} \ccalbH^2 \ccalbC^{-1}) - \arccos(\bbC^{-1} \bbH^2 \bbC^{-1})) \odot \ccalbH^2$. The limit in~\eqref{E:third_bound_aux_3} ensures that the first of these two terms vanishes. 
Similarly, it follows that $\| \ccalbC^{-1} \ccalbH^2 \ccalbC^{-1} - \bbC^{-1} \bbH^2 \bbC^{-1} \| \to 0$ which, combined with the fact that $\arccos$ is a uniformly continuous function, we can always find an $N_{\delta'}$ such that $\| \arccos(\ccalbC^{-1} \ccalbH^2 \ccalbC^{-1}) - \arccos(\bbC^{-1} \bbH^2 \bbC^{-1})\| \leq \delta'$ with high probability. Combining this result with~\eqref{E:third_bound_aux_3} and applying the Davis-Kahan Theorem as done to obtain~\eqref{E:second_bound} we get that
\begin{equation}\label{E:third_bound_lim}
\| \bar{\bbW}_K \bbR - \bbW_K \bbQ \|_{\text{F}} \to 0, \qquad \text{as } N \to \infty.
\end{equation}

\vspace{3mm}
Replacing~\eqref{E:first_bound}, \eqref{E:second_bound_lim}, and \eqref{E:third_bound_lim} into~\eqref{E:main_triangle_ineq} our result follows.

\section{Proof of Lemma \ref{lemma_eigs_gd}}\label{proof_lemma_eigs_gd}

Recall that $\ccaltbA=\mathbb{E}[\tbA]$, and define  $\ccaltbH:=\gamma\bbI+(1-\gamma)\ccaltbA$ as the specific graph filter introduced in Sec. \ref{S:upsampling_operator} as a polynomial of $\ccaltbA$.
Let $\sqJacob$ be given by equation \eqref{eq:expected_jac_dec}, and denote by $\barsqJacob$ the expected squared Jacobian using the graph filter $\ccalbH$, i.e.,
\begin{align}\label{eq:expected_E_dec}
    \barsqJacob = 0.5 \left( \mathbf{1} \mathbf{1}^\top - \frac{1}{\pi} \arccos(\ccaltbC^{-1}  \ccalbU\ccalbU^\top\ccaltbC^{-1})\right) \odot \ccalbU\ccalbU^\top
 \end{align}
with $\ccalbU=\ccaltbH\bbP$ and where the matrix $\ccaltbC$ is the counterpart of $\tbC$ in \eqref{eq:expected_jac_dec}, but using $\ccalbU$ in lieu of $\bbU$.
Given the eigendecompositions $\tbA=\bbV\bbLambda\bbV^\top$, $\sqJacob=\bbW\bbSigma\bbW^\top$, $\ccaltbA=\barbV\barbLambda\barbV^\top$, and $\barsqJacob=\barbW\barbSigma\barbW^\top$, analogously to Lemma~\ref{lemma_eigs_gcg}, we bound the difference between $\bbV_K$ and $\bbW_K$ by bounding the three terms in the right hand side of \eqref{E:main_triangle_ineq}.

\vspace{1mm}
\noindent\textit{Bounding $\|\bar{\bbV}_K\bbT-\bar{\bbW}_K\bbR\|$}. 
We have that $\ccalbU\ccalbU^\top=\ccaltbH\bbP\bbP^\top\ccaltbH^\top$.
Since $\bbP$ is a binary matrix indicating to which community belongs each node, $\bbP\bbP^\top$ is a block diagonal matrix that captures the structure of the communities of the SBM.
Then, because $\ccaltbH$ is also block matrix with the same block pattern that the SBM, it turns out that the matrix $\barsqJacob$ is also a block matrix whose blocks coincide with the communities in the SBM graph.
Therefore, the rest of the bound is analogous to that in Lemma~\ref{lemma_eigs_gcg}.

\vspace{1mm}
\noindent\textit{Bounding $\|\bbV_K-\bar{\bbV}_K\bbT\|$}. The relation between $\bbA$ and $\ccalbA$ is the same as in Lemma~\ref{lemma_eigs_gcg} so the bound provided in \eqref{E:second_bound_lim} holds.

\vspace{1mm}
\noindent\textit{Bounding $\|\bar{\bbW}_K\bbR-\bbW_K\bbQ\|$}. To derive this bound we show that $\|\bbU\bbU^\top-\ccalbU\ccalbU^\top\|=\|\tbH\bbP\bbP^\top\tbH^\top-\ccaltbH\bbP\bbP^\top\ccaltbH^\top\|$ goes to 0 as $N$ grows.
From \eqref{E:third_bound_aux_3} we have that $\|\bbH-\ccalbH\|\to0$, as $N\to\infty$, and hence, $\|\tbH-\ccaltbH\|\to0$, as $N\to\infty$.
Therefore, it can be seen that
\begin{equation}
    \|\bbU\bbU^\top-\ccalbU\ccalbU^\top\|  \to 0, \quad \text{as } N \to \infty,
\end{equation}
with $\|\bbU\bbU^\top-\ccalbU\ccalbU^\top\|$ vanishing as $N$ grows. The remainder of the derivation of the bound is analogous to that for \eqref{E:third_bound_lim}.

\ifCLASSOPTIONcaptionsoff
  \newpage
\fi

\bibliographystyle{IEEEtran}
\bibliography{bibliography}

\begin{thebibliography}{10}
\providecommand{\url}[1]{#1}
\csname url@samestyle\endcsname
\providecommand{\newblock}{\relax}
\providecommand{\bibinfo}[2]{#2}
\providecommand{\BIBentrySTDinterwordspacing}{\spaceskip=0pt\relax}
\providecommand{\BIBentryALTinterwordstretchfactor}{4}
\providecommand{\BIBentryALTinterwordspacing}{\spaceskip=\fontdimen2\font plus
\BIBentryALTinterwordstretchfactor\fontdimen3\font minus
  \fontdimen4\font\relax}
\providecommand{\BIBforeignlanguage}[2]{{%
\expandafter\ifx\csname l@#1\endcsname\relax
\typeout{** WARNING: IEEEtran.bst: No hyphenation pattern has been}%
\typeout{** loaded for the language `#1'. Using the pattern for}%
\typeout{** the default language instead.}%
\else
\language=\csname l@#1\endcsname
\fi
#2}}
\providecommand{\BIBdecl}{\relax}
\BIBdecl

\bibitem{kolaczyk2014statistical}
E.~D. Kolaczyk and G.~Cs{\'a}rdi, \emph{Statistical analysis of network data
  with R}.\hskip 1em plus 0.5em minus 0.4em\relax Springer, 2014, vol.~65.

\bibitem{ortega2018graph}
A.~Ortega, P.~Frossard, J.~Kova{\v{c}}evi{\'c}, J.~M.~F. Moura, and
  P.~Vandergheynst, ``Graph signal processing: Overview, challenges, and
  applications,'' \emph{Proc. IEEE}, vol. 106, no.~5, pp. 808--828, 2018.

\bibitem{rey2019sampling}
S.~Rey, F.~J.~I. Garcia, C.~Cabrera, and A.~G. Marques, ``Sampling and
  reconstruction of diffused sparse graph signals from successive local
  aggregations,'' \emph{IEEE Signal Process. Lett.}, vol.~26, no.~8, pp.
  1142--1146, 2019.

\bibitem{wu2020probabilistic}
J.~Wu, C.~Ma, L.~Li, W.~Dong, and G.~Shi, ``Probabilistic undirected graph
  based denoising method for dynamic vision sensor,'' \emph{IEEE Trans.
  Multimedia}, vol.~23, pp. 1148--1159, 2020.

\bibitem{jordan1998learning}
M.~I. Jordan, \emph{Learning in graphical models}.\hskip 1em plus 0.5em minus
  0.4em\relax Springer Science \& Business Media, 1998, vol.~89.

\bibitem{djuric2018cooperative}
P.~Djuric and C.~Richard, \emph{Cooperative and Graph Signal Processing:
  Principles and Applications}.\hskip 1em plus 0.5em minus 0.4em\relax Academic
  Press, 2018.

\bibitem{shuman2013emerging}
D.~Shuman, S.~Narang, P.~Frossard, A.~Ortega, and P.~Vandergheynst, ``The
  emerging field of signal processing on graphs: Extending high-dimensional
  data analysis to networks and other irregular domains,'' \emph{IEEE Signal
  Process. Mag.}, vol.~30, no.~3, pp. 83--98, 2013.

\bibitem{bronstein2017geometric}
M.~M. Bronstein, J.~Bruna, Y.~LeCun, A.~Szlam, and P.~Vandergheynst,
  ``Geometric deep learning: going beyond euclidean data,'' \emph{IEEE Signal
  Process. Mag.}, vol.~34, no.~4, pp. 18--42, 2017.

\bibitem{tenorio2021robust}
V.~M. Tenorio, S.~Rey, F.~Gama, S.~Segarra, and A.~G. Marques, ``A robust
  alternative for graph convolutional neural networks via graph neighborhood
  filters,'' in \emph{Conf. Signals, Syst., Computers}.\hskip 1em plus 0.5em
  minus 0.4em\relax IEEE, 2021, pp. 1573--1578.

\bibitem{scarselli2008graph}
F.~Scarselli, M.~Gori, A.~Tsoi, M.~Hagenbuchner, and G.~Monfardini, ``The graph
  neural network model,'' \emph{IEEE Trans. Neural Netw.}, vol.~20, no.~1, pp.
  61--80, 2008.

\bibitem{gama2018convolutional}
F.~Gama, A.~G. Marques, G.~Leus, and A.~Ribeiro, ``Convolutional neural network
  architectures for signals supported on graphs,'' \emph{IEEE Trans. Signal
  Process.}, vol.~67, no.~4, pp. 1034--1049, 2019.

\bibitem{wu2020comprehensive}
Z.~Wu, S.~Pan, F.~Chen, G.~Long, C.~Zhang, and S.~Y. Philip, ``A comprehensive
  survey on graph neural networks,'' \emph{IEEE Trans. Neural Netw. Learn.
  Syst.}, 2020.

\bibitem{sakhavi2018learning}
S.~Sakhavi, C.~Guan, and S.~Yan, ``Learning temporal information for
  brain-computer interface using convolutional neural networks,'' \emph{IEEE
  Trans. Neural Netw. Learn. Syst.}, vol.~29, no.~11, pp. 5619--5629, 2018.

\bibitem{kipf2016semi}
T.~N. Kipf and M.~Welling, ``Semi-supervised classification with graph
  convolutional networks,'' \emph{arXiv preprint arXiv:1609.02907}, 2016.

\bibitem{li2018deeper}
Q.~Li, Z.~Han, and X.~Wu, ``Deeper insights into graph convolutional networks
  for semi-supervised learning,'' in \emph{AAAI Conf. Artif. Intell.}, 2018.

\bibitem{cui2019traffic}
Z.~Cui, K.~Henrickson, R.~Ke, and Y.~Wang, ``Traffic graph convolutional
  recurrent neural network: A deep learning framework for network-scale traffic
  learning and forecasting,'' \emph{IEEE Trans. Intell. Transp. Syst.},
  vol.~21, no.~11, pp. 4883--4894, 2019.

\bibitem{wang2017mgae}
C.~Wang, S.~Pan, G.~Long, X.~Zhu, and J.~Jiang, ``Mgae: Marginalized graph
  autoencoder for graph clustering,'' in \emph{ACM Conf. Inf. Knowl. Manag.},
  2017, pp. 889--898.

\bibitem{rey2019deep}
S.~Rey, V.~Tenorio, S.~Rozada, L.~Martino, and A.~G. Marques, ``Deep
  encoder-decoder neural network architectures for graph output signals,'' in
  \emph{Conf. Signals, Syst., Computers}.\hskip 1em plus 0.5em minus
  0.4em\relax IEEE, 2019, pp. 225--229.

\bibitem{rey2021overparametrized}
S.~Rey, V.~Tenorio, S.~Rozada, L.~Martino, and A.~G.~Marques,
  ``Overparametrized deep encoder-decoder schemes for inputs and outputs
  defined over graphs,'' in \emph{European Signal Process. Conf.}\hskip 1em
  plus 0.5em minus 0.4em\relax IEEE, 2021.

\bibitem{wang2018graphgan}
H.~Wang, J.~Wang, J.~Wang, M.~Zhao, W.~Zhang, F.~Zhang, X.~Xie, and M.~Guo,
  ``Graphgan: graph representation learning with generative adversarial nets,''
  in \emph{AAAI Conf. Artif. Intell.}, vol.~32, 2018.

\bibitem{liu2019learning}
W.~Liu, P.~Chen, F.~Yu, T.~Suzumura, and G.~Hu, ``Learning graph topological
  features via gan,'' \emph{IEEE Access}, vol.~7, 2019.

\bibitem{schaub2021signal}
M.~T. Schaub, Y.~Zhu, J.-B. Seby, T.~M. Roddenberry, and S.~Segarra, ``Signal
  processing on higher-order networks: Livin' on the edge... and beyond,''
  \emph{Signal Process.}, vol. 187, p. 108149, 2021.

\bibitem{roddenberry2021principled}
T.~M. Roddenberry, N.~Glaze, and S.~Segarra, ``Principled simplicial neural
  networks for trajectory prediction,'' in \emph{Int. Conf. Mach. Learn.},
  2021.

\bibitem{ulyanov2018deep}
D.~Ulyanov, A.~Vedaldi, and V.~Lempitsky, ``Deep image prior,'' in \emph{IEEE
  Conf. Comput. Vision Pattern Recog.}, 2018, pp. 9446--9454.

\bibitem{heckel2018deep}
R.~Heckel and P.~Hand, ``Deep decoder: Concise image representations from
  untrained non-convolutional networks,'' in \emph{Intl. Conf. Learn. Repr.},
  2018.

\bibitem{liu2018generalized}
S.~Liu, M.~Long, J.~Wang, and M.~I. Jordan, ``Generalized zero-shot learning
  with deep calibration network,'' in \emph{Adv. Neural Inf. Process. Syst.},
  2018, pp. 2005--2015.

\bibitem{yaman2021zero}
B.~Yaman, S.~A.~H. Hosseini, and M.~Ak{\c{c}}akaya, ``Zero-shot self-supervised
  learning for mri reconstruction,'' in \emph{Int. Conf. Learning
  Representations}, 2021.

\bibitem{segarra2017optimal}
S.~Segarra, A.~G. Marques, and A.~Ribeiro, ``Optimal graph-filter design and
  applications to distributed linear network operators,'' \emph{IEEE Trans.
  Signal Process.}, vol.~65, no.~15, pp. 4117--4131, 2017.

\bibitem{jain_1988_algorithms}
A.~K. Jain and R.~C. Dubes, \emph{Algorithms for Clustering Data}.\hskip 1em
  plus 0.5em minus 0.4em\relax Prentice Hall Englewood Cliffs, NJ, 1988,
  vol.~6.

\bibitem{carlsson_2018_hierarchical}
G.~Carlsson, F.~Memoli, A.~Ribeiro, and S.~Segarra, ``Hierarchical clustering
  of asymmetric networks,'' \emph{Adv. Data Anal. Classification}, vol.~12,
  no.~1, pp. 65--105, Mar 2018.

\bibitem{carlsson_2013_axiomatic}
G.~Carlsson, F.~Mémoli, A.~Ribeiro, and S.~Segarra, ``Axiomatic construction
  of hierarchical clustering in asymmetric networks,'' in \emph{IEEE Int. Conf.
  Acoustics, Speech and Signal Process.}, 2013, pp. 5219--5223.

\bibitem{do2020graph}
T.~H. Do, D.~M. Nguyen, and N.~Deligiannis, ``Graph auto-encoder for graph
  signal denoising,'' in \emph{IEEE Int. Conf. Acoustics, Speech and Signal
  Process.}, 2020, pp. 3322--3326.

\bibitem{rey2019underparametrized}
S.~Rey, A.~G. Marques, and S.~Segarra, ``An underparametrized deep decoder
  architecture for graph signals,'' in \emph{IEEE Intl. Wrksp. Computat. Adv.
  Multi-Sensor Adaptive Process.}\hskip 1em plus 0.5em minus 0.4em\relax IEEE,
  2019, pp. 231--235.

\bibitem{mataev2019deepred}
G.~Mataev, P.~Milanfar, and M.~Elad, ``Deepred: Deep image prior powered by
  red,'' in \emph{IEEE/CVF Intl. Conf. Comput. Vision Wrksp.}, 2019.

\bibitem{heckel2019denoising}
R.~Heckel and M.~Soltanolkotabi, ``Denoising and regularization via exploiting
  the structural bias of convolutional generators,'' in \emph{iclr}, 2020.

\bibitem{chen2014signal}
S.~Chen, A.~Sandryhaila, J.~M.~F. Moura, and J.~Kovacevic, ``Signal denoising
  on graphs via graph filtering,'' in \emph{IEEE Global Conf. Signal and Info.
  Process.}\hskip 1em plus 0.5em minus 0.4em\relax IEEE, 2014, pp. 872--876.

\bibitem{wang2015trend}
Y.~Wang, J.~Sharpnack, A.~Smola, and R.~Tibshirani, ``Trend filtering on
  graphs,'' in \emph{Artif. Intell. Statistics}.\hskip 1em plus 0.5em minus
  0.4em\relax PMLR, 2015, pp. 1042--1050.

\bibitem{ono2015total}
S.~Ono, I.~Yamada, and I.~Kumazawa, ``Total generalized variation for graph
  signals,'' in \emph{IEEE Int. Conf. Acoustics, Speech and Signal
  Process.}\hskip 1em plus 0.5em minus 0.4em\relax IEEE, 2015, pp. 5456--5460.

\bibitem{pang2017graph}
J.~Pang and G.~Cheung, ``Graph laplacian regularization for image denoising:
  Analysis in the continuous domain,'' \emph{IEEE Trans. Signal Inf. Process.
  Netw.}, vol.~26, no.~4, pp. 1770--1785, 2017.

\bibitem{onuki2016graph}
M.~Onuki, S.~Ono, M.~Yamagishi, and Y.~Tanaka, ``Graph signal denoising via
  trilateral filter on graph spectral domain,'' \emph{IEEE Trans. Signal Inf.
  Process. Netw.}, vol.~2, no.~2, pp. 137--148, 2016.

\bibitem{tay2020time}
D.~Tay and J.~Jiang, ``Time-varying graph signal denoising via median
  filters,'' \emph{IEEE Trans. Circuits Syst., II, Exp. Briefs}, 2020.

\bibitem{segarra2017designmedian}
S.~Segarra, A.~G. Marques, G.~R.~Arce, and A.~Ribeiro, ``Design of weighted
  median graph filters,'' in \emph{IEEE Intl. Wrksp. Computat. Adv.
  Multi-Sensor Adaptive Process.}, 2017, pp. 1--5.

\bibitem{dorfler2012kron}
F.~Dorfler and F.~Bullo, ``Kron reduction of graphs with applications to
  electrical networks,'' \emph{IEEE Trans. Circuits Syst. I, Reg. Papers},
  vol.~60, no.~1, pp. 150--163, 2012.

\bibitem{chen2021graph}
S.~Chen, Y.~C. Eldar, and L.~Zhao, ``Graph unrolling networks: Interpretable
  neural networks for graph signal denoising,'' \emph{IEEE Trans. Signal
  Process.}, 2021.

\bibitem{marques2020signal}
A.~G. Marques, S.~Segarra, and G.~Mateos, ``Signal processing on directed
  graphs: The role of edge directionality when processing and learning from
  network data,'' \emph{IEEE Signal Process. Mag.}, vol.~37, no.~6, 2020.

\bibitem{sandryhaila2013discrete}
A.~Sandryhaila and J.~M.~F. Moura, ``Discrete signal processing on graphs,''
  \emph{IEEE Trans. Signal Process.}, vol.~61, no.~7, pp. 1644--1656, 2013.

\bibitem{sandryhaila2014discrete}
------, ``Discrete signal processing on graphs: Frequency analysis,''
  \emph{IEEE Trans. Signal Process.}, vol.~62, no.~12, pp. 3042--3054, 2014.

\bibitem{chen2015discrete}
S.~Chen, R.~Varma, A.~Sandryhaila, and J.~Kova{\v{c}}evi{\'c}, ``Discrete
  signal processing on graphs: Sampling theory,'' \emph{IEEE Trans. Signal
  Process.}, vol.~63, no.~24, pp. 6510--6523, 2015.

\bibitem{chen2020measuring}
D.~Chen, Y.~Lin, W.~Li, P.~Li, J.~Zhou, and X.~Sun, ``Measuring and relieving
  the over-smoothing problem for graph neural networks from the topological
  view,'' in \emph{AAAI Conf. Artif. Intell.}, vol.~34, no.~04, 2020, pp.
  3438--3445.

\bibitem{daniely2016toward}
A.~Daniely, R.~Frostig, and Y.~Singer, ``Toward deeper understanding of neural
  networks: The power of initialization and a dual view on expressivity,'' in
  \emph{Adv. Neural Inf. Proc. Syst.}, 2016, pp. 2253--2261.

\bibitem{newman2018networks}
M.~Newman, \emph{Networks}.\hskip 1em plus 0.5em minus 0.4em\relax Oxford
  {U}niversity {P}ress, 2018.

\bibitem{schaub2020blind}
M.~T. Schaub, S.~Segarra, and J.~N. Tsitsiklis, ``Blind identification of
  stochastic block models from dynamical observations,'' \emph{SIAM J. Math.
  Data Sc.}, vol.~2, no.~2, pp. 335--367, 2020.

\bibitem{watts1998collective}
D.~J. Watts and S.~H. Strogatz, ``Collective dynamics of
  ‘small-world’networks,'' \emph{nature}, vol. 393, no. 6684, pp. 440--442,
  1998.

\bibitem{zachary1977information}
W.~W. Zachary, ``An information flow model for conflict and fission in small
  groups,'' \emph{J. Anthrop. Res.}, vol.~33, no.~4, pp. 452--473, 1977.

\bibitem{temperatures2020}
``National centers for environmental information,'' \emph{[Online]. Available:
  https://www.ncei.noaa.gov/data/global-summary-of-the-day}, 2020.

\bibitem{watts1999networks}
D.~J. Watts, ``Networks, dynamics, and the small-world phenomenon,''
  \emph{Amer. J. Sociology}, vol. 105, no.~2, pp. 493--527, 1999.

\bibitem{holme2002growing}
P.~Holme and B.~J. Kim, ``Growing scale-free networks with tunable
  clustering,'' \emph{Physical review E}, vol.~65, no.~2, p. 026107, 2002.

\bibitem{velivckovic2017graph}
P.~Veli{\v{c}}kovi{\'c}, G.~Cucurull, A.~Casanova, A.~Romero, P.~Lio, and
  Y.~Bengio, ``Graph attention networks,'' in \emph{Int. Conf. Learning
  Representations}, 2018.

\bibitem{cardoso2020algorithms}
J.~V. D.~M. Cardoso, J.~Ying, and D.~P. Palomar, ``Algorithms for learning
  graphs in financial markets,'' \emph{arXiv preprint arXiv:2012.15410}, 2020.

\bibitem{yu2015useful}
Y.~Yu, T.~Wang, and R.~J. Samworth, ``A useful variant of the {D}avis--{K}ahan
  theorem for statisticians,'' \emph{Biometrika}, vol. 102, no.~2, pp.
  315--323, 2015.

\end{thebibliography}

\begin{IEEEbiography}[{\includegraphics[width=1in,height=1.25in,clip,keepaspectratio]{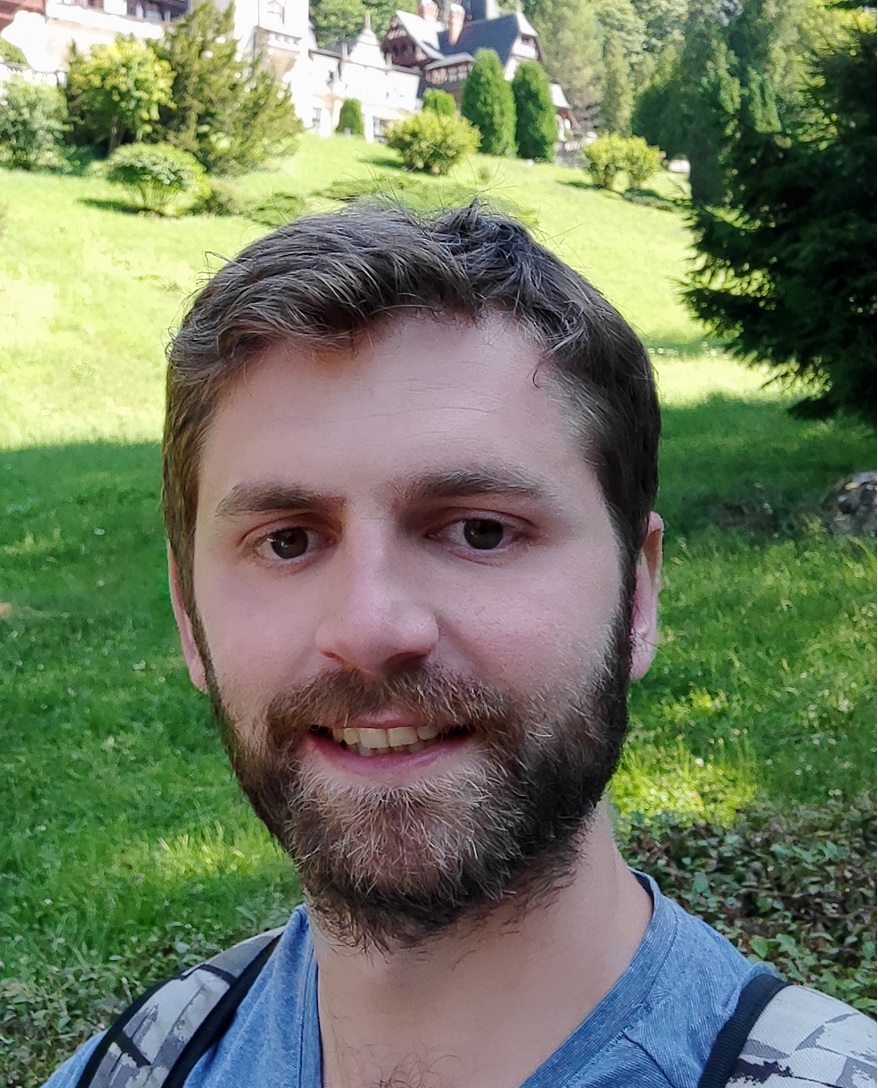}}]{Samuel Rey} (Student Member, IEEE)
received the degree in Telecommunication Engineering in 2016 and the M.Sc. in Telecommunications Engineering in 2018, both with highest honors, from King Juan Carlos University (URJC), Madrid, Spain. He is currently working towards his Ph.~D. thesis with the Department of Signal Theory and Communications of King Juan Carlos University.
His current research focuses are graph signal processing, graph neural networks, non-convex optimization, and data science over networks. 
He received the “Best Young Investigator Award” across all M. Sc. students at URJC in 2018. He was awarded with the Spanish Federal FPU Scholarship for Ph.~D. studies in 2018, and with the Mobility Grant for Ph.~D. FPU students in 2021.
\end{IEEEbiography}

\begin{IEEEbiography}[{\includegraphics[width=1in,height=1.25in,clip,keepaspectratio]{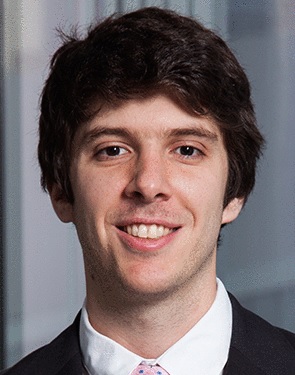}}]{Santiago Segarra} (Senior Member, IEEE)
received the B.Sc. degree in Industrial Engineering with highest honors (Valedictorian) from the Instituto Tecnológico de Buenos Aires (ITBA), Argentina, in 2011, the M.Sc. in Electrical Engineering from the University of Pennsylvania (Penn), Philadelphia, in 2014 and the Ph.D. degree in Electrical and Systems Engineering from Penn in 2016.
From September 2016 to June 2018 he was a postdoctoral research associate with the Institute for Data, Systems, and Society at the Massachusetts Institute of Technology. Since July 2018, Dr. Segarra is a W. M. Rice Trustee Assistant Professor in the Department of Electrical and Computer Engineering at Rice University.
His research interests include network theory, data analysis, machine learning, and graph signal processing. Dr. Segarra received the ITBA’s 2011 Best Undergraduate Thesis Award in Industrial Engineering, the 2011 Outstanding Graduate Award granted by the National Academy of Engineering of Argentina, the 2017 Penn’s Joseph and Rosaline Wolf Award for Best Doctoral Dissertation in Electrical and Systems Engineering, the 2020 IEEE Signal Processing Society Young Author Best Paper Award, the 2021 Rice’s School of Engineering Research + Teaching Excellence Award, and five best conference paper awards.
\end{IEEEbiography}

\begin{IEEEbiography}[{\includegraphics[width=1in,height=1.25in,clip,keepaspectratio]{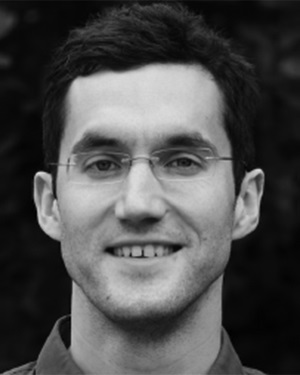}}]{Reinhard Heckel} (Member, IEEE)
received the Ph.D. degree in electrical engineering from ETH Zurich. He was a Visiting Ph.D. Student with the Department of Statistics, Stanford University. He is currently a Rudolf Moessbauer Assistant Professor with the Department of Electrical and Computer Engineering (ECE), Technical University of Munich, and an Adjunct Assistant Professor with the Department of Electrical and Computer Engineering (ECE), Rice University, where he was an Assistant Professor, from 2017 to 2019. Before that, he was a Post-Doctoral Scholar with UC Berkeley—sharing an office with Ilan Shomorony—and a Researcher with the Cognitive Computing and Computational Sciences Department, IBM Research Zurich.. He is working in the intersection of machine learning and signal/information processing with a current focus on deep networks for solving inverse problems, learning from few and noisy samples, and DNA data storage.
\end{IEEEbiography}

\begin{IEEEbiography}[{\includegraphics[width=1in,height=1.25in,clip,keepaspectratio]{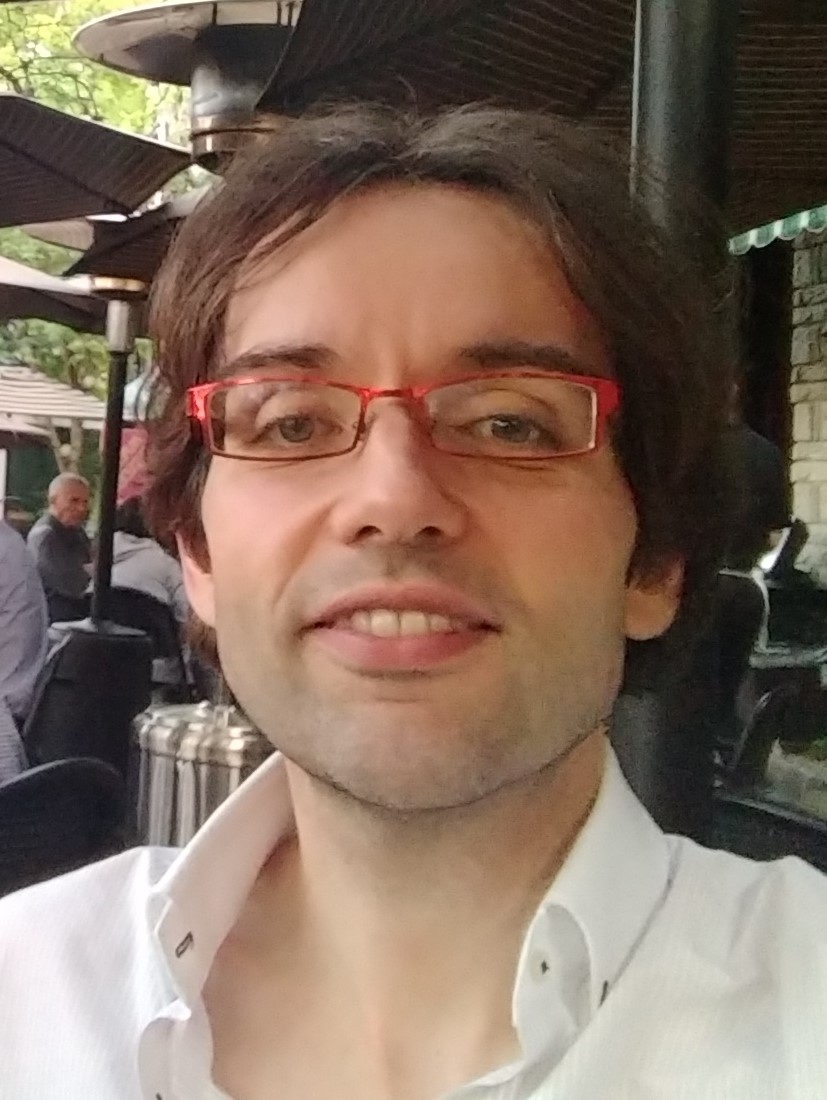}}]{Antonio G. Marques} (Senior Member, IEEE)
received the Telecommunications Engineering degree and the Doctorate degree, both with highest honors, from Carlos III University of Madrid, Spain, in 2002 and 2007, respectively. In 2007, he became a faculty of the Department of Signal Theory and Communications, King Juan Carlos University, Madrid, Spain, where he currently develops his research and teaching activities as a full professor. From 2005 to 2015, he held different visiting positions at the University of Minnesota, Minneapolis. In 2015, 2016 and 2017 he was a visitor scholar at the University of Pennsylvania, Philadelphia. 
His current research focuses on signal processing, machine learning, data science and artificial intelligence over graphs, and nonlinear and stochastic optimization of wireless, power and transportation networks. 
Dr. Marques has served the IEEE in a number of posts, including as an associate editor and the technical / general chair of different conferences, and, currently, he is a Senior Area Editor of the IEEE Transactions on Signal Process. a member of the IEEE Signal Process. Theory and Methods Tech. Comm. His work has been awarded in several journals, conferences and workshops, with recent ones including IEEE SSP 2016, IEEE SAM 2016, IEEE SPS IEEE Y.A. Best Paper Award 2020, and CIT 2021. He is the recipient of the ``2020 EURASIP Early Career Award'' and a member of IEEE, EURASIP and the ELLIS society.
\end{IEEEbiography}

\vfill

\end{document}